\documentclass[%
reprint,
 amsmath,amssymb,
 aps,
prb,
]{revtex4-2}

\newcommand{\mi}{{\mathrm i}}
\newcommand{\cmt}[2]{{[}#1,#2{]}}
\newcommand{\acmt}[2]{{\{}#1,#2{\}}}

\newcommand{\eas}[0]{\begin{eqnarray*}}
\newcommand{\eae}[0]{\end{eqnarray*}}
\newcommand{\les}[0]{\begin{equation}}
\newcommand{\lee}[0]{\end{equation}}
\newcommand{\leas}[0]{\begin{eqnarray}}
\newcommand{\leae}[0]{\end{eqnarray}}
\newcommand{\mchss}[4]
{
\left\{
\begin{array}{cc}
#1 & #2   \\
#3 & #4
\end{array}
\right.
}
\newcommand{\mchsss}[6]
{
\left\{
\begin{array}{cc}
  #1 & #2   \\
  #3 & #4   \\
#5 & #6
\end{array}
\right.
}
\newcommand{\mchsssl}[6]
{
\left\{
\begin{array}{cl}
  #1 & #2   \\
  #3 & #4   \\
#5 & #6
\end{array}
\right.
}

\newcommand{\matthree}[9]
{
\left[
\begin{array}{ccc}
#1 & #2 & #3 \\
#4 & #5 & #6 \\
#7 & #8 & #9 
\end{array}
\right]
}
\newcommand{\mvecfour}[4]
{
\left(
\begin{array}{c}
#1  \\
#2  \\
#3  \\
#4  
\end{array}
\right)
}

\newcommand{\local}{{G}}
\newcommand{\largeG}{{LG}}
\newcommand{\SUQ}{{SU(Q)}}
\newcommand{\QN}{{Q}}
\newcommand{\tw}{{\rm tw}}
\newcommand{\op}{{\rm op}}
\newcommand{\pe}{{\rm pe}}

\usepackage{xcolor}

\newcommand{\cbin}[1]{{\color{blue}#1}}
\renewcommand{\cbin}[1]{}

\usepackage{graphicx}
\usepackage{dcolumn}
\usepackage{bm}

\usepackage[subrefformat=simple]{subcaption}
\begin{document}


\title{
    Topological pump of $\SUQ$ quantum chain and
  Diophantine equation
}%
\author{Yasuhiro Hatsugai}
 \homepage{https://patricia.ph.tsukuba.ac.jp/~hatsugai}
\affiliation{
 Department of Physics, University of Tsukuba,
 Tsukuba 305-8571
}%
\author{Yoshihito Kuno}
\affiliation{
 Graduate School of Engineering Science, Akita University, Akita 010-8502
}%

\date{\today}

\begin{abstract}
  A topological pump of the $\SUQ$ {quantum} chain is proposed 
  associated with a current due to a local $[U(1)]^{\otimes \QN}$
  gauge invariance of colored fermions.
  The $\SUQ$ invariant dimer phases
  are characterized by
  the $Z_\QN$ Berry phases as
  a topological order parameter
  with a $d$-dimensional
  twist space ($d=\QN-1$) as a synthetic Brillouin zone.
  By inclusion of the symmetry breaking perturbation
specified by a rational parameter $\Phi=P/Q$, 
the pump, that encloses around the phase boundary,
is characterized by the $\QN$ Chern numbers
associated with the currents due to
uniform infinitesimal twists.
The analysis of the systems under the open/periodic/twisted boundary conditions
clarifies the bulk-edge correspondence of the pump
where the large gauge transformation generated by
the center of mass (CoM)
plays a central role.
An explicit formula for the Chern number
is given by using the Diophanine equation.

Numerical demonstration by the exact diagonalization and the DMRG for
finite systems ($\QN=3,4$ and $5$) have been
presented to confirm the general discussions for low energy spectra,
edge states, CoM's, Chern numbers and the bulk-edge correspondence.
  A modified Lieb-Schultz-Mattis type argument for the general $\SUQ$
  quantum chain is also mentioned. 
\end{abstract}

\maketitle


\section{Introduction}
\label{sec:introduction}
$U(1)$ gauge invariance
is a key ingredient for the quantum Hall effects as pointed out by Laughlin
 \cite{Laughlin81}
and is true for the Chern insulators
where
the conserved
current associated with the gauge field is closely related with the
topological numbers:
the Chern number of the gapped many-body state
and the TKNN integers of the one-particle bands
 \cite{TKNN,Avron83,NTW,Kohmoto85,haldaneM88}.
It guarantees topological stability of the phase
without any further
symmetry protection.
Existence of the non-trivial phases with non-zero
Chern number is only allowed with time-reversal symmetry breaking
that suggests chiral nature of the phase.

As for most of the topological phases, the bulk is hidden,
in a sense that the topological number of the bulk without boundaries is not
a physical observable.
What have been observed experimentally are low energy localized modes as
the chiral edge states localized
near the boundaries \cite{Halperin82,MacD84,Hatsugai93Rm}.
The edge states reflect the topological number of the bulk
as the bulk-edge correspondence \cite{Hatsugai93Chern,qibec06,Schulz99,graf13,prodan16,mathai16,Tauber20,yatsugi22}
where $U(1)$ local gauge field associated with the Aharonov-Bohm flux
is crucial due to the Laughlin argument.
The effective theory also justifies the bulk-edge correspondence for
topologically non-trivial systems \cite{wen90}.

Focused studies in the decades reveal 
various chiral modes in quantum and
non-quantum phenomena in quite different phenomena
have a topological origin and under the control of the bulk-edge correspondence
where the
Chern number of the bulk predicts the
direction and the number of the chiral modes.
The first non-quantum example can be 
one-way propagating modes of
a gyromagnetic photonic crystal \cite{HR08,Wang08,Phexp09},
mechanical chiral modes in 
microtubes \cite{PrPr09} and coupled optical resonators \cite{Hafezi11}
are also governed by
the bulk-edge correspondence. 
The concept is applied for a wide variety of phenomena
in
photonics \cite{ochiai12,hu15,ozawa19} and topological circuits \cite{Ningyuan15,Albert15}.
It also includes mechanical systems \cite{Nash15,huber15}
and cold atoms \cite{Fallani15}.
The bulk-edge correspondence is also
a key concept in the focused studies of
topological insulators in the decades \cite{KaneMele05,bern06,konig07,moore09,hassan10}.
One of the recent surprises is that
equatorial waves 
near the equator of the
earth,
that is well known in geophysics,
are the chiral edge modes associated with the non zero-Chern number \cite{PD17}.
There exist chiral edge modes
in evolutionary game theory \cite{yoshida-game21} and
biological flows in 
neural progenitor cells \cite{kgc20}.
The bulk-edge correspondence is universal.
In these classical phenomena, chiral edge states are topologically stable and
protected by
the bulk gap, although the Chern numbers of the bulk are never observed but
guarantee the existence of the edge states.
Even in these phenomena, one can introduce $U(1)$ gauge fields
to the governing equation
by the minimal coupling to the spatial derivative.
The 
$U(1)$ gauge field is fictitious and never observed but it
predicts the chiral edge states associated with the Laughlin argument,
which implies
the bulk-edge correspondence in classical systems as well.

The local $U(1)$ gauge invariance in one-dimension implies
an adiabatic charge transport of a gapped quantum chain
associated with time as an additional synthetic dimension.
This is a topological pump originally proposed by
Thouless \cite{Thouless83,NiuThouless84,avron88,Brouwer98,alt99} in (1+1) dimensions
where the transported charge is quantized as is written
by the Chern number.
 Similar pumping for other degrees of freedom have been
 also proposed \cite{shindou05,Berg11,Kraus12}.
Here we do not need time-reversal symmetry breaking in the 1D system
 for the non-trivial topological pump. 
This idea of the topological pump
is old but only
after more than three decades' experimental trials, the topological pump
has been finally realized in 
cold atom experiments \cite{Nakajima16,Lohse16}.
The pump is real and topological. These discoveries motivate
to clarify the effects of edge states in the
topological pump \cite{Hatsugai16pm}.
 The topological pump is
an adiabatic transfer of the charge.
As for an open system with boundaries, the adiabatic cycle of the
pump implies everything is going back to the original state after
the period. It implies nothing is transported in total.
 What occurs
is that the contributions due to the bulk and edges are cancelled.
That is, the pumped charge due to bulk is given by a back action
of the edge states.
In a suitable normalization,
a localization length of the edge states scales to zero
in a large system size limit.
It implies a quantization of contribution due to edge states
in a large system.
It guarantees quantization of the pumped charge due to bulk. 
This is the key idea
of a topological nature of the topological pump and
the bulk-edge correspondence of the pump \cite{Hatsugai16pm}.
The bulk-edge correspondence is special in the topological pump, that is,
the edge is hidden and the bulk gives a physical observable,
a center of mass that is a time integral of the
current \cite{wang13,Naka18,Hatsugai16pm,watanabe18}.
The pumped $U(1)$ charge due to the edge states is never
experimentally observed due to the gapless nature of the edge states.
It implies the breakdown of the adiabaticity. In other words,
contribution of the edge states can not be measured experimentally in
a finite speed pump since the adiabatic condition can not be satisfied.
What is measured is that of the bulk.
In the experiment, one measures a motion of the center of mass of the system.
Its derivative is the current.

 Recently this bulk-edge correspondence of the topological pump is also investigated for interacting fermions \cite{Naka18,Kuno20charge}, quantum spins \cite{Schweizer16,Kuno21spin} and bosons \cite{Zeng16,Goz,sheng16,Kuno21boson}. This cancellation mechanism is  applied for  fractional quantum Hall states \cite{Kudo21adia}.
A phase transition point 
  between gapped
  symmetry protected topological phases (SPT) \cite{Gu09,Pollmann10}
  is the source of non-trivial topology.
  This topological transition  is characterized by
  the quantized
  Berry phases \cite{Hatsugai06qb,Hirano08,Chepiga16,Kariyado18,Fubasami19,araki20}.
  The role of the edge states of spin pumping
  is also discussed in a mathematically rigorous way
  for the AKLT Hamiltonian \cite{aklt} and its modifications \cite{tasaki22}.
Note that experimental studies for the topological pump using cold atoms
are rapidly developing \cite{Fabre22,Li22,citro22}.
 
We here investigate topological pump of the
$\SUQ$ {quantum} chain. It is surprising rich
and the bulk-edge correspondence clarifies all details of the rich structures.
One of the surprises is the Diophantine equation,
that has been successfully explains
the TKNN integer of the quantum Hall effect (Harper equation) on a lattice,
also used to explain the Chern numbers of the $\SUQ $ pump analytically.

 The paper 
 is organized as follows.
 After the introduction,
 in Sec.\ref{sec:suq}, the $\SUQ$ quantum chain due to Affleck
 is described as
 a generalization of the $S=1$ bilinear-biquadratic {quantum} chain.
 By using a colored fermion representation, symmetries of the system
 are described. Especially $Z_\QN$ and large gauge transformation due to
 $\QN$ gauge symmetries
 are introduced. A gapped $\SUQ$ symmetric dimer phase is discussed
 for a periodic system that is a source of the non-trivial topological pump.
 By using a time as a synthetic dimension and introducing the 
 $\SUQ$ symmetry breaking term, 
an $\SUQ$ topological pump is proposed, that goes around the gap closing
 dimer transition of the $\SUQ$ {quantum} chain. 
 In Sec.\ref{sec:current}, $\QN$ currents associated with the 
 gauge symmetries are introduced and the center of mass (CoM),
 that generates the large gauge transformation is defined
 where open/twisted/periodic boundary conditions are carefully discussed.
 The $d=\QN+1$ dimensional
 synthetic Brillouin zone as a parameter space to define
 the current and $Q$-closed paths passing through the
 $Z_\QN$ symmetric point are introduced where the averaged currents
 along the paths are used to define the topological pump in
 the adiabatic approximation.
 The averaged CoM along the path for the open boundary condition
 is not continuous and a topological number of the edge states is
 defined by using
 the discontinuities.
 The bulk-edge correspondence is proposed for the $\QN$ different topological
 numbers of bulk and edges.
 In Sec.\ref{sec:qBerryGap},
 $Z_\QN$ quantization of the Berry phases
 is defined by the loop in the synthetic Brillouin zone and
discussed in details. A modified Lieb-Schultz-Mattis type
 argument is also given associated with the (anti)-translational symmetry
 of the uniform problem.
 In Sec.\ref{sec:emergent}, $Z_\QN$ symmetry of systems
 with odd number of sites are discussed in relation to the
 edge states with numerical justifications.
 Also $Z_\QN\times Z_\QN$ emergent symmetry for systems in the open
 systems after taking an infinite size limit is discussed
 based on the numerical calculations of
 low energy spectra. The topological numbers of
 the system with edges are given due to this emergent symmetry.
 Based on the bulk-edge correspondence,
 an explicit analytical formula of the $\QN$ Chern numbers
 of bulk is given by using the Diophantine equation of the TKNN for
 the quantum Hall effect on the lattice.
 In Sec.\ref{sec:num}, numerical evaluation of
 the topological numbers of edges and bulk are explicitly given
 by using the exact diagonalization and DMRG calculation
 for $\QN=3,4$ and $5$ systems,
 that justify the consistency of the discussion.

 
\section{$\SUQ$ quantum chain}
\label{sec:suq}
\subsection{Fermion representation}
\label{sec:fermion}

Let us start considering an $S=1$ quantum spin chain
with nearest neighbor bilinear-biquadratic interaction
\begin{align*}
 H_S(\omega_S ) &= \sum_{j} 
 \big[\cos \omega_S (\bm{S} _j\cdot \bm{S} _{j+1})
  + \sin \omega_S (\bm{S} _j\cdot \bm{S} _{j+1})^2\big],
\end{align*}
where $\cmt{S_{j,\alpha}}{S_{j,\beta} } = i \sum_{\gamma }\epsilon_{\alpha \beta \gamma } S_{j, \gamma }$,
($\alpha,\beta =1,2,3$)
and $\bm{S}_j^2= S(S+1)$, ($S=1$). 
It has a long history of study \cite{Chubukov91,Fathe95,kawashima02,Lauchli06,Yang22}.
We discuss its
non-uniform 
$\SUQ$ extension 
by a fermion representation due to Affleck \cite{AFFLECK1986409,AFFLECK1988582,PhysRevLett.54.966,PhysRevB.55.8295} (See also appendix \ref{sec:color})
\begin{align}
 \bm{S} _j\cdot \bm{S} _{j+1} &= H^{(1)}(\{1\})\big|_{\QN=3}+ (const),
 \\
 (\bm{S} _j\cdot \bm{S} _{j+1})^2 &=  H^{(2)}(\{1\})\big|_{\QN=3} + (const),
 \label{eq:Ham12}
\end{align}
where $H^{(1,2)}$ are defined for general $\QN$ as
\begin{align} 
H^{(1,2)}(\{J^{(1,2)}_{j,\alpha \beta }\}) &= \sum_{j} h^{(1,2)}_{j,j+1},
\nonumber \\
 h_{j,j+1}^{(1)}
 &= \sum_{\alpha \beta }^\QN
 J^{(1)}_{j,\alpha\beta}
c_{j, \alpha } ^\dagger
c_{j+1, \beta } ^\dagger 
 c_{j+1, \alpha   }
 c_{j, \beta }
 \nonumber \\
 &= \sum_{\alpha \beta }^\QN
J^{(1)}_{j,\alpha\beta}
  h^{ex}_{j,\alpha ;j+1,\beta },
 \label{eq:Ham1}
\\
 h_{j,j+1}^{(2)}
 &= \sum_{\alpha \beta }^\QN
 J^{(2)}_{j,\alpha\beta}
 c_{j, \alpha } ^\dagger
 c_{j+1 ,\alpha } ^\dagger
 c_{j+1, \beta  }
 c_{j, \beta }
 \nonumber \\
 &= \sum_{\alpha \beta }^\QN
 J^{(2)}_{j,\alpha\beta} h^{ph}_{j,\alpha ; j \beta },
 \\
 h^{ph}_{j,\alpha ; j \beta } &=  
 \psi _{j,\alpha ;j+1,\alpha} ^\dagger
  \psi _{j,\beta ;j+1,\beta } ,
  \label{eq:Ham2}
\end{align}
where
$({\bm{J}^{(1,2)}_{j}}) ^\dagger = \bm{J}^{(1,2)}_{j} $,
$\big(\bm{J}^{(1,2)}_{j} \big)_{\alpha, \beta }
=
J^{(1,2)}_{j,\alpha \beta }$
and
$c_{j,\alpha }$, ($\alpha =1,\cdots,\QN$) is a canonical
fermion annihilation operator for a color $\alpha =1,\cdots,\QN$ at the site $j$, 
$\acmt{c_{j,\alpha } }{c_{j ^\prime ,\beta } ^\dagger }=\delta _{jj ^\prime }\delta_{\alpha \beta } $
with a constraint 
$\sum_\alpha \hat n_{j,\alpha} =1$, $\hat n _{j,\alpha} =c_{j, \alpha }^\dagger c_{j,\alpha } $
at each site $^\forall j$.
 The exchange of colors, $h^{ex} _{i,\alpha ;j, \beta }$ 
 and the pair hopping, $h^{ph}_{i,\alpha ;j, \beta }$ 
 at the link $i,j$
 are defined as 
 \begin{align} 
 h^{ex}_{i,\alpha ;j,\beta } &= 
  c_{i, \alpha } ^\dagger
c_{j, \beta } ^\dagger 
 c_{j, \alpha   }
 c_{i, \beta },
\\
h^{ph}_{i,\alpha ;j,\beta } &=
\psi_{i,\alpha ;j,\alpha } ^\dagger \psi_{i,\beta ;j,\beta} ,
 \end{align}
 where
$\psi_{i,\alpha ;j,\beta} = c_{i,\alpha }c_{j,\beta }$ 
is 
a pairing amplitude.
They operate as
 \begin{align*}
 h^{ex}_{i,\alpha ; j ,\beta } |\beta _i \alpha _{j} \rangle &=
  |\alpha _i \beta _{j} \rangle,
 \\
 h^{ph}_{i,\alpha ; j, \beta } |\beta _i \beta _j \rangle &=
 |\alpha _i \alpha _j \rangle ,
\end{align*}
where 
$  | \alpha_i \beta_{j} \rangle = c_{i,\alpha}^\dagger c_{j,\beta }^\dagger 
|0 \rangle $.

  Up to constant, $H_S$ reduces to the sum of $H^{(1)}$ and $H^{(2)}$ 
  when $Q=3$. 
The spin-1 operators at the site $j$ is
 written by a generator of the $SO(3)$ spatial rotation as
$S_{j,\alpha } =\sum_{\beta \gamma } c_{j,\beta } ^\dagger S^\alpha_{\beta\gamma } c_{j,\gamma }$, where
$S^\alpha_{\beta\gamma } =-i \epsilon_{\alpha \beta \gamma } $, ($\alpha =1,2,3)$.
See also appendix \ref{sec:color}.
 
As for the boundary condition, we  discuss both of
the open boundary condition 
and the periodic boundary condition ($c_{L+1,\alpha }\equiv c_{1,\alpha }$) assuming the lattice sites
are labeled as $j=1,\cdots,L$ unless otherwise specified.
We discuss each of the $H^{(1,2)}$ separately or $H$ at $\omega_S =0, \frac {\pi}{2} $ since the transformation properties are different. 

\subsection{$\SUQ$ and $Z_\QN$ Symmetries}
\label{sec:symm}
When the coupling is color independent,
 $J^{(1,2)}_{j,\alpha \beta }=J_j^{(1,2)}$, 
the Hamiltonian $H^{(1,2)}$
is invariant for the global $SU(Q)$ transformation respectively,
\begin{align*}
 {\cal U}^{(1)} H^{(1)}( {\cal U}^{(1)} ) ^\dagger  &= H^{(1)},
 \\
 {\cal U}^{(2)} H^{(2)}( {\cal U}^{(2)} ) ^\dagger  &= H^{(2)},
\end{align*}
\begin{align}
 {\cal U}^{(1)} &= e^{ -i\sum_{j,\alpha \beta } c_{j,\alpha  } ^\dagger u_{\alpha \beta }c_{j,\beta }},
 \\
   {\cal U}^{(2)} &= e^{ -i\sum_{j,\alpha \beta } (-1)^{j-1} c_{j,\alpha } ^\dagger u_{\alpha \beta }c_{j,\beta }},
   \label{eq:suq}
\end{align}
where $u$ is a generator of $Q\times Q $ traceless hermitian matrix
($ {\rm Tr}\, u=0$ and $u^\dagger =u$)
\footnote{
When $Q=3$, it is spanned by the Gell-Mann matrices
\begin{align*}
 \lambda _1 &= \matthree
     {0}{1}{0}
     {1}{0}{0}
     {0}{0}{0},
 \lambda _2 = \matthree
     {0}{-i}{0}
     {i}{0}{0}
     {0}{0}{0}=S^3,
 \lambda _3 = \matthree
     {1}{0}{0}
     {0}{-1}{0}
     {0}{0}{0},
     \\
 \lambda _4 &= \matthree
     {0}{0}{1}
     {0}{0}{0}
     {1}{0}{0},
 \lambda _5 = \matthree
     {0}{0}{-i}
     {0}{0}{0}
     {i}{0}{0}=-S^2,
 \lambda _6 = \matthree
     {0}{0}{0}
     {0}{0}{1}
     {0}{1}{0},
     \\
 \lambda _7 &= \matthree
     {0}{0}{0}
     {0}{0}{-i}
     {0}{i}{0}=S^1,
 \lambda _8 = \matthree
     {1}{0}{0}
     {0}{1}{0}
     {0}{0}{-2}/\sqrt{3}. 
\end{align*}
}.
Note that the fermions transform as ($g=(g^*) ^{-1}= e^{i u}\in \SUQ$
\footnote{
 For an Hermite matrix $\bm{G} $, let us define
 ${\cal G}=  \bm{c} ^\dagger   \bm{G} \bm{c} $
 where $\bm{c} ^\dagger =(c_1 ^\dagger ,c_2 ^\dagger ,\cdots)$.
Assuming
 that $\bm{G} $ is diagonalized by a unitary matrix 
 $\bm{U} $ as
 $\bm{G}= \bm{U} \bm{g} \bm{U} ^\dagger $, $\bm{g}={\rm diag} \,(g_1,g_2,\cdots)$, $g_i\in\mathbb{R}$, we have
 ${\cal G}=\sum_i g_i d_i ^\dagger d_i $
where $\bm{d} \equiv\bm{U} ^\dagger \bm{c}, \bm{c} =\bm{U} \bm{d} $.
Then for 
$
{\cal U} = e^{-i \theta {\cal G} },\ {\cal U} d_i {\cal U} ^\dagger =
e^{-i \theta g_i d_i ^\dagger d_i}
d_i
e^{i \theta g_i d_i ^\dagger d_i}=
e^{ i \theta g_i} d_i$ ($\theta \in\mathbb{R}$), it reads
$
   {\cal U} \bm{c} \, {\cal U}  ^\dagger =  \bm{U} {\cal U} \bm{d} \, {\cal U}  ^\dagger =
   \bm{U} e^{ i \theta \bm{g} } \bm{d} = \bm{U} e^{ i \theta \bm{g} } \bm{U} ^\dagger \bm{U}  \bm{d} =
   e^{i \theta \bm{G} }\bm{c}
   $.
   Also it gives
$   {\cal U} \bm{c}^\dagger {\cal U}  ^\dagger =    \bm{c} ^\dagger  e^{-i \theta \bm{G} }
   $
})

\begin{align}
 {\cal U}^{(1)} c_{j,\alpha }( {\cal U}^{(1)} ) ^\dagger  &=\sum_\beta g_{\alpha \beta }c_{j,\beta },
 \\
  {\cal U}^{(2)} c_{j,\alpha }( {\cal U}^{(2)} ) ^\dagger  &=
  \mchss
    {\sum_\beta g_{\alpha \beta }c_{j,\beta }}{j:\text{{odd}}}
    {\sum_\beta g^*_{\alpha \beta }c_{j,\beta }}{j:\text{{even}}}.
\end{align}


%

 Especially $Z_\QN\subset\SUQ$ symmetry is important for the following 
 discussion from topological view points.
  Although the twists introduced later break $SU(\QN)$ symmetry
  in general, 
  this $Z_\QN$ still remains as a symmetry at the high symmetric twists
(denoted by $G$, see below)  
  which is a generalized anti-periodic boundary condition for
  $\QN=2$ \cite{Kitazawa97}.

 $Z_\QN$ is given by the global cyclic shift of the fermion colors
 as
 \begin{align}
  c_{j,\alpha } &\to   c_{j,\alpha -1,\text{ mod }\QN}=
  {\cal U}_{Z_\QN}
  c_{j,\alpha } 
  {\cal U}_{Z_\QN} ^\dagger , \\
\mvecfour
  {c_{j,1}}
  {c_{j,2}}
  {\vdots}
  {c_{j,\QN}}  
  &\to 
\mvecfour
  {c_{j,\QN}}
  {c_{j,1}}
  {\vdots}
  {c_{j,\QN-1}},\nonumber
  \\
  &=
  Z_\QN
  \mvecfour
  {c_{j,1}}
  {c_{j,2}}
  {\vdots}
  {c_{j,\QN}}  
  =
  {\cal U}_{Z_\QN}
  \mvecfour
  {c_{j,1}}
  {c_{j,2}}
  {\vdots}
  {c_{j,\QN}}  
  {\cal U}_{Z_\QN}^\dagger ,\nonumber
 \end{align}
 where
  \begin{align}
Z_\QN &=   
  {  \left(
\begin{array}{cccc}
 0 &0&\cdots& 1 \\
 1 &0&0 & \vdots \\
 \vdots  &\ddots &\ddots &0 \\
0  &\cdots &1& 0
\end{array}
\right)
}
=e^{i z_\QN}\in SU(Q),
\\
{\cal U} _{Z_\QN} &= \prod_j e^{-i c_{j,\alpha } ^\dagger (z_\QN)_{\alpha \beta } c_{j,\beta }}.
 \end{align}
See Sec.\ref{sec:zq}.

 It implies that all of the eigen states are
 labeled by the eigen values of the unitary transformation ${\cal U}_{Z_\QN} $
as
 \begin{align*}
  {\cal U} _{Z_\QN} | \omega ^n \rangle &=
| \omega ^n \rangle \omega ^n.
 \end{align*}
 

\subsection{Gauge symmetry}
\label{sec:local-gauge}
Further the Hamiltonians have a (local) $[U(1)]^\QN$ gauge invariance
for $\varphi _{j,\alpha }\in \mathbb{R}$, ($\forall j$ and $^\forall \alpha $), as 
\begin{align}
 {\cal U}_\local^{(1)} H^{(1)}
 (\{J_{j,\alpha \beta }^{(1)}\})
 ( {\cal U}_\local^{(1)} ) ^\dagger  &= H^{(1)} (\{\bar J_{j,\alpha \beta }^{(1)}\}),
 \nonumber
 \\
 {\cal U}_\local^{(2)} H^{(2)}
 (\{J_{j,\alpha \beta }^{(2)}\})
 ( {\cal U}_\local^{(2)} ) ^\dagger  &= H^{(2)} (\{\bar J_{j,\alpha \beta }^{(2)}\}),
 \nonumber
 \\
  {\cal U}_\local^{(1)} c_{j,\alpha }( {\cal U}_\local^{(1)} ) ^\dagger  &= e^{i \varphi _{j,\alpha }}c_{j,\alpha },
 \\
  {\cal U}_\local^{(2)} c_{j,\alpha }( {\cal U}_\local^{(2)} ) ^\dagger  &= e^{i (-1)^{{j-1}}\varphi _{j,\alpha }}c_{j,\alpha },
\end{align}
where
\begin{align}
 {\cal U}_\local^{(1)} &= e^{ -i\sum_{j,\alpha } \varphi _{j,\alpha }\hat n_{j,\alpha }},
 \\
  {\cal U}_\local^{(2)} &= e^{ -i\sum_{j,\alpha } (-1)^{{j-1}}\varphi _{j,\alpha }\hat n_{j,\alpha }}.
\end{align}
Then, the couplings in the Hamiltonian of Eq.(\ref{eq:Ham12})
are transformed as
\begin{align}
  \bar J_{j,\alpha \beta }^{(1)} &=
  e^{i \Omega _{j,\alpha ;j+1,\beta }} J_{j,\alpha \beta }^{(1)},
  \\
    \bar J_{j,\alpha \beta }^{(2)} &=
  e^{i (-1)^{{j-1}}\Omega _{j,\alpha ;j+1,\beta }} J_{j,\alpha \beta }^{(2)},
  \\
  \Omega _{j,\alpha ;j+1,\beta } &=
  -(\varphi _{j,\alpha }-\varphi _{j,\beta })+(\varphi _{j+1,\alpha }-\varphi _{j+1,\beta }),
\end{align} 
 where $j=1,\cdots,L-1$ for the open boundary condition and 
{ $ \varphi _{L+1,\alpha }\equiv \varphi _{1,\alpha }$ for the periodic boundary condition. 
We always assume
the system size $L$ is even for the discussion of the periodic boundary condition.

Taking all local gauge parameters constant, $\varphi _{j,\alpha } =\phi _\alpha $,
one has
\begin{align}
 e^{-i \phi_\alpha N_\alpha } H^{(1)}  e^{i \phi_\alpha N_\alpha } &= H^{(1)},
 \\
 e^{-i \phi_\alpha \bar N_\alpha } H^{(2)}  e^{i \phi_\alpha \bar N_\alpha } &= H^{(2)},
\end{align}
where
$ N_\alpha = \sum_j \hat n_{j,\alpha }$ and 
$\bar N_\alpha= \sum_j (-1)^{j{-1}} \hat n_{j,\alpha }$.
Differentiation by $\phi _\alpha $ implies $Q$
conservation laws ($\alpha =1,\cdots,\QN$)
\begin{align}
 \cmt{N_\alpha }{H^{(1)}} &= 0,\quad \cmt{\bar N_\alpha }{H^{(2)}} = 0,
 \label{eq:cons-law}
\end{align}
where $\cmt{N_\alpha }{N_\beta }=0$ and $\cmt{\bar N_\alpha }{\bar N_\beta }=0$.

 Since the $Z_\QN$ operation shifts these quantum numbers
as $\bar N_\alpha \to \bar N_{\alpha -1}$, it results in degeneracy if
\begin{align*}
 (\bar N_1,\cdots,\bar N_\QN )& \ne
 (\bar N_\QN,\bar N_1, \cdots,\bar N_{\QN-1} ).
\end{align*}

\subsection{Large gauge transformation}
\label{sec:large-gauge}
Taking the gauge parameters as 
\begin{align}
 \varphi _{j,\alpha } &= x_j\varphi_\alpha,
 \\
 x_j &= \frac {j- j_0}{L}\in[-\frac 1 2 ,\frac 1 2 ], 
\quad j_0=\frac {L+1}{2},
\\
 \Omega _{j,\alpha ;j+1 \beta }&=
 \mchss
   {\frac 1 L({\varphi_\alpha -\varphi_\beta }) }{j=1,\cdots,L-1}
   {(\frac 1 L -1) ({\varphi_\alpha -\varphi_\beta })}{j=L},
\end{align}
the $[U(1)]^Q$ large gauge transformation ${\cal U}_\largeG^{(1,2)} $ is defined by 
\begin{align} 
 {\cal U}_\largeG^{(1)} &= e^{ -i\sum_\alpha \varphi_\alpha {\cal P}_\alpha ^{(1)}},
 \\
 {\cal U}_\largeG^{(2)} &= e^{ -i\sum_\alpha \varphi_\alpha {\cal P}_\alpha ^{(2)}},
 \label{eq:defG}
\end{align}
where
\begin{align} 
 {\cal P}_\alpha ^{(1)}&=  \sum_j x_j\hat n_{j,\alpha },
   \ \
 {\cal P}_\alpha ^{(2)}=  \sum_j (-1)^{j{-1}} x_j\hat n_{j,\alpha },
    \label{eq:defP2}
\end{align}
are the center of mass (CoM) \cite{KSV93,resta94,resta98}.
They are
generators of the large gauge transformations \cite{Hatsugai16pm,watanabe18}.
They induce  changes in the couplings as
\begin{align*} 
 {\cal U}_\largeG^{(1)} H^{(1)}
 (\{J_{j,\alpha \beta }^{(1)}\})
 ( {\cal U}_\largeG^{(1)} ) ^\dagger  &= H^{(1)} (\{\bar J_{j,\alpha \beta }^{(1),G}\}),
\\
 {\cal U}_\largeG^{(2)} H^{(2)}
 (\{J_{j,\alpha \beta }^{(2)}\})
 ( {\cal U}_\largeG^{(2)} ) ^\dagger  &= H^{(2)} (\{\bar J_{j,\alpha \beta }^{(2),G}\}),
 \end{align*} 
\begin{align}
 \bar J_{j,\alpha \beta }^{(1),G} &=
 e^{i \frac {\varphi_\alpha-\varphi_\beta }{L}}
 J_{j,\alpha \beta }^{(1)}
 \mchss
 {1}{j=1,\cdots,L-1}
 {e^{-i(\varphi_\alpha-\varphi_\beta )}} {j= L},
 \\
 \bar J_{j,\alpha \beta }^{(2),G} &=
 e^{i(-1)^{{j-1}} \frac {\varphi_\alpha-\varphi_\beta }{L}}
 J_{j,\alpha \beta }^{(2)}
 \mchss
 {1}{j=1,\cdots,L-1}
 {e^{+i(\varphi_\alpha-\varphi_\beta )}} {j= L:\text{even}},
\end{align}
where
\begin{align}
 {\cal U}_\largeG^{(1)} c_{j,\alpha}( {\cal U}_\largeG^{(1)} ) ^\dagger
&= e^{i\varphi_\alpha x_j}c_{j,\alpha},
  \\
 {\cal U}_\largeG^{(2)} c_{j,\alpha}( {\cal U}_\largeG^{(2)} ) ^\dagger
&= e^{i\varphi_\alpha (-1)^{{j-1}} x_j}c_{j,\alpha}.
\end{align}

 Note that the constraint $\sum_\alpha \hat n_{j,\alpha }=1$ implies 
\footnote{
 When $L$ is even,
 $1-2+-\cdots+(L-1)-L =(1-2)+(3-4)+\cdots+((L-1)-L) =
 -1 \frac {L}{2}=-L/2$
 and
$
 \sum_j (-1)^{j-1}x_j = L ^{-1}   \sum_j (-1)^{j-1}(j-j_0)
 = -\frac 1 2 $.
 When $L$ is odd,
 $1-2+-\cdots-(L-1)+L =
 (1-2)+(3-4)+\cdots+(L-2)-(L-1))+L =
 -\frac {L-1}2+L=\frac {L+1}{2} $ 
 and
$
 \sum_j (-1)^{j-1}x_j = L ^{-1}   (\sum_j (-1)^{j-1}j-j_0)
 = L ^{-1} (\frac {L+1}{2} -\frac {L+1}{2})=0$
}
\begin{align} 
\sum_{\alpha=1}^\QN  {\cal P}_\alpha ^{(1)}&=  \sum_j x_j = 0,
\label{eq:sumP}
\\
\sum_{\alpha=1}^\QN  {\cal P}_\alpha ^{(2)}&= 
   \mchss
     {-\frac 1 2 }{L:\text{even}}
     {0}{L:\text{odd}}. 
    \label{eq:sumPs}
\end{align}

\subsection{Periodic system: gapped ground state of dimers} 
 \label{sec:two}

To realize a topological pump,
we require a gapped unique ground state
for a periodic boundary condition and also 
 with non-trivial edge states for a system with edges.
See examples \cite{Hatsugai16pm,Kuno20charge,Kuno21spin,Nakagawa18,Greschner20,Lin20}

As for the $\QN=3$ case, the spectra and eigen states of the two site systems
$H^{(1,2)}_{a,b}$ 
are
listed in  Tables
\ref{table:33} and \ref{table:3b3} ($J_{\alpha \beta ,1}=J$) respectively
($|0 \rangle $ is a fermion vacuum).
Generic $\QN$ case for $H^{(2)}_{a,b}$ is summarized in
 Table \ref{table:QbQ}.
The sites are labeled by $a$ (:odd) and $b$ (:even).
They are consistent with the decomposition of the representations,
$3\otimes 3=\bar 3\oplus 6$ and
$3\otimes \bar 3=1\oplus 8$.

\renewcommand{\arraystretch}{1.15}
\begin{table}[t]
 \caption{Energies and eigen states of $H^{(1)}$ ($\QN=3$).
  where
  $|\alpha_a \beta_b \rangle =c_{i\alpha} ^\dagger c_{j\beta} ^\dagger | 0 \rangle   $.
The sites are labeled by $a$ (:odd) and $b$ (:even). }
 \label{table:33}
 \centering
 \begin{tabular}{|c|c|c|}
  \hline
  state & $E$ & $(N_1,N_2,N_3)$
  \\
  \hline 
  $|1_a 1_b \rangle $ & $J$ & $(2,0,0)$ \\
  \hline 
  $|2_a 2_b \rangle $ & $J$ & $(0,2,0)$ \\
  \hline 
  $|3_a 3_b \rangle $ & $J$ & $(0,0,2)$ \\
  \hline
  $(|1_a 2_b \rangle+|2_a 1_b \rangle )/\sqrt{2} $ & $J$ & $(1,1,0)$ \\
  \hline 
  $(|1_a 2_b \rangle-|2_a 1_b \rangle )/\sqrt{2} $ & $-J$ & $(1,1,0)$ \\
  \hline
  $(|2_a 3_b \rangle+|3_a 2_b \rangle )/\sqrt{2} $ & $J$ & $(0,1,1)$ \\
  \hline 
  $(|3_a 2_b \rangle-|2_a 3_b \rangle )/\sqrt{2} $ & $-J$ & $(0,1,1)$ \\
  \hline
  $(|3_a 1_b \rangle+|1_a 3_b \rangle )/\sqrt{2} $ & $J$ & $(1,0,1)$ \\
  \hline 
  $(|3_a 1_b \rangle-|1_a 3_b \rangle )/\sqrt{2} $ & $-J$ & $(1,0,1)$ \\
  \hline
 \end{tabular}
\end{table}

\begin{table}[t]
 \caption{Energy and eigen states of $H^{(2)}$ ($\QN=3$).
  The sites are labeled by $a$ (:odd) and $b$ (:even)
  ($\omega =\frac {-1+i\sqrt{3}}{2}$).
 }
 \label{table:3b3}
 \centering
 \begin{tabular}{|c|c|c|}
  \hline
  state & $E$ & $(\bar N_1,\bar N_2,\bar N_3)$
  \\
  \hline 
  $|S_{ab} \rangle =(|1_a1_b \rangle +|2_a2_b \rangle +|3_a3_b \rangle )/\sqrt{3} $ & $3J$ & $(0,0,0)$ \\
  \hline 
  $|\omega _{ab} \rangle =(|1_a1_b \rangle +\omega |2_a2_b \rangle + \omega ^2 |3_a3_b \rangle )/\sqrt{3} $ & $0$ & $(0,0,0)$ \\
  \hline 
  $|\omega_{ab} ^{2}  \rangle =(|1_a1_b \rangle +\omega ^{2} |2_a2_b \rangle + \omega ^4 |3_a3_b \rangle )/\sqrt{3} $ & $0$ & $(0,0,0)$ \\
  \hline
  $|1_a2_b \rangle$ & $0$ & $(1,-1,0)$ \\
  \hline 
  $|2_a1_b \rangle $ & $0$ & $(-1,1,0)$ \\
  \hline
  $|2_a3_b \rangle$ & $0$ & $(0,1,-1)$ \\
  \hline 
  $|3_a2_b \rangle $ & $0$ & $(0,-1,1)$ \\
  \hline
  $|3_a1_b \rangle$ & $0$ & $(-1,0,1)$ \\
  \hline 
  $|1_a3_b \rangle $ & $0$ & $(1,0,-1)$ \\
  \hline
 \end{tabular}
\end{table}

\begin{table}[t]
 \caption{{
   Energies and eigen states of $H^{(2)}$ 
   where $|\omega _{ij}^n \rangle$, $n=1,\cdots,Q-1$, 
   $|\alpha_a \beta_b \rangle$, $\alpha \ne \beta =1,\cdots,Q$,
   ($\omega =e^{i \frac {2\pi}{Q} })$.
The sites are labeled by $a$ (:odd) and $b$ (:even).   
  Generic case.}
}
 \label{table:QbQ}
 \centering
 \begin{tabular}{|c|c|c|}
  \hline
  state & $E$ & $\bar N_\alpha$
  \\
  \hline
  $|S_{a,b} \rangle =\psi_{a,b} ^\dagger |0 \rangle $ &
  $QJ$ & $^\forall \bar N_\alpha =0$ \\
  \hline 
  $|\omega _{ab}^n \rangle =Q ^{-1/2} \sum_{\alpha =1}^Q\omega ^{\alpha n}|\alpha _a \alpha _b \rangle $ & $0$ & $^\forall \bar N_\alpha =0$ \\
  \hline
  $|\alpha _a \beta _b \rangle,\ \alpha \ne \beta $ & $0$ & $\bar N_\alpha =-\bar N_\beta =1,\bar N_{\gamma \ne \alpha ,\beta } =0$ \\
  \hline
 \end{tabular}
\end{table}

Since we need a unique (singlet) ground state for the two site problem
as a dimer
in the following, we discuss $H^{(2)}$ or $H$ of $\omega_S =\frac {\pi}{4} $
\begin{align*}
 J_j &= \mchss
 {J_o}{j:\text{odd}}
 {J_e}{j:\text{even}} ,\quad 
J_o,J_e\le 0 .
\end{align*}

The extension to the $\SUQ$ case (Table \ref{table:QbQ}) is
straightforward and due to the decomposition $\QN\otimes \bar \QN=1\oplus(\QN ^2-1)$
 \cite{Slansky}. It is 
a SPT
 protected by $Z_Q$ symmetry \cite{Hatsugai06qb,Hatsugai11}.
The two site Hamiltonian for $J_j=J<0$ is
\begin{align*}
 H_{ab} &= J \psi_{a,b} ^\dagger \psi_{a,b}
\end{align*}
where $\psi_{a,b}=\QN ^{-1/2} \sum_\alpha \psi_{a,\alpha ;b,\alpha } $, ($a\ne b$).

The singlet is given by $|S_{ab}\rangle =
\psi_{a,b} ^\dagger | 0 \rangle $ with its energy
$\QN J$ and $\bar N _\alpha =0, ^\forall \alpha $,
since $\cmt{\psi_{a,b} }{\psi_{a,b} ^\dagger }| 0 \rangle =1$
\footnote{
{
 $
\cmt{\psi_{a,b} }{\psi_{a,b} ^\dagger }
  =1-\QN ^{-1} \sum_\alpha (n_{a,\alpha }+n_{b,\alpha })
$
}}.
The rest of zero energy $\QN^2-1$ states are given by the $\QN-1$ states, 
$| \omega _{ij}^n \rangle = \QN ^{-1/2} \sum_\alpha \omega ^ {n\alpha } |\alpha _i \alpha _j \rangle $, $n=1,\cdots,\QN-1$ with
$\bar N_\alpha =0$ and
$\QN(\QN-1)$ states,
$|\alpha _i \beta _j \rangle $, ($\alpha \ne \beta $) with $\bar N_\alpha =1, \bar N_\beta =-1, \bar N_\gamma =0,(\gamma \ne \alpha ,\beta )$ where we assume $a$ is odd and $b$ is even.


 Noting this two-site problem,
 we have two different unique gapped ground states for the
periodic system ($c_{L+1,\alpha }\equiv c_{1,\alpha }$)
with
 different dimer limits $J_o=0$ and $J_e=0$ as
\begin{align}
 |g_{\pe, eo} \rangle &= \prod_{j=1}^{L/2} \psi_{2j+1, 2j} ^\dagger | 0 \rangle,\ (J_o=0),
 \label{eq:dimer_eo}
 \\
 |g_{\pe,oe} \rangle &= \prod_{j=1}^{L/2} \psi_{ 2j-1,2j} ^\dagger | 0 \rangle,\ (J_e=0).
 \label{eq:dimer_oe} 
\end{align}
 Note that both states are labeled by the occupations,
 $\bar N_\alpha =0$, $ \alpha =1,\cdots,\QN$.

Due to the adiabatic continuity,
the ground state is gapped and unique if the interaction
between the dimers ($J_{j,\alpha \beta }$, $j$:even) is finite but weak enough ($J_2\ne 0, |J_2|\ll |J_1|$).
It is a SPT phase protected by
$Z_\QN $ symmetry
associated with the quantized Berry phase \cite{Hatsugai11,Kariyado18}.
See Sec.\ref{sec:zq}.
As is clear in the Table, we do not require  $\SUQ$ symmetry.
We may allow $Z_\QN$ invariant twists at any links as is introduced later.
At $J_o=J_e$,
the energy gap (of a finite system) closes as is clear from the discontinuous change of the
quantized Berry phase (discussed later).
This gap closing point
is a source of the
non-trivial topology which we discuss in this paper. 


\section{Currents, center of mass and bulk-edge correspondence}
\label{sec:current}

\subsection{Current and Synthetic Brillouin zone (twist space)}
\label{sec:zq}
Noting that the large gauge transformation,
let us start considering a dimerized Hamiltonian
$H^{(2)}(\{J ^{(2)}_{j,\alpha \beta } \})$
by
\begin{align}
 J_{j,\alpha \beta } &= \mchsssl
 {J_j}{j=1,\cdots,L-1}
 {J_j e^{-i(\varphi_\alpha -\varphi_\beta )}}{\text{PBC},\ j=L:\text{even}}
 {0}{\text{OBC}, \ j=L},
 \label{eq:dimerJ-twist}
 \\
 J_i &= J_0+\delta J (-1)^j\cos \frac {2\pi t}{T} \in\mathbb{R},
 \label{eq:dimerJ}
\end{align}
where $\ J_0,\delta J\in \mathbb{R}$ and $t$ is a time with a period $T$.
To be explicit, we take $J_0<0$ and $\delta J>0$
 in the following numerical demonstration.
We also include a
symmetry breaking term to realize the topological pump.
To be concrete, let us consider a following term for the generic $\SUQ$ case
(it reduces to the staggered potential for $SU(2)$ case)
 \begin{align}
 H_B (t) &= \sum_{j,\alpha } \hat n_{j,\alpha } \Delta _\alpha (t),
 \\
 \Delta_\alpha (t) &= \Delta  \sin 2\pi(\frac {t}{T}+ \Phi {\alpha } ),
  \label{eq:symb}
\end{align}
where $ \Phi = \frac P Q$ 
and $\Delta $ is a strength of a symmetry breaking. The integers $P$ and $Q$ are mutually co-prime.
We omit the superscript ``$^{(2)} $'' unless explicitly specified and both of the
periodic/open systems are discussed.
Noting that Eq.(\ref{eq:dimerJ-twist}),
the Hamiltonian $H_{\op}(\{J_{j,\alpha \beta }\})$ for the open system is $\varphi_\alpha$-independent
and $H_{\pe}(\{J_{j,\alpha \beta }\})\equiv H_\tw$, for the periodic system, is with a twisted boundary condition.
The difference between $H_\op$ and $H_\tw$ is only at the boundary link, $j=L$.
Then the large gauge transformation by ${\cal U}_\largeG $ induces 
${\cal O}(L ^{-1} ) $ twists for each link as
\begin{align}
 \bar H_{\op,\pe}
 &= H_{\op,\pe}(\{\bar J_{j,\alpha \beta }\}),
 \\
 \bar J_{j,\alpha \beta } &= e^{i{(-1)^{j-1}} \frac {\varphi_\alpha -\varphi_\beta }{L} }J_j\
 \mchss{(j=1,\cdots,L-1)}{: \text{open}} 
    {(j=1,\cdots,L)}{: \text{periodic}}.
\end{align}
Explicitly for the open/periodic cases, they are written as
\begin{align}
 \bar H_\op &= {\cal U} _\largeG H_\op {\cal U} _\largeG ^\dagger ,
 \label{eq:g-open}
 \\
 \bar H_\pe &= {\cal U} _\largeG H_\tw {\cal U} _\largeG ^\dagger ,
 \label{eq:g-periodic}
 \\
  {\cal U}_\largeG &=  e^{-i \sum_\alpha \varphi_\alpha {\cal P}_\alpha }.
 \label{eq:def-ug}  
\end{align}
The twists are uniform both for the
periodic $\bar H_\pe $ and the open $\bar H_\op$ Hamiltonians.
The periodic system is translational invariant by the period $2$ with dimerization.

Without dimerization, uniformity of the $H_\pe$ is written as
 $J_j=J$ ($j$-independent).
 This is inherited as {\em anti-translation} invariance of $\bar H_\pe$ as
 \begin{align}
  {\cal A} _{\cal T}  \bar H_\pe  {\cal A} _{\cal T} ^{-1} &= \bar H_\pe,
\\
 {\cal A} _{\cal T} &=
  {\cal U} _{\cal T} {\cal K} ,
   \\
    {\cal U} _{\cal T}  c_{j,\alpha }  {\cal U} _{\cal T} ^\dagger &= c_{j+1,\alpha },
    \label{eq:anti}
 \end{align}
 where $ {\cal U} _{\cal T} $ is unitary and
 ${\cal A}_{\cal T} $ is anti-unitary (${\cal K} $ is a complex conjugate).
  It implies $\bar J_{j,\alpha \beta }=(\bar J_{j\pm 1,\alpha \beta })^*  $.
 
In the following, the Hamiltonian is extended by adding a symmetry breaking term
$H_B$, that is gauge invariant as ${\cal U}_\largeG H_B {\cal U}_\largeG ^\dagger =H_B $.
See Table \ref{table:HandS}.
%

\begin{widetext}
 \begin{center}
        \begin{table}[t]
 \caption{Hamiltonians, energies and states with different boundary conditions.}
 \label{table:HandS}
 \centering
 \begin{tabular}{|c||c|c|c|c|c|}
  \hline
  & $\theta $ dependence & on $T^d (\Theta )$ & Energy on
  $T^d (\Theta )$ & ground state & current 
  \\
  \hline
  $H_\op$ & independent & independent& independent & $|g_{0,\op} \rangle $& -
  \\
  \hline
  $\bar H_\op$ & uniform & ill-defined$^{*1}$ & independent &$|g_\op \rangle ={\cal U}_\largeG|g_{0,\op} \rangle $&
  $j_\op =  \hbar ^{-1}  \langle g_\op| \partial_\theta \bar H_\op | g_\op \rangle $
  \\
  \hline
  $ H_\tw$ & at the boundaries & well-defined & well-defined$^{*2}$ &$|g_{0,\tw} \rangle$ & -
  \\
  \hline
  $ \bar H_\pe$ & uniform & ill-defined$^{*1}$ & well-defined$^{*2}$ & $|g_\pe \rangle
 ={\cal U}_\largeG|g_{0,\tw} \rangle  $ &
  $j_\pe =  \hbar ^{-1}  \langle g_\pe| \partial_\theta \bar H_\pe | g_\pe \rangle $  \\
  \hline 
 \hline
 \end{tabular}
 \\
 $^{*1}$: Does not satisfy the periodicity in $\Theta $.
$^{*2}$: Assuming the unique ground state on $\Theta $,  $E_\tw=E_\pe$. 
        \end{table}
        \end{center}

        \end{widetext}

To define a current, let us
introduce a
$d$-dimensional twist space ($d$-dimensional torus),
$T^d=\{(\theta _1,\theta _2,\cdots,\theta _d)|\, \theta _\alpha \in[0,2\pi], \alpha =1,\cdots, d\}$ ( $d=Q-1$).
This is a synthetic Brillouin zone \cite{Hatsugai11},
which introduces twist for the Hamiltonian $H_\tw$ by
\begin{align*}
 \varphi_1 &= \theta _1,\\
 \varphi_2 &= \theta _1 +\theta _2, \\
 &\vdots \\
 \varphi _\alpha &= \theta _1+\cdots + \theta _\alpha , \\
 &\vdots \\ 
 \varphi_d &= \theta _1 +\theta _2+\cdots+\theta _d, \\
 \varphi_Q &\equiv \varphi _0=0
\end{align*}
that is,
\begin{align*}
 \varphi_1-\varphi_0 &= \theta _1,\\
 \varphi_2-\varphi_1 &= \theta _2, \\
 &\vdots
 \\
 \varphi_\alpha -\varphi_{\alpha -1} &= \theta _\alpha , \\ 
 &\vdots \\ 
 \varphi_d -\varphi_{d-1} &= \theta _d, \\
 \varphi_Q - \varphi_{Q-1} &= -\theta _1-\theta _2\cdots -\theta _d\equiv \theta _Q,
\end{align*}
where
$\theta _\alpha $
is defined in modulo $2\pi$, that is, $\theta _\alpha =0$ and $2\pi $ are identified.
 It implies a formal relation
 $\theta _{1}+\cdots+\theta _d+\theta _Q =0 $.
 Note that the Hamiltonian depends on
 \begin{align*}
  \varphi_\alpha -\varphi_\beta &= \theta _\alpha +\cdots+\theta _{\beta +1},\   ^\forall \alpha ,\beta ,\
\alpha \ge \beta +1.
  \end{align*} 
 Since the Hamiltonian is invariant
for the 
$Z_\QN$ shift of the fermions, $c_{j,\alpha }\to c_{j,\alpha -1}$,
which induces a shift $\varphi_\alpha \to\varphi_{\alpha +1} $,
and also for the constant shift of $ ^\forall \varphi_\alpha $ by
subtracting $\theta _1$ denoted by $\sim$ as
\begin{align*}
 \varphi_1 &\to \varphi_{2}= \theta _1+ \theta _2\sim \theta _2,\\
 \varphi_2 &\to\varphi_3= \theta _1 +\theta _2+\theta _3 \sim \theta _2+\theta_3 , \\
 &\vdots \\ 
 \varphi _\alpha &\to\varphi_{\alpha +1}= \theta _1+\cdots + \theta _{\alpha+1}
 \sim \theta _2+\cdots + \theta _{\alpha+1}, \\
 &\vdots \\ 
 \varphi_{d-1} &\to \varphi_{d} = \theta _1 +\theta _2+\cdots+\theta _d
 \sim\theta _2+\cdots+\theta _d,
 \\ 
 \varphi_d &\to \varphi_{d+1} = \varphi_Q=\theta _1 +\theta _2+\cdots+\theta _d
 +\theta _Q,\\
  &\qquad \sim
 \theta _2+\cdots+\theta _d +\theta _Q,
 \\
 \varphi_Q &= \varphi _0\to \varphi_1 =\theta _1\sim 0.
\end{align*}
It is given by the cyclic shift of the parameter space supplemented by $\theta _Q$ 
\begin{align}
 (\theta _1,\cdots,\theta _d,\theta _Q)
 &\to  (\theta _2,\cdots,\theta _Q,\theta _1).
 \label{eq:theta-shift}
\end{align}
It implies $Z_\QN$ equivalence of loops $\ell_{V_\alpha G V_{\alpha +1}}$, $\alpha =1,\cdots,Q$ as shown later.


It is useful to express this parameter space as shown in
Fig.\ref{fig:param} and Fig.\ref{fig:2Dparam}.
Let us start $Q=d+1$ equivalent points $V_\alpha $,
($\alpha=1,\cdots,Q\equiv 0$) on a $(d-1)$-dimensional sphere $S^{d-1}$,
which is constructed recursively from the $0$-dimensional sphere (2 points).
$G=(\frac {2\pi}Q,\cdots,\frac {2\pi}Q)$ is a center of mass of all vertices $V_0,\cdots,V_d$, which is a
center of the sphere on which all vertices lie.
See ref. \cite{Hatsugai11} for the details. Its low dimensional examples are
2 vertices of a line ($S^0$, $d=1$, Fig. \ref{fig:2Dparam}(a)),
3 vertices of a triangle on a circle ($S^1$, $d=2$, Fig. \ref{fig:2Dparam}(b))
and 4 vertices of a tetrahedron on 
a sphere in 3 dimension ($S^2$, $d=3$).
$T^d$ is spanned by the $d$ vectors, $\vec e_j=\overrightarrow{V_0V_\alpha }/2\pi$,
$\alpha =1,\cdots,d$
as $\Theta =\theta _1 \vec{e}_1+\cdots+ \theta _d \vec{e}_d $, which is abbreviated as
$\Theta =(\theta _1,\cdots,\theta _d)$.

\begin{figure}[t]
 \includegraphics[width=80mm]{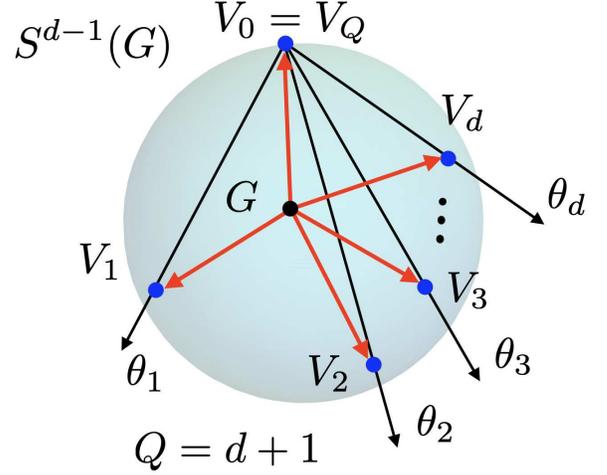}
\caption{\label{fig:param} 
 $d$-dimensional parameter space of the twist and $Q$ loops
 $\ell_{V_0GV_1},\ell_{V_1GV_2},\cdots,\ell_{V_dGV_0}$
on torus $T^d$.}
\end{figure}

\begin{figure}[t]
\includegraphics[width=90mm]{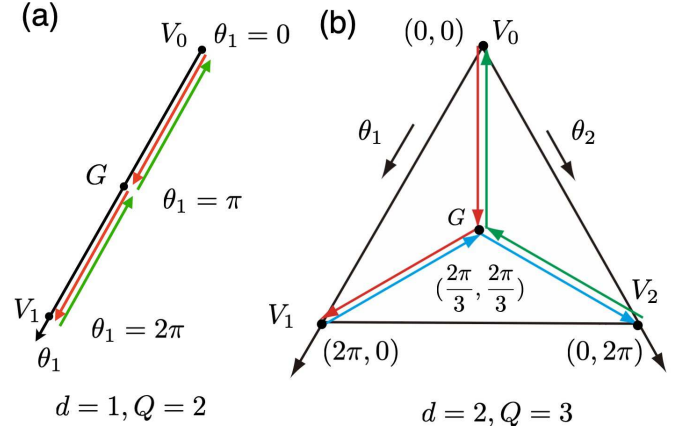}
\caption{\label{fig:2Dparam} 
 Examples for the paths in the parameter spaces for (a) $d=1, \QN=2$ and (b) $d=2, \QN=3$
where
 $G$ is a $Z_\QN$ symmetric point. The twist at $G$, $\QN=2$ corresponds to
 the anti-periodic boundary condition.}
\end{figure}

It defines a synthetic Brillouin zone for the twisted Hamiltonian
\begin{align}
 H_\tw(\Theta)=H_\tw(\Theta + 2\pi \vec e_\alpha ), \alpha =1,\cdots,d.
\end{align}
that is, all vertices $^\forall V_\alpha $ are identified.

Note that the $Q$ paths
\begin{align}
 \ell_{V_\alpha GV_{\alpha +1}} &= -\overrightarrow{GV_\alpha }+\overrightarrow{GV_{\alpha +1}}, \
 \alpha =1,\cdots,Q,
\end{align}
in Fig.\ref{fig:param}, forms loops for $H_\tw(\Theta )$.
Note that $\bar H_\pe(\Theta )$ does not satisfies this periodicity.

By taking any path $\ell=\{\Theta |(\Theta=\Theta (\theta ),
\theta\in[0,2\pi] \}$ parameterized by $\theta $
in the parameter space, 
let us define a current 
\begin{align}
 {\cal J} ^\ell &\equiv \hbar ^{-1} \partial _\theta \bar H
=\hbar ^{-1} \partial _\theta 
 H (\{\bar J_{j,\alpha \beta }\}).
\end{align}

Eqs.(\ref{eq:g-open})and (\ref{eq:g-periodic}) imply respectively
\begin{align}
 {{\cal J}}_{\op} &= \hbar ^{-1} \partial _\theta \bar H_\op = -i \hbar ^{-1} 
\sum_\alpha \partial _\theta \varphi_\alpha \cmt{{\cal P}_\alpha }{\bar H_\op },
 \label{eq:open-current}
 \\
  {{\cal J}}_{\pe} &= \hbar ^{-1}
  \partial _\theta \bar H_\pe = -i \hbar ^{-1} \sum_\alpha \partial _\theta \varphi_\alpha \cmt{{\cal P}_\alpha }{\bar H_\pe} \nonumber \\
  &\qquad\qquad\qquad\qquad
  + \hbar ^{-1} {\cal U} _\largeG \partial _\theta H_\tw {\cal U} _\largeG ^\dagger, 
  \label{eq:periodic-current}
\end{align}

Then the 
evaluation by the time-dependent state $|G_{\op}(t) \rangle $
for the open boundary condition that
obeys the Schr\"odinger equation, $i \hbar \partial _t|G_{\op}(t) \rangle
= \bar H_{\op} |G_{\op} (t) \rangle $,
($\langle G_{\op}|G_{\op}\rangle =1$)
gives
\begin{align}
{j} _\op^G &\equiv \langle G_{\op}(t)|{{\cal J}}_\op |G_{\op}(t) \rangle
=  \sum_\alpha \partial _\theta \varphi_\alpha \partial _t{ P}^G_{\op,\alpha},
\end{align}
where 
\begin{align}  P^G_{\op,\alpha} (t)&= \langle G_{\op}(t)|{\cal P} _\alpha | G_{\op} (t)\rangle ,
 \label{eq:currentG}
\end{align}
is a center of mass of $| G _\op (t)\rangle $.
Note that the similar discussion
for the periodic boundary condition is not simple.
Eq.(\ref{eq:periodic-current})
has extra term 
($H_\tw $ depends on $\varphi_\alpha $)
and also the operator ${\cal P}_\alpha $ itself
does not have a physical meaning 
since the origin of the reduced coordinate $x_j$ is arbitrary
for the periodic boundary condition.

\subsection{Adiabatic current}
\label{sec:adia-current}

Let us start by a general discussion of the time-dependent
evolution of a state by the adiabatic approximation
assuming the initial state $|G(0)\rangle $ is 
a gapped unique ground state $|g(0) \rangle $
of the snapshot Hamiltonian $\bar H(t)$,
\begin{align*}
 | G(0) \rangle &= |g(0) \rangle, 
 \\
 \bar H(t) | n (t)\rangle &= |n(t) \rangle E_n(t),
\end{align*}
where $n=0,1,2,\cdots,$ and $|g \rangle =|n \rangle, n=0$, $\langle n| n ^\prime \rangle = \delta _{nn ^\prime }$.
We further assume the snapshot Hamiltonian is always
gapped $E_n(t)>E_g(t)$, ($n\ne g$).
When
the time modulation of the Hamiltonian is slow enough, the adiabatic
approximation is justified (See appendix \ref{sec:adia}) as
\begin{align}
 |G \rangle &\approx C\big[|g \rangle + i \hbar \sum_{n\ne g}\frac {|n \rangle \langle n| \partial _t g \rangle }{E_n-E_g} \big],
\end{align}
where $C$ is a time dependent phase factor ($|C|=1$).
In the present discussion, $\bar H$ is $\bar H_\op$ or $\bar H_\pe$.

Under this adiabatic approximation, the observed
current is given by \cite{Thouless83} (See also appendix \ref{sec:adia})
\begin{align}
 j^G &= \langle G| {\cal J} | G \rangle \approx j,
 \label{eq:jG-j}
 \\
 j &= \hbar ^{-1} \langle g|\partial _\theta \bar H |g \rangle - i B,
 \label{eq:j-by-B}
 \\
 B &= \partial _\theta A_t -\partial _t A_\theta ,
 \label{eq:def-B}
\end{align}
where $A_\mu $ ($\mu=\theta ,t$) is the Berry connection
\begin{align}
 A_\mu &= \langle g|\partial _\mu g \rangle .
\end{align}
Here we assume the ground state $|g \rangle $ 
as a function of $\theta $ and $t$ by
$ \bar H(\theta ,t)|g(\theta ,t)\rangle =|g(\theta ,t)\rangle E_g(\theta ,t)$.
It is covariant for the phase transformation,
$| g ^\prime \rangle = | g \rangle e^{i\chi}$, ($\chi=\chi(\theta ,t)$)
that induces the gauge transformation for
the Berry connection
\begin{align}
 A ^\prime _\mu &= A_\mu + \partial _\mu \chi,
\end{align}
although the field strength $B$ and the current $j$ are gauge invariant. 

Let us here define the average current along the path $\ell$ such as
$\ell_{V_\alpha GV_{\alpha +1}}$
connecting
two equivalent points in the parameter space (parameterized
by $\theta \in[0,2\pi]$)
\begin{align}
 \bar j^\ell(t) &= \int_0^{2\pi} \frac {d \theta}{2\pi} \, j(\theta ,t).
 \label{eq:ad-current}
\end{align}
Due to the Feynmann's theorem, the first term Eq.(\ref{eq:j-by-B}) is written as
$\langle g|\partial _\theta \bar H| g \rangle
=\partial _\theta E(\theta) $,
$E_\theta =\langle g|\bar H | g \rangle $.
 It vanishes for the open boundary condition,
 since 
 $E_\op$ is $\theta $ independent.
 As for the periodic boundary condition, again 
$\int_0^{2\pi} d \theta \partial _\theta E_\pe(\theta )=0$
 since $E_\pe(\theta )=E_\tw(\theta ) $
 is periodic for any closed path connecting the equivalent
 points in the period $T$. See Table \ref{table:HandS}.
It results in
\begin{align}
 \bar j^\ell(t) &= -i \int _\ell \frac {d\theta}{2\pi} \, B.
\label{eq:bj-by-B}
\end{align}
Note that this is valid both for open and periodic boundary conditions.
Since the current is carried by bulk, the effect of the boundaries
is ${\cal O}(L ^{-1} ) $. One can expect, in the infinite
size limit,
\begin{align}
 \bar j^\ell_\pe &=  \bar j^\ell_\op,\ (L\to\infty).
\label{eq:j-bed}
\end{align}
  {\em This is the bulk-edge correspondence for the adiabatic current.}

  \subsection{Temporal gauge and discontinuity of CoM}
  \label{sec:com}
  Adiabatic pump is a periodic transfer of charge
  between the time period $[0,T]$
    assuming that the periodicity of the Hamiltonian as $\bar H(t+T)=\bar H(t)$
    under the adiabatic condition.
    Here $\bar H$ denotes
$\bar H_\pe$ or $\bar H_\op$ respectively.
    As for the gauge fixing of the Berry connection, 
    let us take a temporal gauge
$ A_t^{(t)}(\theta ,t) = 0$ by
  \begin{align}
   A_\theta ^{(t)}(\theta ,t) &= -\int_0^t d \tau\, B(\theta ,\tau).
  \end{align}
This is apparently gauge invariant and
  one may check $B=\partial _\theta A_t^{(t)}-\partial _t A_\theta ^{(t)}  =-\partial _t A_\theta ^{(t)}  $.
  Note that $A_\theta ^{(t)}$ is not periodic in time,
  $A^{(t)}(\theta ,t+T)\ne A^{(t)}(\theta ,t)$,
  although $\bar H$
  is periodic in time (period $T$).
  It is further
  written by the Berry connection in a generic gauge as \cite{Hatsugai16pm}
  \begin{align}
   A_\theta ^{(t)}(\theta ,t) &= -\int_0^t d \tau \, \big[\partial _\theta A_t(\theta ,\tau)
    -\partial _t A_\theta (\theta ,\tau)\big]\nonumber
   \\
   &= A_\theta (\theta ,t)
   -\partial _\theta \int_0^t d \tau \, A_t(\theta ,\tau)
   - A_\theta (\theta ,0).
  \end{align}

  The average current in the temporal gauge 
  is written as 
  \begin{align}
   \bar j^\ell(t) &=-i \int_\ell \frac {d \theta }{2\pi} B=
   i \partial _ t\bar A_\theta ^{(t)}(t),
   \\
   \bar A_\theta ^{(t),\ell}(t)&= \int_\ell \frac{d \theta}{2\pi} \,   A_\theta ^{(t)}(\theta ,t)\nonumber
   \\
   &=
   \frac {-1}{2\pi } \int_\ell d \theta \,\int_0^t d\tau B(\theta , \tau).
  \end{align}
  This is valid both for open and periodic boundary condition.
  
  As for the open boundary condition, the parameter dependence is only
  due to the large gauge transformation ${\cal U}_\largeG $
  as $|g_\op \rangle = {\cal U}_\largeG | g_{0,\op} \rangle $ where
  $ | g_{0,\op} \rangle $, 
  is a ground state of $H_\op$,
($ H_\op| g_{0,\op} \rangle = |g_{0,\op} \rangle  E_g$).
  We safely assume $|g_{0,\op}\rangle $
  is $\theta $  independent, $\partial _\theta |g_{0,\op} \rangle = 0$.
  It implies
  \begin{align}
   A_{t,\op} &=\langle g_\op| \partial _t g_\op \rangle =
   \langle g_{0,\op}| \partial _t g_{0,\op} \rangle,\ \partial _\theta A_{t,\op} =0,
   \\
   A_{\theta,\op} &=\langle g_\op| \partial _\theta g_\op \rangle =   \langle g_{0,\op}|
   {\cal U}_\largeG ^\dagger \partial_\theta {\cal U}_\largeG |g_{0,\op} \rangle
   \nonumber
   \\
   &\qquad\qquad\qquad
   = -i\sum_\alpha \partial _\theta \varphi_\alpha P_\alpha(t),
  \end{align}
  where
  \begin{align}
   P_\alpha(t) &= \langle g_{0,\op}|{\cal P} _\alpha | g_{0,\op} \rangle
   =
   {\sum_j (-1)^{j{-1}} x_j n_{j,\alpha}(t)},
   \label{eq:CoM}
  \end{align}
  is a center of mass of the $\alpha $-particle without twists
  ($n_{j,\alpha }(t)
  =\langle g_{0,\op}(t)| \hat n_{j,\alpha }| g_{0,\op}(t) \rangle $).
  The Berry connection in the temporal gauge is written as
  \begin{align}
   A_{\theta,\op} ^{(t)}(\theta ,t) &= 
-i \sum_\alpha \partial _\theta \varphi_\alpha \big[P_\alpha (t)-P_\alpha (0)\big].
  \end{align}
  
  The current averaged over the path $\ell$ is written as
  \begin{align}
   \bar j^\ell &= \bigg[
    \int_\ell \frac {d \theta }{2\pi} \sum_\alpha \partial _\alpha \varphi_\alpha \bigg]\partial _t P_\alpha (t) \nonumber
   \\
   &= \partial_t \bar P^\ell(t),
  \end{align}
  where
  \begin{align}
   \bar P^\ell(t) &= \sum_\alpha \frac {\Delta \varphi_\alpha }{2\pi} P_\alpha (t),\\
   \Delta \varphi_\alpha &= \varphi_\alpha (2\pi)-\varphi_\alpha (0).
  \end{align} 

  Along the path $\ell_{V_0GV_1}$, $\theta _1:0\to 2\pi$ and $\theta_\alpha :0\to 0$,
  $\alpha =2,\cdots,d$. It implies
  $\varphi_\alpha :0\to 2\pi$,
  $\Delta \varphi_\alpha =2\pi$ for all $\alpha =1,\cdots,d$.
  Also $\varphi_0=\varphi_Q= 0$. Therefore the averaged current
$ \bar j_\op^{\ell_{V_0GV_1}} $  
  is written as
  \begin{align}
   \bar j_\op^{\ell_{V_0GV_1}} &= \partial _t  \bar P^{\ell_{V_0GV_1}},
   \\
   \bar P^{\ell_{V_0GV_1}}&=   -P_0,
  \end{align}
  where
   $P_0 \equiv -\sum_{\alpha =1}^d P_\alpha$.
   (Note that $P_0=P_\QN$, ($L$: odd) and $P_0=P_\QN+1/2$, ($L$: even)
   due to the constraint Eq.(\ref{eq:sumPs}). 
   As for $\QN=3$, $P_0=-P_1-P_2$.
   (See Fig.\ref{fig:2Dparam}).
  Along the path $\ell_{V_1GV_2}$,
  $\theta_1:2\pi \to 0$,
  $\theta_2:0\to 2\pi $,
  $\theta _\alpha :0\to 0$, ($\alpha =3,\cdots,d$).
It implies
  $\Delta \varphi_1=-2\pi$, $\Delta \varphi_\alpha =0$, ($\alpha =2,\cdots,d$) and
  \begin{align}
   \bar j_\op^{\ell_{V_1GV_2}} &= \partial _t \bar P^{\ell_{V_1GV_2}}, \\
\bar P^{\ell_{V_1GV_2}} &= -   P_1.
  \end{align}
Similarly 
along the path $\ell_{V_\alpha GV_{\alpha +1}}$, ($\alpha =1,\cdots,d-1$),
$\theta _\alpha :2\pi\to 0$ and $\theta_{\alpha +1}:0\to 2\pi$.  It implies
  $\Delta \varphi_\alpha =-2\pi$, $\Delta \varphi_\beta =0$, ($\beta\ne \alpha $). Then, in general,
  \begin{align}
   \bar j_\op^{\ell_{V_\alpha GV_{\alpha +1}}} &= \partial_t \bar P^{\ell_{V_\alpha GV_{\alpha +1}}},
   \label{eq:jCoMP}
   \\
   \bar P^{\ell_{V_\alpha GV_{\alpha +1}}} &= -P_\alpha,\ \alpha =1,\cdots,d-1,d.
   \label{eq:CoMP}
  \end{align}
   Note that 
  this is justified also for $\alpha =d$, since
  along the last path $\ell_{V_dGV_Q}=\ell_{V_dGV_0}  $,
  $\theta_d:2\pi\to 0 $ and $\Delta \varphi_d=-2\pi$.
Since the center of mass $P_\alpha $
is a physical observable of the snapshot ground state for the open boundary condition,
it is periodic in time, $P_\alpha (t+T)=P_\alpha (t)$,
as the Hamiltonian is periodic.
  An important observation is that $\bar P^\ell(t)$ is not continuous ($L\to\infty$)  and has discontinuities at $t=t_i$, $i=1,2,\cdots$ of the jump $\pm \frac 1 2 $ due to
  edge states (as shown later). 
  Then the pumped charge  $Q^\ell_\op$
  in the cycle $T$ due to the current $\bar j^\ell_\op$
  is written as
  \begin{align}
   Q^\ell_\op &= \int_0^Tdt\, \bar j^\ell_\op
 =\sum_i
   \int_{t_i}^{t_{i+1}}dt\,\bar j_\op^\ell
\nonumber   \\
   &= \sum_i  \int_{t_i+0}^{t_{i+1}-0}dt\,
\partial _t \bar P^\ell(t)
=
\sum_i \bar P^\ell(t)\big|_{t_i+0}^{t_{i+1}-0}
\nonumber
\\
&=
- \sum_i\bar P^\ell(t)\big|_{t_i-0}^{t_{i}+0}
=-\sum_i \Delta \bar P^\ell (t_i)
=I^\ell.
\label{eq:charge-by-disc}
  \end{align}
The discontinuities $I^\ell$ is defined by
  \begin{align}
   - I^\ell &= \sum_i \Delta \bar P^\ell (t_i),
\\
   \Delta \bar P^\ell (t_i) &= \bar P^\ell(t)\big|_{t_i-0}^{t_{i}+0}
   = \pm \frac 1 2,
   \label{eq:disc-p}
  \end{align}
  where 
  $\bar P^\ell(t)$ is not continuous at
  $t=t_i$ ($i=1,\cdots$).
  The sign is determined by the behavior of the edge state that causes
  the jump.
  Since the localization length (typical length scale)
  of the edge states
  is finite,
  it scales to zero in the rescaled coordinate $x_j$. It implies
  the contribution of the edge states localized near one of the boundaries
  is $\pm \frac 1 2 $
  (See appendix \ref{sec:disc}).
  Due to the conservation of the charge, the number of the discontinuities
  is even. It implies the sum of the discontinuities,
  $I$, is an integer. This is the quantization of the pumped charge.

  The physical current
  $\bar j_\op^\ell $ is carried by the bulk
  even with the open boundary condition and is 
determined by the 
discontinuities due to the edge states by the back action
based on the periodicity of
$\bar P^\ell$ in time. 
  
  \subsection{Bulk-edge correspondence}
  \label{sec:bec}
  As for the periodic boundary condition, $\bar A_\theta ^{(t),\ell}$ is
  smooth and the 
  pumped charge averaged along the path $\ell$
  is given by
  \begin{align}
   Q_\pe^\ell &\equiv
   \int_0^Tdt\, \bar j^\ell_\pe\nonumber
   \\
   &= 
   \frac {1}{2\pi i}
   \int_0^T dt\,
   \int_\ell d \theta \,
   B_\pe(\theta ,t ),
   \\
   B_\pe(\theta ,t ) &=
   \partial _\theta A_{t,\pe}-\partial _t A_{\theta,\pe},
   \\
   A_{\mu,\pe} &= \langle g_\pe|\partial _\mu g_\pe \rangle,
  \end{align}
  where $\bar H_\pe | g_\pe \rangle = |g_\pe \rangle E_g$.
  As for the periodic boundary condition,
  the Berry connection $A_{\mu,\pe} $ and thus $B_\pe$ is also
  defined by the Hamiltonian $\bar H_\pe$, Eq.(\ref{eq:g-periodic}),
  that is not periodic/invariant by the shift
  $\Theta \to \Theta + 2\pi \vec{e}_\alpha  $.
  See Table \ref{table:HandS}.
  It implies that the periodic Hamiltonian $\bar H_\pe$, Eq.(\ref{eq:g-periodic}),
  is not defined on the torus $T^d$.
  On the other hand,
  the ground state,
  $|g_0 \rangle $, of the twisted Hamiltonian, $ H_\tw$, Eq.(\ref{eq:g-open}),
  is periodic by the shift and well defined on the torus $T^d$.
  Noting that
  $|g_\pe \rangle = {\cal U}_\largeG | g _{0,\tw} \rangle  $ and
  $ H_\tw | g_{0,\tw}\rangle = | g_{0,\tw} \rangle E_g^\tw $, 
one has
  \begin{align}
   A_{\theta,\pe} &= -i \sum_\alpha \partial_\theta \varphi_\alpha P_{\alpha,\tw} 
   +A_{\theta,\tw},
   \\
A_{t,\pe} &= A_{t,\tw} ,
  \end{align}
  where $ P_{\alpha,\tw} = \langle g_{0,\tw }| {\cal P}_\alpha | g_{0,\tw} \rangle $.

It results in
  \begin{align}
   B _\pe &= B_\tw
+i \sum_\alpha \partial_\theta \varphi_\alpha \partial _t P_{\alpha,\tw}.
  \end{align}
  Since $P_{\alpha,\tw}$
  is smooth and
  periodic in time, the last term does not contribute to the total pumped charge. Then
  \begin{align}
   Q^\ell_\pe &= \frac {1}{2\pi i} \int_\ell d \theta \,\int_0^T dt\,
   B_\tw(\theta ,t )\equiv C^\ell.
   \label{eq:bec-qandC}
  \end{align}
  This integral is over
  a torus $T^2=\{(\theta ,t)|\theta\in \ell, t\in[0,T]\}$
  without boundaries. It gives the Chern number $C$ that is integer.
  Now due to Eq.(\ref{eq:charge-by-disc}), we have
  \begin{align}
   Q^\ell_\pe &=    Q^\ell_\tw = Q^\ell_\op,
   \label{eq:bec-charge}
   \\
   I^\ell &= C^\ell.
   \label{eq:bec-topo}
  \end{align}
    {\em   This is the bulk-edge correspondence of the topological pump. }

     As for the canonical path $\ell_{V_{\alpha }GV_{\alpha+1} }$, $\alpha =0,\cdots d=\QN-1$,
     it is given by
    \begin{align}
     I^\alpha &= C^\alpha,\ 
     \alpha =0,\cdots,\QN-1,
     \label{eq:becP}
    \end{align}
    where
    \begin{align}
     I^\alpha &\equiv I^{\ell_{V_{\alpha }GV_{\alpha+1} }}=\sum_i \Delta P_{\alpha}(t_i),
     \label{eq:CanI}
    \end{align} 
     is a sum of the discontinuities 
    of the $\alpha $-particle in the cycle and 
    $C^\alpha \equiv C^{\ell_{V_{\alpha }GV_{\alpha+1} }}$
    is the Chern number defined on a torus
    $T^2=[0,T]\times V_{\alpha }GV_{\alpha+1} $. 
    
     
 \section{$Z_\QN$ Berry phase, symmetry and gap closing}
 \label{sec:qBerryGap}
 \subsection{$Z_\QN$ quantization}
 \label{sec:qBerry}
    In this section, let us discuss the Hamiltonian without symmetry breaking term ($\Delta =0$).
    Using the Hamiltonian $H_\tw$,
    the Berry phase $\gamma_\ell $ is defined
    since the path $\ell$ forms a loop for $H_\tw$ as
    \begin{align}
     i \gamma_\ell &= \int_\ell d \theta\, A_\theta.
    \end{align}

    The $Z_\QN$ shift ${\cal U}_{Z_\QN} $ as a shift
    in the parameter space as shown in Sec.\ref{sec:symm}
    induces a map of the Hamiltonians
    in the parameter space 
    $\varphi_{\alpha }\to\varphi_{\alpha+1 } $
(See Eq.(\ref{eq:theta-shift}))    
and the canonical loop     
    $\ell_{V_\alpha G V_{\alpha +1}}$ 
    \begin{align}
     {\cal U}_{Z_\QN}
     \bar H_\tw(\bm{\theta} )
        {\cal U}_{Z_\QN} ^\dagger
        &=
        \bar H_\tw(\bm{\theta }^\prime ),
    \end{align}
    where 2 points $\bm{\theta } $ and $\bm{\theta }^\prime $
    in $\Theta $ are parameterized by the same $\theta $ as 
    \begin{align*}
     \bm{\theta} &\in \ell_{V_{\alpha-1} G V_{\alpha }}(\theta ),
     \\
     \bm{\theta}^\prime &\in \ell_{V_{\alpha} G V_{\alpha+1} }(\theta ).
    \end{align*}
    It implies that we may take
    \begin{align}
     |g_{0,\tw} (\bm{\theta } ^\prime) \rangle &=
{\cal U}_{Z_\QN}      |g_{0,\tw} (\bm{\theta } ) \rangle ,
    \end{align}
    and
    \begin{align}
     \gamma _{\ell_{V_{\alpha-1} G V_{\alpha }}}
&=     
     \gamma _{\ell_{V_\alpha G V_{\alpha +1}}} \equiv\gamma _\QN,\ \alpha=1,\cdots \QN
    \end{align}
    since ${\cal U}_{Z_\QN} $ does not depend on the parameter and thus
    $
  \langle g_{0,\tw} (\bm{\theta } ^\prime) | \partial _\theta g_{0,\tw} (\bm{\theta } ^\prime) \rangle =
  \langle g_{0,\tw} (\bm{\theta }) | \partial _\theta g_{0,\tw} (\bm{\theta })\rangle $.
  Then using the fact,
  \begin{align}
\sum_{\alpha =1}^Q \ell_{V_\alpha G V_{\alpha +1}} &= 0,
  \end{align}
 it results $Q \gamma _Q=0$, (mod $2\pi$). It implies $Z_\QN$ quantization
  \begin{align}
     \gamma _\QN &=   \frac {2\pi n}{Q},\ \ n\in \mathbb{Z}.
  \end{align}
    This $Z_\QN$ Berry phase characterizes
    a symmetry protected topological phase \cite{Hatsugai06qb,Hatsugai10,Hatsugai11,Kariyado18}.
    It is
    a generalized the $Z_2 $ Berry phase characterizes a singlet pair or a covalent bond \cite{Hatsugai06qb}
    \footnote{
    This $Z_2$ quantization for the Berry phase for the non-interacting systems was    also discussed before \cite{zak89,yh09}.}.
    Similarly $\gamma _\QN$ characterizes the $\SUQ$ dimer phase.
    The dimer limit is characterized by this $Z_\QN$ Berry phase and the symmetry protection and adiabatic continuity
    guarantee the quantization. 
    Unless the gap closes by the deformation to the dimer limit $J_i=0$ ($i$:odd),
the Berry phase is given by
(See Fig.\ref{fig:znberry} and 
appendix \ref{sec:dimer})
\begin{align}
 \gamma _\QN &= +\frac {2\pi}{\QN},\ \text{mod }2\pi.
\end{align}
It is topologically stable unless the gap closes against
finite coupling $J_p<J_e<0$ and $J_e<J_o<0$.
It also gives
\begin{align} 
 \gamma _{V_{\alpha }G V_{\beta ^\prime }} &=  + \frac {2\pi  } \QN (\alpha -\beta  ),\ \text{mod }2\pi.
\end{align}


\begin{figure}[t] 
 \begin{center}
 \includegraphics[width=85mm]{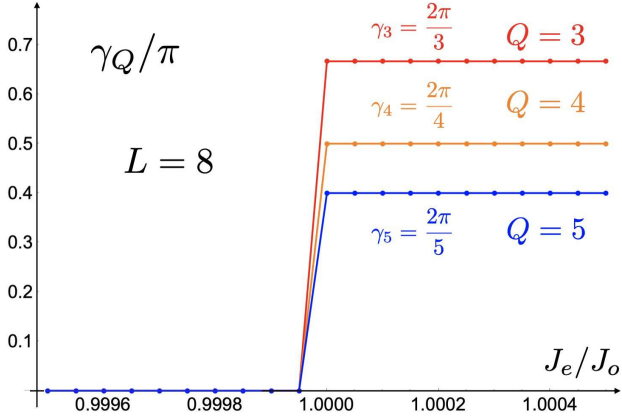}
 \end{center} 
\caption{\label{fig:znberry} 
  $Z_\QN$ Berry phase $\gamma_\QN $ for $\QN=3,4$ and $5$
 as a function of $J_e/J_o$ ($L=8$). We have confirmed that the results
 are reproduced by the symmetry indicators Eq.(\ref{eq:symind}).}
\end{figure}

\subsection{Symmetry indicators}
\label{sec:indc}
Let us first note that the Berry phase
needs to be evaluated by a single gauge fixing 
\cite{Hatsugai04gaugefix,Hatsugai07char,Hatsugai06qb,Hatsugai10} (see Appendix \ref{sec:Berry-gauge}).
Here we assume that $| g_\phi \rangle $ is gauge fixed over the loop
$\ell_{V_{\alpha -1}G V_\alpha }=\ell_{V_{\alpha -1}G}-\ell_{V_{\alpha }G}$ by a single gauge fixing by $|\phi \rangle $
as
\begin{align}
 | g_\phi \rangle &= P|\phi \rangle /\sqrt{N_\phi} ,
\end{align}
where
$ P = |g_{0,\tw} \rangle \langle g_{0,\tw} |$ and
$ N_\phi=|\langle \phi| g_{0,\tw} \rangle |^2\ne 0$
    \footnote{If $N_\phi=   |\langle \phi| g_{0,\tw} \rangle |=0$,
    slightly modify $|\phi \rangle  $, then $N_\phi\ne 0$.
  It is always possible if
the dimension of the Hilbert  space is larger than $1$.}.

Noting that
\begin{align}
 H_\tw(\bm{\theta }^\prime ) &=
 {\cal U}_{Z_\QN}
 H_\tw(\bm{\theta } ) 
{\cal U}_{Z_\QN} ^\dagger  ,
\end{align}
where
   $   \bm{\theta} \in \ell_{V_{\alpha-1} G}(\theta )$,
and
$   \bm{\theta}^\prime \in \ell_{V_{\alpha} G }(\theta )$,
one has
\begin{align}
     |g_\phi (\bm{\theta } ^\prime) \rangle 
 &= 
 {\cal U}_{Z_\QN}      |g_\phi (\bm{\theta } ) \rangle
 e^{-i ^\exists \Omega (\bm{\theta } )}.
 \label{eq:ind}
\end{align}
 This extra phase factor is due to the fact that
   the phase convention of the state
$     |g_\phi (\bm{\theta } ^\prime) \rangle $ by
$|\phi \rangle $ at $\bm{\theta } ^\prime $
 is, in general, different from that of
 $ {\cal U}_{Z_\QN}      |g_\phi (\bm{\theta } ) \rangle$.

Since ${\cal U}_{Z_\QN} $ is independent of the parameter,
the Berry phase is written as
\begin{align}
 \gamma _Q &=  \gamma _{\ell_{V_{\alpha -1}G V_{\alpha} }} =-i\int_{\ell_{V_{\alpha -1}G V_ \alpha }}
 d \theta \, \langle g_\phi |  \partial _\theta g_\phi \rangle
\nonumber 
 \\
 &= 
-i \int_{\ell_{V_{\alpha -1}G}}d \theta \, \langle g_\phi |  \partial _\theta g_\phi \rangle 
 +i
 \int _{\ell_{V_{\alpha }G}} d \theta \, \langle g_\phi | \partial _\theta g_\phi \rangle
 \nonumber
 \\
 &=
 {+} \int _{\ell_{V_{\alpha-1 }G}}
 d \theta \, \partial _\theta \Omega
 = \Omega (G)-\Omega (O ),
\end{align}
where all vertices $V_\alpha $ are identified to the origin $O $ in modulo $2\pi$.
Supplementing $\theta _Q=-\sum_{\alpha =1}^d \theta _\alpha $,
at the vertices $O$ and $G$, the parameters are 
\begin{align}
 \bm{\theta } = { O} &: (\theta_1,\cdots,\theta_Q)=(0,\cdots,0),\qquad \text{mod}\, 2\pi
\\
 \bm{\theta } = G &: (\theta_1,\cdots,\theta_Q)=(\frac {2\pi}{Q} ,\cdots,\frac {2\pi}{Q} ),\ \text{mod}\, 2\pi.
\end{align}
It implies the Hamiltonian is invariant by the shift of the fermions by ${\cal U}_{Z_\QN} $
at  $O$ and $G$ as
\begin{align}
 \cmt{H_\tw(\bm{\theta } )}{{\cal U}_{Z_\QN} } &= 0,\ \ \bm{\theta } =O,G.
\end{align}
Therefore ${\cal U}_{Z_\QN} $ is a symmetry of the Hamiltonian
and {$e^{i\Omega} $} is an eigen value of the symmetry operation (symmetry indicator)
as
\begin{align}
 {\cal U}_{Z_\QN} | g_\phi(O) \rangle &= | g_\phi(O) \rangle e^{{+}i \Omega(O) },
 \label{eq:symO}
 \\
 {\cal U}_{Z_\QN} | g_\phi(G) \rangle &= | g_\phi(G) \rangle e^{+{}i \Omega(G) } .
 \label{eq:symG}
\end{align}
Since ${\cal U} _{Z_\QN}$ is trivial at $O$ and $\Omega (O)=0$, it results
in
\begin{align}
 \gamma _Q &= \Omega (G)=\text{Arg}\,
  \langle g_\phi(G) |{\cal U} _{Z_\QN}  | g_\phi(G) \rangle .
 \label{eq:symind}
\end{align}
Physical meaning of these quantities is clear by the adiabatic deformation to the dimer limit
by the $Z_\QN$ Berry phase. It also implies 
the gap closing at $G$ associated with a topological transition due to the discrete change of
the $Z_\QN$ Berry phase.

It is also directly observed by the discretized formula of the Berry phase
(discretizing the path $V_0G$ into $M$ segments
$\theta _m={\frac {2\pi}{\QN} } \frac {m}{M}$, ($ m=0,\cdots,M $) as
 \begin{align*}
 \gamma _\QN &=
  \lim_{M\to\infty}\text{Arg}\,\big(
 \langle \theta_0 |\theta _{1} \rangle
\langle \theta_1 |\theta _2 \rangle\cdots 
\langle \theta_{M-1} |\theta _M \rangle
\\
&\qquad\qquad
\times
 \langle \theta_M|\theta _{M-1} ^\prime \rangle
 \cdots \langle \theta_2 ^\prime |\theta _1 ^\prime \rangle
  \langle \theta_1 ^\prime |\theta _0 \rangle \big).
\end{align*}
where $|\theta _m ^\prime \rangle =
{\cal U}_{Z_\QN}|\theta _m \rangle $, 
$|\theta _m \rangle = |g_{{\phi}}(\theta_m) \rangle $.
Since
  {
 ${\cal U}_{Z_\QN} | \theta _M \rangle = | \theta _M \rangle e^{i \Omega (G)}$ and
   ${\cal U}_{Z_\QN} | \theta _0 \rangle = | \theta _0 \rangle e^{i \Omega (O)}$,  }
  it is written as
     \begin{align*}
 \gamma _\QN &= \lim_{M\to\infty}\text{Arg}\,\big(
 \langle \theta_0 | \theta _{1}  \rangle
\langle \theta_1 |\theta _2 \rangle\cdots 
\langle \theta_{M-1} |\theta _M \rangle
\\
&\qquad\qquad
\times
 \langle \theta_M|{\cal U}_{Z_\QN}|\theta _{M-1} \rangle
 \cdots \langle \theta_2|\theta _1 \rangle
 \langle \theta_1| {\cal U}^\dagger _{Z_\QN}|\theta _0 \rangle\big) 
 \\
 &=
\Omega (G)-\Omega (O),
\end{align*}
 where
 ${\cal U}_{Z_\QN} ^\dagger | \theta _0 \rangle
 = | \theta _0 \rangle e^{-i \Omega (O)}$ and  $\langle \theta _M |{\cal U}_{Z_\QN} = e^{i \Omega (G)}\langle \theta _M |  $ due to Eqs.(\ref{eq:symO}) and (\ref{eq:symG}).
  
\subsection{Modified Lieb-Schultz-Mattis (LSM) argument}
\label{sec:mLSM}
As is clear, the system of the dimer limit is gapped.
This gap is stable for inclusion of finite coupling between
the dimers, at least, for a finite size system.
One may naturally expect this gap converges to some finite
values by taking an infinite size limit $L\to\infty$
assuming the ground state is
adiabatically connected to a set of disconnected dimers.
 As for a uniform system, existence of the gap is
unclear and the problem has a long history of studies.
Some of the recent studies are topological.
Especially in relation with the Haldane conjecture for
the $S=1$ Heisenberg model \cite{haldane83}.
Since 
$H^{(2)}$ for $\QN=2$, is
equivalent to the standard $S=1/2$ Heisenberg model,
$H^{(1)}$ (appendix \ref{sec:su2}),
well-known Lieb-Schultz-Mattis (LSM) theorem \cite{lsm61,affleck86}
is applied and
the energy gap of the finite system
with the periodic boundary condition vanishes when $L\to\infty$.
This is consistent with the existence of the
gapless excitation as the
des Cloizeaux and Pearson mode of the $\QN=2$ case \cite{dp62}.
Note that the LSM theorem also allows existence of the
finite size gap between the states
which become degenerate in the thermodynamic limit
associated with the symmetry breaking.
On the other hand, for the $\QN=3$ case,
a series of studies \cite{Parkinson87,bb89,Klumper89}
has clarified that the uniform system has doubly
degenerate dimerized ground states in the $L\to\infty $ limit.
The case, $\QN>3$, is also discussed by Affleck suggesting a 
similar conclusion
(double degeneracy due to dimerization) \cite{Affleck90}.
In this subsection, we give a topological argument
for the gap closing for even $\QN\ge 2$ of the finite size system.

The gap of the finite system under the twist
is strongly constrained by considering the Berry phase,
which works as a topological order parameter responding to
the local twist as an external perturbation
 \cite{Hatsugai06qb,Hatsugai11,Chepiga2013,Kariyadofrac15,Maru18,Kudo21}. 
In Ref. \cite{Hirano08Deg},
the standard $S=1/2$ Heisenberg model, $H^{(1)}$ ($\QN=2$)
was considered.
If the gap remains open for all values
of the twist, one can prove that the Berry phase pattern,
associated with the local twist at the link, needs to be
alternating in this $S=1/2$ case. This clearly contradicts with the uniformity of
the system. It results in that the Berry phase can not be defined, that is,
the gap closing of the system
at some twist \cite{Hirano08Deg}.
The argument can be extended to the present system $\QN>2$ as shown here.
The claim is that,
 as for a finite system of even $\QN\ge 2$,
 the energy gap
between the ground state and the next one
under the twisted boundary condition
vanishes at some twist.

Up to this point,
 we have discussed Berry phases associated with the twist at the boundary link $L$ and $1$. Let us write it as $\gamma ^L_\ell$.
In a similar way, one may also define the Berry phase $\gamma ^{L-1}$
associated with the twist at
$L-1$ and $L$.
Let us write the Hamiltonians with the twists as
\begin{align}
H_\tw^L &= J_e \sum_{\alpha, \beta }e^{-i(\varphi_\alpha -\varphi_\beta )}c_{L,\alpha } ^\dagger c_{1,\alpha } ^\dagger
c_{1,\beta } c_{L,\beta }
\nonumber
\\
 &+J_o \sum_{\alpha, \beta }c_{L-1,\alpha } ^\dagger c_{L,\alpha } ^\dagger
c_{L,\beta } c_{L-1,\beta }+ \cdots,
\\
H_\tw^{L-1} &= 
J_e \sum_{\alpha, \beta }c_{L,\alpha } ^\dagger c_{1,\alpha } ^\dagger
c_{1,\beta } c_{L,\beta }
\nonumber\\
 &+J_o \sum_{\alpha, \beta }e^{{+}i(\varphi_\alpha -\varphi_\beta )}c_{L-1,\alpha } ^\dagger c_{L,\alpha } ^\dagger
c_{L,\beta } c_{L-1,\beta }+ \cdots,
\end{align}
where $\cdots$ does not include $c_{L,\alpha }$.
 {\it Note that the sign of the twist is reversed.}
They are related with each other by the
the gauge transformation ${\cal U}_L=e^{{-}i \sum_\alpha \varphi_\alpha \hat n_{L,\alpha }} $,
${\cal U}_L ^\dagger c_{L,\alpha }{\cal U} _L =e^{{+}i \varphi_\alpha \hat n_{L,\alpha } }
c_{L,\alpha }$
(See also appendix \ref{sec:dimer})
as
\begin{align}
H_\tw^{L} &= {\cal U} _L H_\tw^{L-1} {\cal U} _L ^\dagger. 
\end{align}

 The Berry phases are defined by $| g ^L \rangle $ and $| g ^{L-1} \rangle $,
 which are the ground states of $H^{L}_\tw$ and $H^{L-1}_\tw$ respectively
 as
\begin{align} 
 i \gamma ^L &= \int_\ell d \theta A^L,\ A^L = \langle g^L| \partial _\theta g^L \rangle,
\\
 i \gamma ^{L-1}
&= \int_\ell d \theta A^{L-1},\ A^{L-1} = \langle g^{L-1}| \partial _\theta g^{L-1} \rangle ,
\end{align}
where $H^L| g^L \rangle =| g^L \rangle E$ and
$H^{L-1}| g^{L-1} \rangle =| g^{L-1} \rangle E$.
Noting that $| g ^L \rangle ={\cal U}_L |g^{L-1} \rangle $
\footnote{
 We take a global single gauge for $| g^{L-1} \rangle $ (appendix \ref{sec:Berry-gauge}).
 Then the phase of 
  $| g^{L} \rangle\equiv {\cal U}_L | g^{L-1} \rangle $
 is also globally unique. 
It implies Eq.(\ref{eq:lsm})
is without $2\pi$ ambiguity.
}, it induces
\begin{align}
 A^L
 &= A^{L-1}
 {-} i \sum_\alpha( \partial_\theta \varphi_\alpha ) \langle g^{L-1}| \hat n_{L,\alpha }| g^{L-1} \rangle.
 \label{eq:twist}
\end{align}
Generically, with the twist, $\SUQ$ symmetry is (slightly)
broken even without
explicit symmetry breaking term $H_B$, that is,
the fermions with different colors are not equivalent
and
$ \langle g^{L-1}| \hat n_{L,\alpha }| g^{L-1} \rangle \ne \frac {1}{\QN} $.
 However, 
this symmetry breaking effect due to the twist
is not localized at the twisted link.
The large gauge transformation Eqs.(\ref{eq:defG}) and (\ref{eq:def-ug}), 
${\cal U}_\largeG $, maps the periodic system with ${\cal O}(1/L) $ twist
to the system with twisted boundary condition
preserving 
the local charge density remains unchanged 
because ($\cmt{{\cal U}_{\largeG}}{\hat n_{j,\alpha }}=0)$.
Then 
$ \langle g^{L-1}| \hat n_{j,\alpha }| g^{L-1} \rangle$ is $j$ independent
both for the periodic/twisted system. 
It implies the effects are of the order of $L ^{-1} $ as
$ \langle g^{L-1}| \hat n_{L,\alpha }| g^{L-1} \rangle =\frac {1}{\QN} +{\cal O}(L ^{-1} ) $.
Then integrating Eq.(\ref{eq:twist}) over the loop $\ell_{V_\alpha G V_{\alpha +1}}$,
we have for a sufficiently large system
\begin{align}
 \gamma_{\ell_{V_\alpha G V_{\alpha +1}}} ^L &=  \gamma_{\ell_{V_\alpha G V_{\alpha +1}}} ^{L-1}
{+} \frac {2\pi}{\QN},
\label{eq:lsm} 
\end{align}
since $\Delta \varphi_\beta =-2\pi (\beta=\alpha ),0 (\beta\ne\alpha )$.
 Note that the possible ${\cal O}(L ^{-1} ) $ extra
 term vanishes after the integration to be consistent with
 the $Z_\QN$ quantization of the Berry phases.
This constraint needs to be satisfied for any systems even with
 site dependent $J_i$'s.
The two dimer limits,
  $  \gamma_{\ell_{V_\alpha G V_{\alpha +1}}} ^L = 0$,
  $  \gamma_{\ell_{V_\alpha G V_{\alpha +1}}} ^{L-1} = {-}\frac {2\pi}{Q} $
  and
  $  \gamma_{\ell_{V_\alpha G V_{\alpha +1}}} ^L = {+}\frac {2\pi}{Q} $,
  $  \gamma_{\ell_{V_\alpha G V_{\alpha +1}}} ^{L-1} =0$,
  are consistent with Eq.(\ref{eq:lsm}).

   If the system is uniform, $J_o=J_e$, the anti-translation invariance
of the system with the twist,
$H^{L}_\tw={\cal A}_{{\cal T} } H_\tw^{L-1} {\cal A} _{{\cal T} } ^{-1}$,
implies
 \begin{align} 
  \gamma_{\ell_{V_\alpha G V_{\alpha +1}}} ^L &= 
  - \gamma_{\ell_{V_\alpha G V_{\alpha +1}}} ^{L-1},\
   \text{mod } 2\pi.
  \label{eq:lsm-uni}
\end{align}
  By Eqs.(\ref{eq:lsm}) and (\ref{eq:lsm-uni}), we have constraints
  for the Berry phase for a uniform system as
  \begin{align}
   \gamma_{\ell_{V_\alpha G V_{\alpha +1}}} ^L &= 
   - \gamma_{\ell_{V_\alpha G V_{\alpha +1}}} ^{L-1}={+}\frac {\pi}{Q},  \ \text{mod } 2\pi
   \label{eq:case1}
  \end{align}
   or
  \begin{align}
   \gamma_{\ell_{V_\alpha G V_{\alpha +1}}} ^L &= \pi\frac {Q{+}1}{Q},
   \ \gamma_{\ell_{V_\alpha G V_{\alpha +1}}} ^{L-1}=\pi\frac {Q{-}1}{Q},  \  \text{mod } 2\pi.
   \label{eq:case2}
  \end{align}

  As for the even $\QN$, these constraints,
   Eq.(\ref{eq:case1}) and Eq.(\ref{eq:case2}),
   contradict the $Z_\QN$ quantization of the Berry phases
  $\gamma =2\pi\frac {n}{Q},\ n\in\mathbb{Z}$.
   This contradiction implies that the Berry phase can not be well-defined.
  It is only possible when the gap of the (finite size) system closes.
  A level crossing between the ground state and the next one occurs
  at some twist parameter.
  Assuming the degenerate dimer states for the infinite size system,
  the gap between the linear combinations of the dimer states of the finite size system closes at the twisted parameters.
  As for the odd $\QN$, the second case, Eq.(\ref{eq:case2}),
   is compatible with the $Z_\QN$ quantization ($Q\pm 1$ is even), 
   although these quantized values are
   different from that of the dimer limit
   \footnote{ 
   It implies the gap closing at some twist
     by inclusion of the coupling  between the dimers.}.
   In principle, it
   allows a unique gapped ground state of the uniform system
   for any value of the twist.
     Although it does not occur in
     the present numerical calculations shown in Fig.\ref{fig:znberry},
     inclusion of long range
  couplings and additional terms which respect $Z_\QN$ symmetry may realize such a ground state.

  \section{Emergent $Z_\QN\times Z_\QN$ symmetry and explicit Chern numbers}
  \label{sec:emergent}

 \subsection{Open system: edge states and low energy spectrum}

\begin{figure}[t] 
 \begin{center}
 \includegraphics[width=85mm]{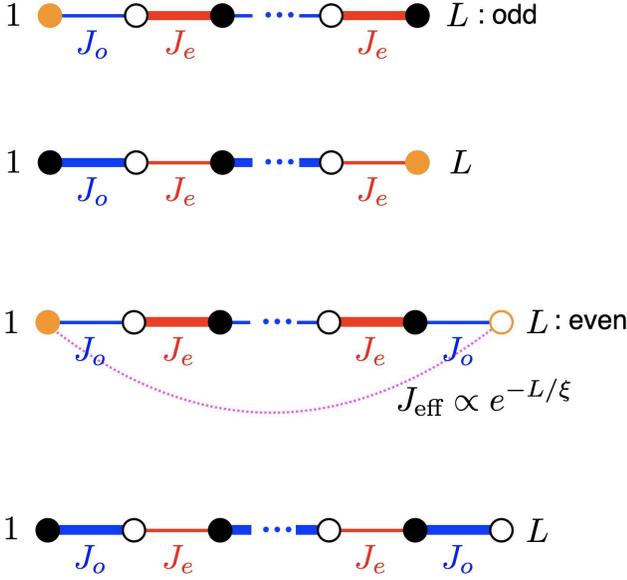}
 \end{center} 
\caption{\label{fig:dimers} 
 Dimer configurations of open systems for even/odd sites.
}
\end{figure}

 
 Although most of the discussion in the paper is for even $L$ systems,
 let us consider, in this section,   even/odd  $L$ systems separately,
 especially 
 near the dimer limits $|J_o|\ll |J_e|$
 and $|J_e|\ll |J_o|$.
 
 \underline{ $L$:odd. }
When the system size $L$ is odd,
the ground states are given for each dimer limits $J_o=0$ and $J_e=0$
as (See Fig.\ref{fig:dimers})
\begin{align}
 |g_{1,eo}^{L:\text{odd}},\alpha \rangle 
 &= c_{1,\alpha } ^\dagger
 \prod_{j=1}^{(L-1)/2} \psi_{ 2j+1,2j} ^\dagger | 0 \rangle,\ (J_o=0),
 \\
 |g_{L,eo}^{L:\text{odd}},\alpha \rangle 
 &= c_{L,\alpha } ^\dagger
 \prod_{j=1}^{(L-1)/2} \psi_{ 2j-1,2j} ^\dagger | 0 \rangle,\ (J_e=0),
\end{align}
where 
$\bar N_\beta = {1} $ for $\beta =\alpha$ and ${0}$ otherwise.
 It implies $\QN $-fold degeneracy of the ground states.
Their charge distributions are
\begin{align}
\langle g_{1,eo}^{L:\text{odd}},\alpha |n_{ j,\beta }
|g_{1,eo}^{L:\text{odd}},\alpha \rangle
&=
\mchsss
  {\frac {1}{Q} }{j\ne 1}
  {0}{j= 1, \beta \ne \alpha}  
  {1}{j= 1, \beta =\alpha},
  \\
\langle g_{L,eo}^{L:\text{odd}},\alpha |n_{ j,\beta }
|g_{L,eo}^{L:\text{odd}},\alpha \rangle
&=
\mchsss
  {\frac {1}{Q} }{j\ne L}
  {0}{j= 1, \beta \ne \alpha}  
  {1}{j= 1, \beta =\alpha},
\end{align}
{where}  $ |g_{1,eo}^{L:\text{odd}},\alpha \rangle $ is a product of the bulk and
 completely localized state at $j=1$ with the color $\alpha $.
{Similarly} $ |g_{L,eo}^{L:\text{odd}},\alpha \rangle $ is a product of the bulk and
 completely localized state at $j=L$ with the color $\alpha $.
This degeneracy is stable for inclusion of a finite coupling $J_o$ and $J_e$
since finite matrix elements
with different quantum numbers $\bar N_\alpha $'s are prohibited due to
the symmetry.
 It implies the charge distributions are modified continuously 
 for a finite coupling.
The numerical results for
$L=9$ systems with dimerization
 $\bar N_\alpha :(1,0,0,0,0)$
are obtained by the exact diagonalization
and shown in Fig.\ref{fig:charge-oddL}.
They are consistent with the present picture.
The ground state is given by
the gapped bulk and boundary states
(edge states at both ends) localized near the boundaries.

 \begin{figure}[t]
\includegraphics[width=90mm]{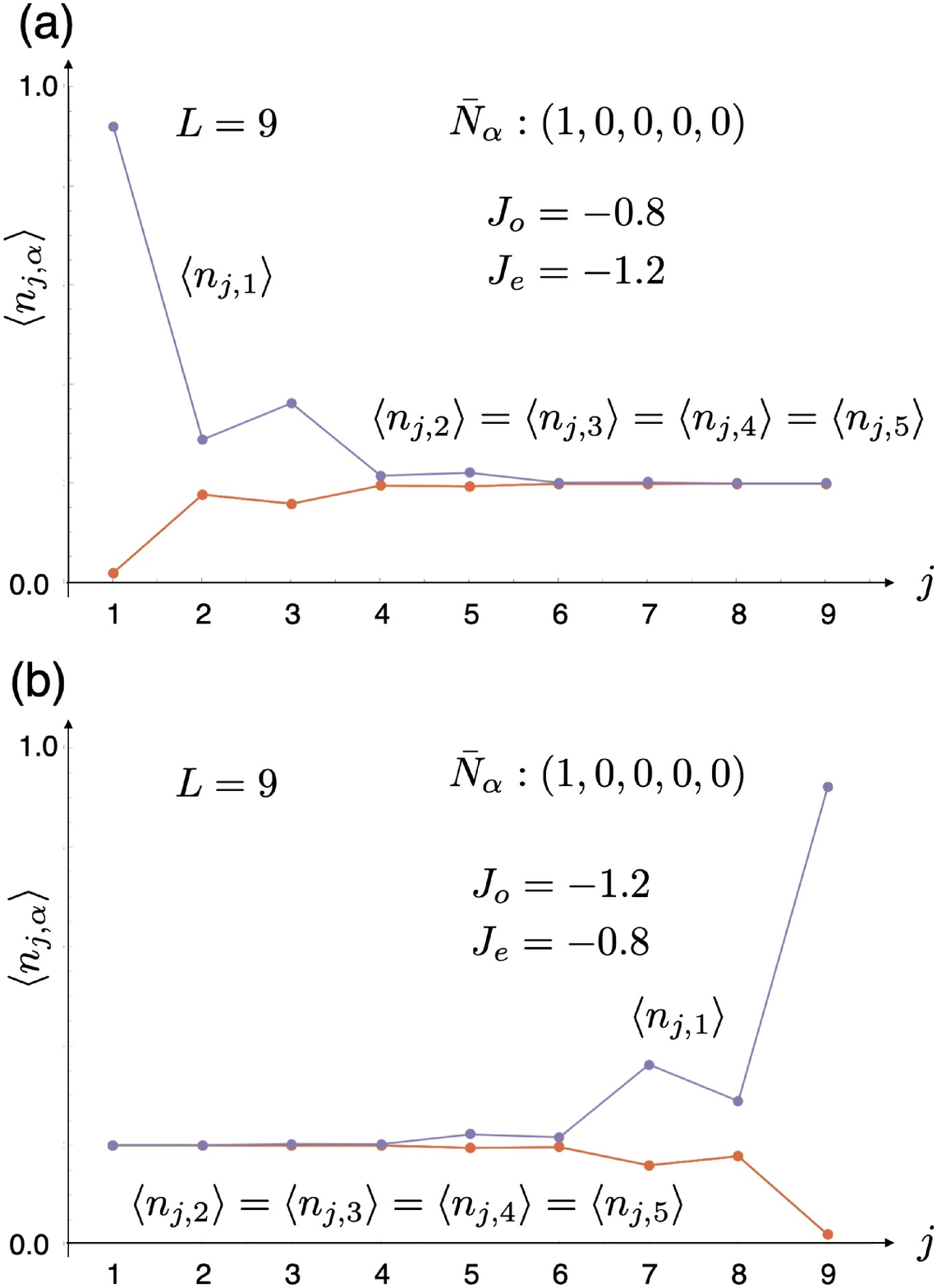}
\caption{\label{fig:charge-oddL}
 Charge distribution $\langle n_{j,\alpha } \rangle $ of the
unique ground state of the odd $L$ system ($L=9$) with $\QN=5$, $\bar N_\alpha :(1,0,0,0,0)$
 (a): $J_o=-0.8$, $J_e=-1.2$ and (b): $J_o=-1.2$, $J_e=-0.8$.
}
\end{figure}

\underline{ $L$:even.}
As for the system with $L$: even, the ground states of the dimer limit
are again given by
\begin{align}
 |g_{\op, eo }^{L:\text{even}} (\alpha ,\beta) \rangle 
 &= c_{1,\alpha } ^\dagger
  c_{L,\beta } ^\dagger
  \prod_{j=1}^{L/2-1} \psi_{2j+1, 2j} ^\dagger | 0 \rangle, (J_o=0)
  \label{eq:geo}
 \\
 |g_{\op,oe}^{L:\text{even} } \rangle  &=
 \prod_{j=1}^{L/2} \psi_{ 2j-1,2j} ^\dagger | 0 \rangle,\ (J_e=0)
 \label{eq:goe}
\end{align}
where $\bar N_\gamma =0, ^\forall\! \gamma $ for both cases.
It implies $\QN^2$-fold degeneracy for $J_o=0$ and
gapped unique ground state for $J_e=0$.
If $|J_o|>|J_e|$, the unique gapped ground state,
$ |g_{eo}^{L:\text{even} } \rangle  $ is stable for inclusion of the
finite coupling $J_e$.
However, as for $|J_o|<|J_e|$,
in contrast to the $L$:odd case, 
the $\QN^2$-fold degeneracy of the ground states
is unstable
for the finite size systems. 
The degeneracy is lifted for the finite coupling
due to the residual interaction between the edge states at both ends.
As for a chain of the finite length, we expect an effective coupling between
 the two boundary states at both ends.
 It is a generalization of the Kennedy's discussion
 \cite{Kennedy90,expKennedy,arikawa09,chepiga18}.
 This effective coupling, $J_{\rm eff}$, is expected to behave as $e^{-L/\xi}$
for $L$   where
$\xi$ is a correlation length between the edge states,
 which can be proportional to the inverse of the bulk energy gap.
 This is confirmed numerically for $\QN=3$ and $4$ in Fig.\ref{fig:gap3} and Fig.\ref{fig:gap5}.
 

\begin{figure}[t] 
 \begin{center}
 \includegraphics[width=80mm]{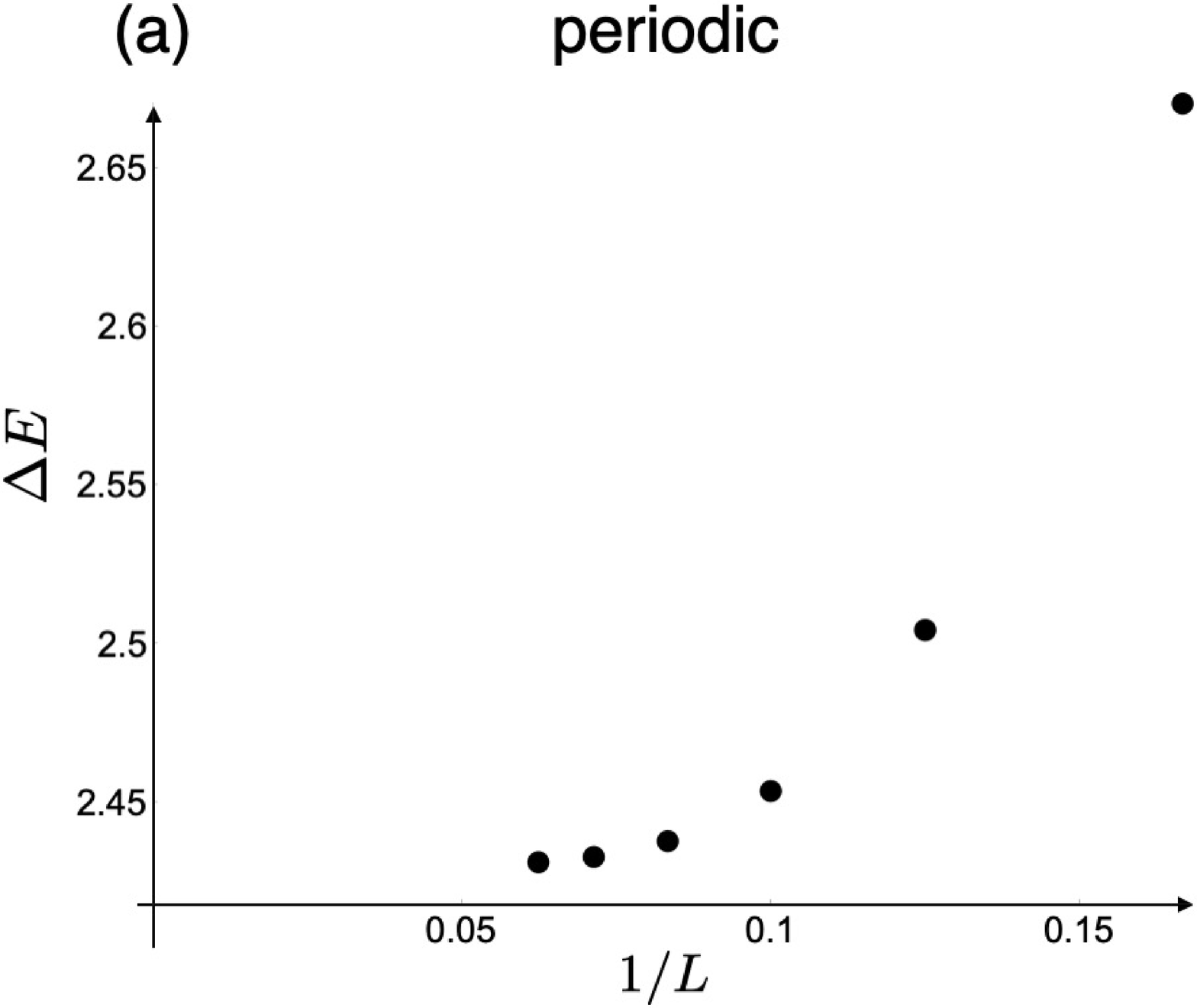}
 \end{center} 
 \begin{center}
  \includegraphics[width=80mm]{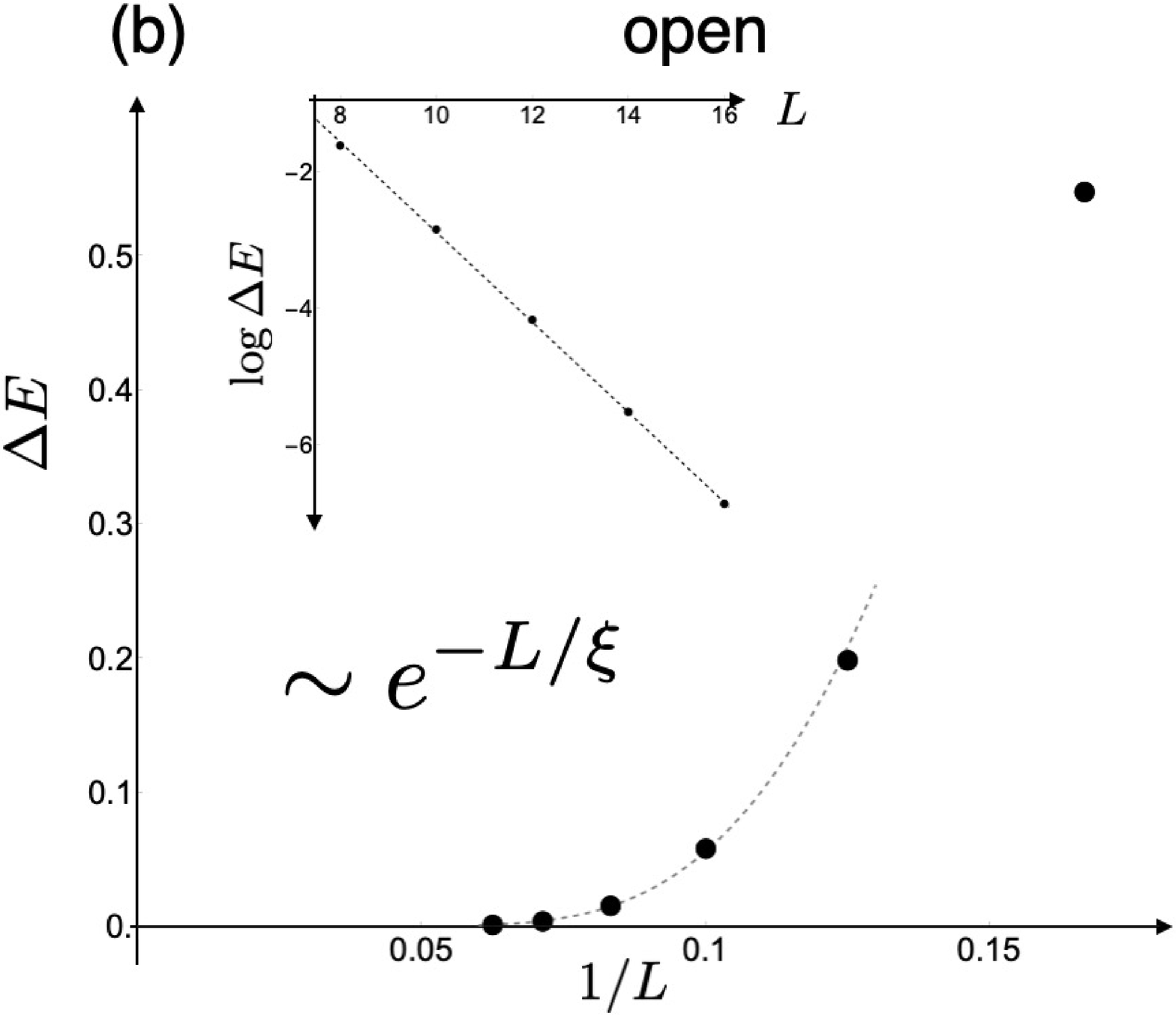}
\end{center}
\caption{\label{fig:gap3} 
 Energy gap of the $SU(3)$ symmetric Hamiltonian $H^{(2)}$,
 (a): periodic boundary condition and (b) open boundary condition. The system sizes are $L=6,8,10,12,14,16$ and $J_j=-0.8 (j:\text{odd})$ and $J_j=-1.2 (j:\text{even})$.
The data of the open boundary condition for $L\ge 10$ are fitted by the localization length $\xi=1.52$. 
}
\end{figure}

\begin{figure}[t] 
 \begin{center}
 \includegraphics[width=80mm]{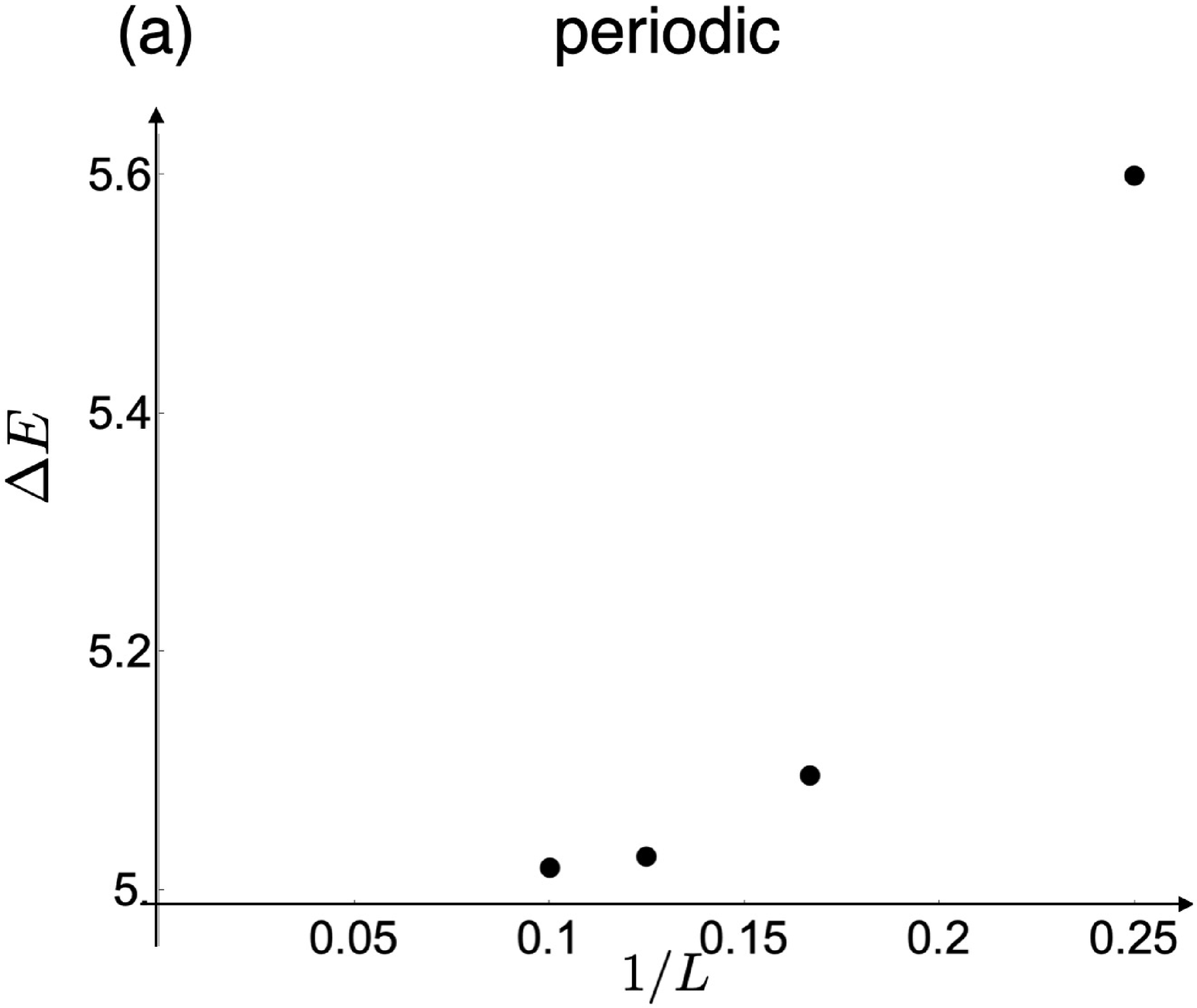}
 \end{center} 
 \begin{center}
  \includegraphics[width=80mm]{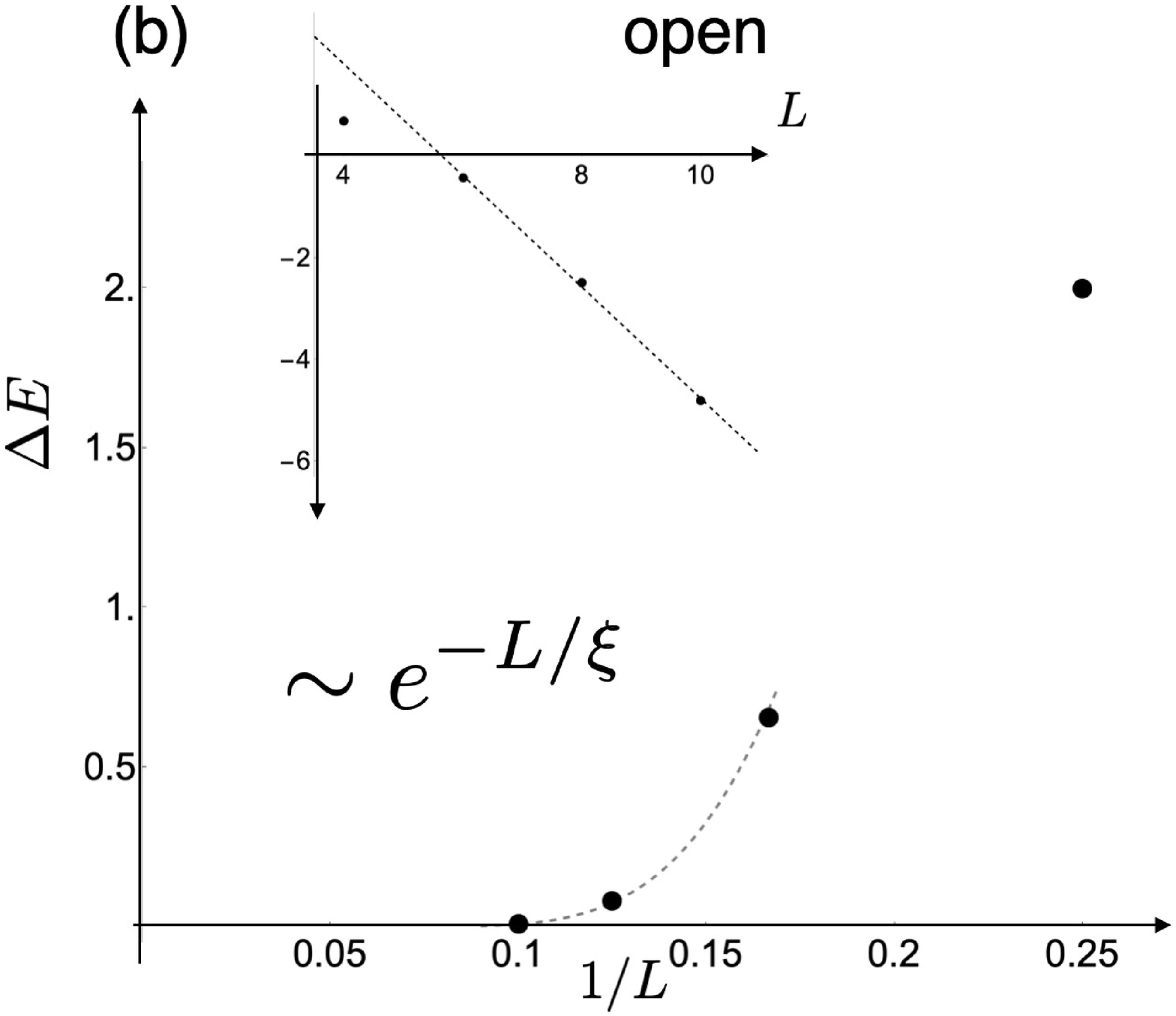}
\end{center}
\caption{\label{fig:gap5} 
 Energy gap of the $SU(5)$ symmetric Hamiltonian $H^{(2)}$,
 (a): periodic boundary condition and (b) open boundary condition. The system sizes are $L=4,6,8,10$ and $J_j=-0.8 (j:\text{odd})$ and $J_j=-1.2 (j:\text{even})$.
The data of the open boundary condition for $L\ge 6$ are fitted by the localization length $\xi=0.91$. 
}
\end{figure}

\renewcommand{\arraystretch}{1.}
        \begin{table}[t]
 \caption{Lowest 15 energies of $H^{(2)}$ with $\QN=3$ ($L=6$ and $8$).}
 \label{table:ene}
 \centering
\scalebox{0.7}{
 \begin{tabular}{|c|c|c|c|c|c|c|}
  \hline
& \multicolumn{2}{|c|}{Periodic} &\multicolumn{2}{|c|}{Open} &\multicolumn{2}{|c|}{Open} \\
  \hline
  $(J_1,J_2)$ & \multicolumn{2}{|c|}{$(-0.8,-1.2),(-1.2,-0.8)$} &\multicolumn{2}{|c|}{$(-1.2,-0.8)$} &\multicolumn{2}{|c|}{$(-0.8,-1.2)$} 
  \\
  \hline 
&  $ L=6 $ & $ L=8 $ &  $ L=6 $ & $ L=8 $ &  $ L=6 $ & $ L=8 $ 
 \\
 \hline
1 & -12.00000000 & -15.91335471 & -11.54504463 & -15.51898367 & -9.19845500 & -12.87602132 \\
2 & -9.32998146 & -13.40936964 & -8.73554793 & -12.84032668 & -8.65117764 & -12.67776804 \\
3 & -9.32998146 & -13.40936964 & -8.73554793 & -12.84032668 & -8.65117764 & -12.67776804 \\
4 & -9.32998146 & -13.40936964 & -8.73554793 & -12.84032668 & -8.65117764 & -12.67776804 \\
5 & -9.32998146 & -13.40936964 & -8.73554793 & -12.84032668 & -8.65117764 & -12.67776804 \\
6 & -9.32998146 & -13.40936964 & -8.73554793 & -12.84032668 & -8.65117764 & -12.67776804 \\
7 & -9.32998146 & -13.40936964 & -8.73554793 & -12.84032668 & -8.65117764 & -12.67776804 \\
8 & -9.32998146 & -13.40936964 & -8.73554793 & -12.84032668 & -8.65117764 & -12.67776804 \\
9 & -9.32998146 & -13.40936964 & -8.73554793 & -12.84032668 & -8.65117764 & -12.67776804 \\
10 & -8.40000000 & -12.45492049 & -8.01759677 & -12.29865018 & -7.32249080 & -11.45644523 \\
11 & -8.04273842 & -12.45492049 & -8.01759677 & -12.29865018 & -6.72949546 & -10.84747170 \\
12 & -8.04273842 & -12.45492049 & -8.01759677 & -12.29865018 & -6.72949546 & -10.84747170 \\
13 & -8.04273842 & -12.45492049 & -8.01759677 & -12.29865018 & -6.72949546 & -10.84747170 \\
14 & -8.04273842 & -12.45492049 & -8.01759677 & -12.29865018 & -6.72949546 & -10.84747170 \\
15 & -8.04273842 & -12.45492049 & -8.01759677 & -12.29865018 & -6.72949546 & -10.84747170 \\
 \hline
 \end{tabular}
}
  \end{table}

That is, we expect an
  exponentially small coupling between the both ends
which operates for the $\QN^2$-fold low energy multiplet
of the edge states.
It is described by the effective Hamiltonian of the effective
boundary fermions (assuming that they live at $j=1$ and $L$)
$c_{1, \alpha }$ and $c_{L, \alpha }$
($
\alpha =1,\cdots, N
$
) as
\begin{align}
 h_{\text{eff}} &=
 J_{\text{eff}} 
 c_{1, \alpha } ^\dagger c_{1, \beta  } 
 c_{L, \alpha  } ^\dagger c_{L, \beta  }
=
\QN J_{\text{eff}} \psi_{1,L} ^\dagger \psi_{1,L} .
\end{align}

 The energy spectrum of the
 low energy multiplet is given by the decomposition
of the tensor product of $\SUQ$ representation
as $\QN\otimes \bar \QN=1 \oplus (\QN^2-1)$
as a generalization of singlet-triplet decomposition for
the Kennedy's case.
 The unique ground state among the multiple
is a singlet approximately given by 
\begin{align}
|S_{1,L} \rangle &= \psi_{1,L} ^\dagger \otimes \prod_{j}^{L/2-1}\psi_{2j+1,2j} ^\dagger |0 \rangle ,
\end{align}
with its energy $\QN J_{\text{eff}}$. This is a generalization of
the Kenndey's singlet and triplet
for the $S=1$ Haldane chain 
 \cite{Kennedy90,expKennedy,arikawa09,chepiga18}.
Note that the state $| S_{1,L} \rangle $ is interpreted as a tensor product of
a gapped bulk and edge states.
  Due to the uniqueness and the $\SUQ$ invariance,
it implies that the one point function is constant as
$\langle S_{1,L} | n_{j \alpha } | S_{1,L} \rangle = 1/Q$ for all $\alpha $
when the average is defined by the trace over the degenerate states.
Anything localized is not observed in the charge distribution. 
This is to be compared with the results shown in Fig.\ref{fig:charge-oddL} for the {odd} $L $ case. 
The other $\QN^2-1(=8, N=3)$ are at the zero energy.
The lowest 15 energies of $\QN=3$ are listed in Table \ref{table:ene}.
System size dependencies of
the gap for the $\QN=3$ and $\QN=5$ are
shown in the insets of
Fig.\ref{fig:gap3} and Fig.\ref{fig:gap5}.
They
show $J_{eff}\propto e^{-L/\xi}$, which imply that the low energy multiplets are described by
the edge states.
 Assuming this behavior, we may assume exact $\QN^2$ degeneracy
 for the infinite system with boundaries, that is, taking an infinite size limit
 for the open system.
This $\QN^2$-fold degeneracy is exact only in the $L\to\infty$ limit.
In this sense, the $\QN^2$-fold degeneracy implies
that emergence of
$
Z_\QN^{\text{left}}
\times
Z_\QN^{\text{right}}
$ {symmetry}
in the infinite chain,
which was originally
mentioned in a chiral symmetric fermion system \cite{ryu02}. This corresponds to the $2^2$-fold degeneracy and
$Z_2\times Z_2$ symmetry
 of the Haldane chain \cite{KennedyTasaki92} and dimer phases of $S=1/2$ quantum spin chain \cite{yama93}.
 See Table \ref{table:ene} as well.

\subsection{Low energy multiplet of edge states with symmetry breaking}

 Let us consider a dimerized system
 $|J_e|\ne |J_o|$, ($J_i<0$)
 with open boundary condition.
We assume 
 the system size $L$ {is} even
assuming
it is sufficiently large {(compared with the gap)}.
When
 $|J_o|< |J_e|$, 
the low energy sector of the system is composed of a $\QN$-fold degenerate
multiplet with
edge states localized near $j=1$ and $j=L$
which is spanned by
degenerate $\QN^2$ low energy states Eq.(\ref{eq:geo}),
$\{ |g_{{eo};\alpha ,\beta }^{L:\text{even}} \rangle |\alpha, \beta =1,\cdots, \QN\}$.
This multiplet is separated from the
other states by the bulk gap.
The interaction between the both ends are negligibly small since
we assume the system sized is large.
Then the low energy multiplet is {$\QN^2$}-fold degenerate and the
$Z_\QN$ symmetry breaking term,
$H_B$, Eq.(\ref{eq:symb}),
operates within this multiplet perturbatively
assuming that the symmetry breaking is small compared with the bulk gap.
This perturbative discussion is exact as for the level crossing (selection rule) within the multiplet
assuming the gap between the multiplet and else {(global spectral structure)} is finite
where the energy scale of the splitting is governed by the
gap of the bulk.
When the dimerization pattern is reversed, $|J_o|> |J_e|$,
the system is gapped even for the open boundary condition and the ground state is
adiabatically connected to the unique gapped one, Eq.(\ref{eq:goe}).

Let us extend the time dependence of the pump by
shifting the timing of the dimerization as follows
\begin{align}
 J_i &= J_0+(-1)^j \delta J\cos\big(2\pi(\frac {t-t_0}{T} )\big),\ J_0<0,\delta J>0,
 \\
 \Delta _\alpha  &=\Delta \sin\big(2\pi( \Phi \alpha + \frac {t}{T} )\big).
\label{eq:proto}
\end{align}
As for the open boundary condition,
the edge states only appear when the coupling is weak at the both
boundaries ($L$: even),
that is, $|J_o|<|J_e|$ ($J_0<0, \delta J>0$).
This period is specified by
\begin{align}
  \frac {t_0} T+ \frac {1}{4} + n <\frac {t}{T} < \frac {t_0}T +\frac 3 4+ n,
 \ ^\exists n\in\mathbb{Z}.
 \label{eq:t0edge}
\end{align} 

Let us first discuss energies of 
the symmetry breaking Hamiltonian $H_B$, 
Eq.(\ref{eq:symb}), within the $\QN$-fold degenerate multiplet
$ |\alpha ,\beta \rangle =|g_{\op, eo }^{L:\text{even}} (\alpha ,\beta) \rangle $,
$\alpha ,\beta =1,\cdots,\QN$
\begin{align*}
 H_B(t) |g_{\op, eo}^{L:\text{even}} (\alpha ,\beta) \rangle
 &= 
 |g_{\op, eo}^{L:\text{even}} (\alpha ,\beta) \rangle
 E_{\alpha \beta }(t),
 \\
 E_{\alpha \beta }(t) &= \Delta _\alpha(t)+  \Delta _\beta (t).
\end{align*}
See Table  \ref{table:lowEs} and Fig.\ref{fig:lowspecB}.
Assuming the system size is sufficiently large $L\to\infty$, 
the center of masses (CoM) $P_\alpha $ and the quantum numbers $\bar N_\alpha $ are also shown.
For example,
as for the state, $|1_1,1_L\rangle = |g_{\op, eo}^{L:\text{even}} (1 ,1) \rangle$,
the CoM's are $P_1=x_1\cdot(+ 1)+x_L\cdot (-1)\to -1$,
$P_2=0$ and
$P_3=0$.
As for the state, $|1_1,2_L\rangle = |g_{\op, eo}^{L:\text{even}} (1 ,2) \rangle
$, $P_1=x_1\cdot(+ 1)\to -\frac 1 2$,
$P_2=x_L\cdot(- 1)\to-\frac 1 2 $ and
$P_3=0$.
Generically, as for the state, $|\alpha_1, \alpha_L \rangle $,
$\bar N_\alpha =(0,\cdots,0)$. Its CoM's are
$P_\alpha=-1$ and $P_\gamma =0 $ for any $\gamma \ne \alpha$.
As for the state, $|\alpha, \beta \rangle $, ($\alpha \ne \beta $),
$\bar N_\alpha  =+1$,
$\bar N_\beta  =-1$,
$\bar N_\gamma  =0$ for any
$ \gamma \ne\alpha,\beta$.
Its CoM's are
$P_\alpha =P_\beta =-\frac 1 2 $,
$P_\gamma =0$ $ (\gamma\ne \alpha ,\beta $).
Explicit examples for $\QN=3$ are shown in Table \ref{table:lowEs}.


\renewcommand{\arraystretch}{1.2}
\begin{table}[b]
  \caption{Energies,
    $E_{\alpha \beta }(t) = \Delta _\alpha(t)+  \Delta _\beta (t)$,
    quantum numbers, $\bar N_\alpha $, and CoM's, $P_\alpha $,
    of the symmetry breaking term $H_B$ for the multiplet of the boundary spins,
  $ \Delta _\alpha = \Delta \sin 2\pi (\Phi \alpha + \frac {t}{T} ), \ \alpha =1,2,3$ where $\Phi=P/Q, P=1,Q=3$. The generic $\QN$ case is discussed in a straightforward way.
 }
 \label{table:lowEs}
 \centering
 \begin{tabular}{|c|c|c|c|}
  \hline
  Multiplet & $E_\alpha+E_\beta  $ & $\bar N_\alpha $ & $P_\alpha$, ($L\to\infty$)
  \\
  \hline \hline
  $ |1_1,1_L \rangle $ & $ E_{11}= 2\Delta _1 $ &$ (0,0,0)$&$(-1,0,0)$
 \\
 \hline
 $ |2_1,2_L \rangle $ & $E_{22} =2\Delta _2$ & $(0,0,0)$ &$(0,-1,0)$
 \\
 \hline
 $ |3_1, 3_L \rangle $ & $E_{33}= 2\Delta _3 $ & $(0,0,0)$ &$(0,0,-1))$
 \\
 \hline
$ |1_1, 2_L \rangle $ & $E_{12}= \Delta _1+\Delta _2 $ & $(+1,-1,0)$ &$(-\frac 1 2 ,-\frac 1 2 ,0)$
 \\
 \hline
$ |2_1, 1_L \rangle $ & $ E_{21}=\Delta _2+\Delta _1 $ & $(-1,+1,0)$ &$(-\frac 1 2 ,-\frac 1 2 ,0)$
 \\
 \hline
$ |1_1, 3_L \rangle $ & $ E_{13} =\Delta _1+\Delta _2$ & $(+1,0,-1)$ &$(-\frac 1 2 ,-\frac 1 2 ,0)$
 \\
 \hline
$ |3_1, 1_L \rangle $ & $ E_{31} =\Delta _3+\Delta _1$ & $(-1,0,+1) $&$(-\frac 1 2 ,0,-\frac 1 2 )$
 \\
 \hline
$ |2_1, 3_L \rangle $ & $E_{23} =\Delta _2+\Delta _3 $ & $(0,+1,-1)$&$(0,-\frac 1 2,-\frac 1 2 )$
 \\
 \hline
$ |3_1, 2_L \rangle $ & $ E_{32} =\Delta _3+\Delta _2$  &$ (0,-1,+1) $&$(0,-\frac 1 2 ,-\frac 1 2 )$
\\ \hline
 \end{tabular}
\end{table}

\begin{figure}[t]
 \vskip 1.0cm
 \includegraphics[width=80mm]{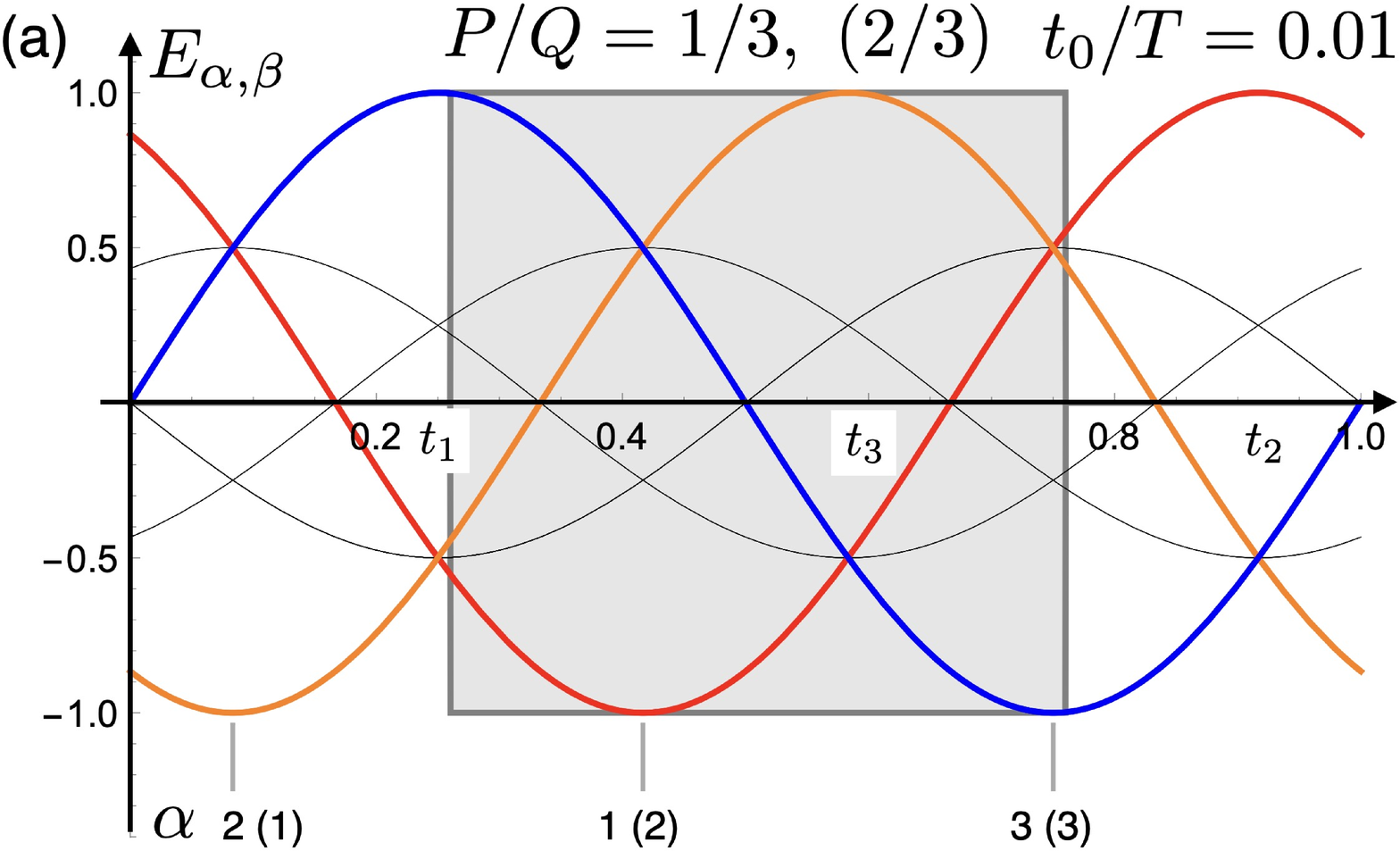}\\
 \includegraphics[width=80mm]{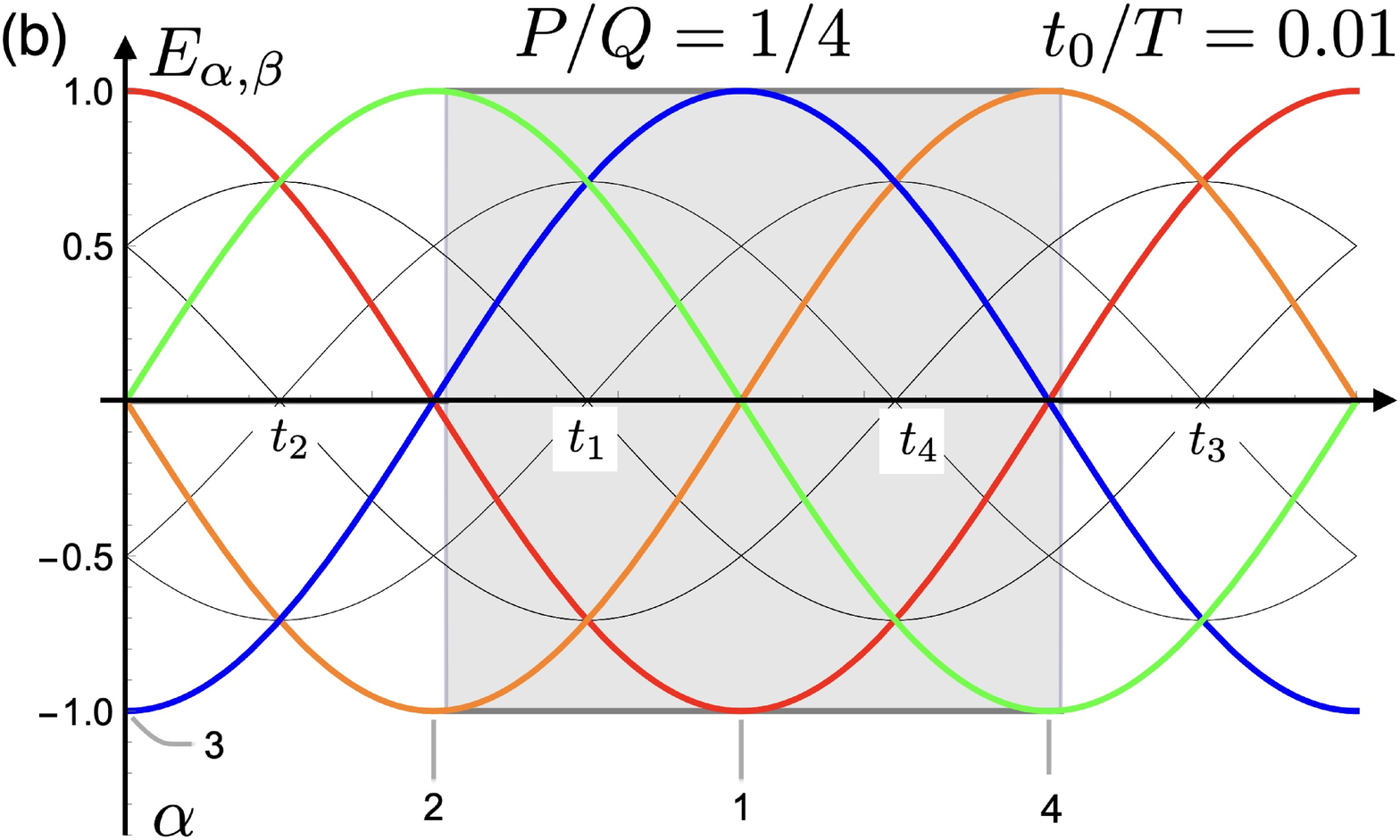}\\
 \includegraphics[width=80mm]{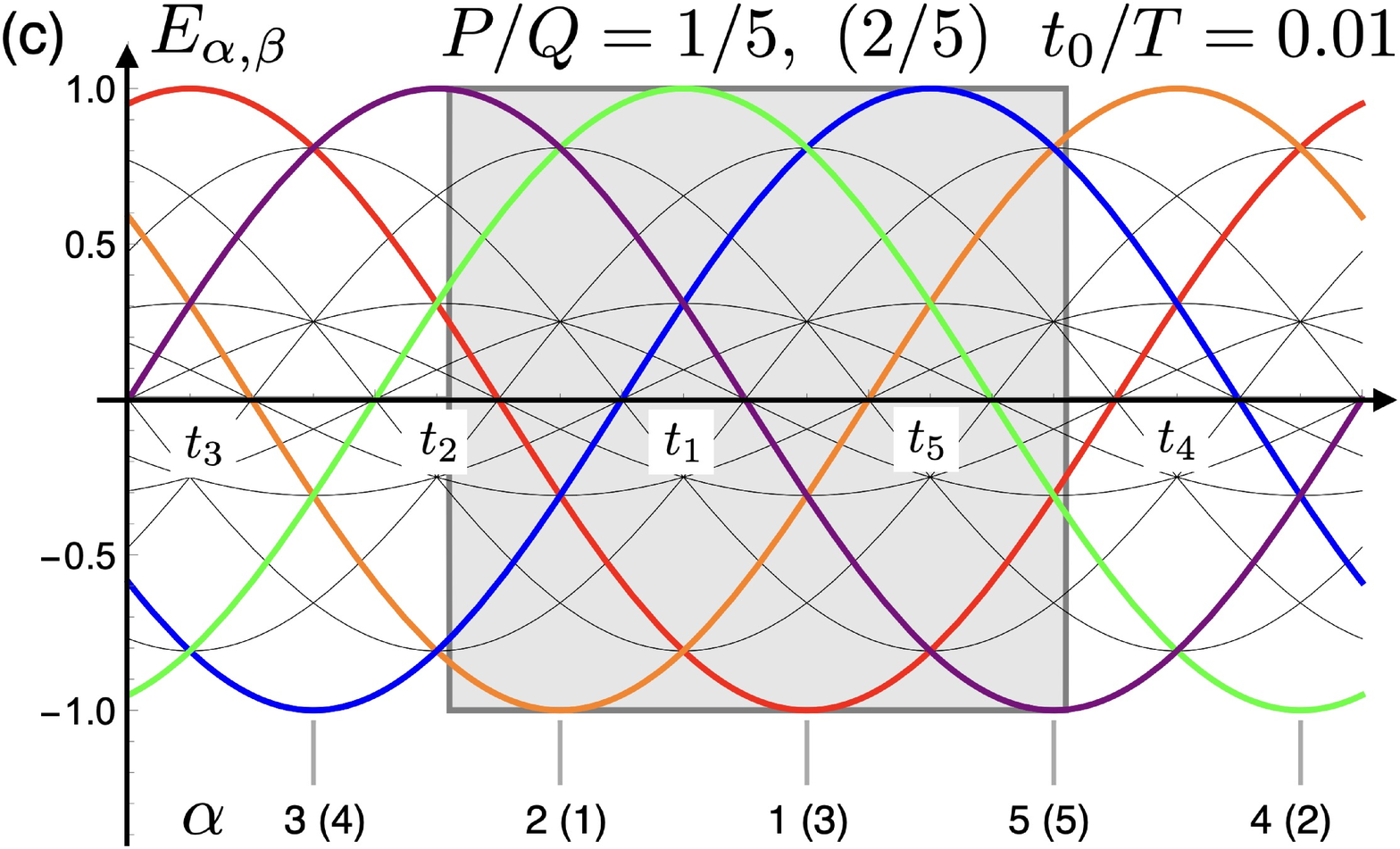}
\caption{\label{fig:lowspecB}
 Normalized low energy spectra of the symmetry breaking term $H_B(t)$ for
 the multiplet $|\alpha _L \beta _R \rangle $ where $L$ and $R$ are effective
 spins at both ends, 
 $E_{\alpha \beta }(t)=\Delta _\alpha (t)+\Delta _\beta (t) $
 as a function of $t$.
 The gray rectangles show the region $|J_o|<|J_e|$,
that is
 specified by Eq.(\ref{eq:t0edge}) for $t_0/T=0.01$. Colored lines are $E_{\alpha \alpha }$ and the numbers shown in the bottom denote $\alpha $. They are for $P=1$, $\Phi=1/Q$ and the ones
 in the parenthesis is for $P=2$, $\Phi=2/Q$.
(a) $\QN=3$ (b) $\QN=4 $ and (c) $\QN=5$.
}
\end{figure}

\begin{figure}[t]
  \includegraphics[keepaspectratio, scale=0.25]{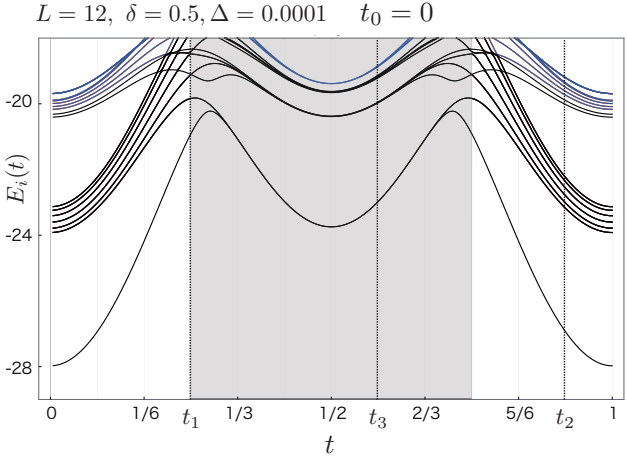}
 \caption{\label{fig:L12SPT-q3}
  Lowest 15 energy levels of the $SU(3)$ symmetric Hamiltonian
  of the 12 site system with open boundary condition within $\bar N_\alpha $ sectors, where $\delta =0.5, \Delta =0.00001 $.
  To avoid numerical instability of the diagonalization,
  $\Delta =0.0$ is not used.
  The black lines are for $\bar N_\alpha :(0,0,0)$,
  the red ones are  for $(1,-1,0)$ and $(-1,1,0)$,
  the orange ones are  for $(0,1,-1)$ and $(0,-1,1)$,
  and
  the blue ones are  for $(-1,0,1)$ and $(1,0,-1)$.
  The colored lines are degenerate {with the black ones
  and hidden} within the width of the lines.
  The gray region is for $|J_0(t)|<|J_e(t)|$ where 
  the low energy spectrum is composed of approximately degenerate
  $3^2$ edge states of the multiplet.
  {This justifies
    that the calculation for the sector, $\bar N_\alpha=0 $
    ($^\forall \alpha $) can be enough as for the low energy spectra. }
}

\end{figure}


Examples of the time dependence of $E_{\alpha \beta }(t)$ for $\QN=3$, $\QN=4$ and $\QN=5$ are shown in Fig.\ref{fig:lowspecB}.
Generically the lowest energies are always given
by $E_{\alpha \alpha }(t)$ if $\Delta >0$.
Then the $j$-th level crossing 
between $E_{\alpha \alpha }(t)$ and
 $E_{\beta \beta }(t)$, ($\alpha \ne \beta $), at $t=t_j$, ($j=1,\cdots,Q$)
 occurs when
\begin{alignat}{1}
\sin 2\pi(\frac {P}{Q}\alpha + \frac {t_j}{T})
  &=
  \sin 2\pi(\frac {P}{Q} \beta + \frac {t_j}{T}), 
\\
 {P} \alpha  &\equiv j\qquad ( \text{mod}\, Q ),
 \\
  {P}\beta &\equiv j+1\ ( \text{mod}\, Q ).
\end{alignat}
Let us write as
$\alpha =\tau_j$ and
$\beta=\tau_{j+1}$ where
the Diophantine equation due to TKNN \cite{TKNN,avron-dio} is
\begin{align}
 j &= \tau_j P + s _j Q,
\end{align}
where $\tau_j,s_j\in \mathbb{Z}$ is used in modulo $Q$.

It implies that the level crossing 
from the energy
 $E_{\tau_j,\tau_j}$
 to $E_{\tau_{j+1},\tau_{j+1}}$
occurs at $t_j$
(see Fig.\ref{fig:lowspecB})
\footnote{
 \begin{align*} 
& \sin 2\pi(\frac {j}{Q} + \frac {t_j}{T})
-  \sin 2\pi(\frac {j+1}{Q} + \frac {t_j}{T})
\\
&= 2 \sin 2\pi \frac {1}{2Q} \cos
2\pi(\frac {j+\frac 1 2 }{Q} + \frac {t_j}{T})=0.
 \end{align*}
 It implies 
$\frac 1 4 \text{ or } \frac 3 4 = \frac {j+\frac 1 2 }{Q} + \frac {t_j}{T}$.
Here we choose $\frac 3 4 $ since $\frac 1 4$ gives the level crossing at the positive energies.
}
 \begin{align}
  \frac {t_j}{T} &= \frac 3 4 -\frac {j+\frac 1 2 }{Q} ,\ \text{mod}\, 1.
  \label{eq:crs}
 \end{align}

Let us discuss the shift, $t_0$, dependence of the
low energy spectrum
of
the Hamiltonians with open boundary condition.
The low energy spectrum without symmetry breaking perturbation
for $\QN=3$ is shown in Fig.\ref{fig:L12SPT-q3}. It shows
(approximate) $\QN^2=9$ fold degeneracy due to edge states when the coupling
at the boundary is weak {as specified in the period
 by Eq.(\ref{eq:t0edge}).}
Generically the degeneracy is $\QN^2$.
Assuming the emergent $\QN\times\QN$ symmetry of the infinite chain
with boundaries,
this degeneracy is lifted by the symmetry breaking perturbation $H_B$.
The low energy spectra for $\QN=3$ and $\QN=5$
are shown in Fig.\ref{fig:L12-levelc-q3} and Fig.\ref{fig:L10-levelc-q5}.

 As the results indicate,
the hybridization of the edge states
at both boundaries is negligibly small.
Then approximate level crossings in the figures
are identified only by the spectrum of the
symmetry breaking Hamiltonian $H_B $ between
the edge states,
which are explicitly shown in
 Fig.\ref{fig:L12-levelc-q3} and Fig.\ref{fig:L10-levelc-q5}.
For example, in
Fig.\ref{fig:L12-levelc-q3}(a), there is a three fold (approximate) level crossing
at $t=t_3$. This should be compared with the level crossing
in Fig.\ref{fig:lowspecB}(a) at $t=t_3$.
As for the spectrum of $H_B$,
the three fold degeneracy is given by the change of the ground state
from the state with the energy $E_{11}$ (red) to
that with $E_{33}$ (blue). At the level crossing, some other states with energies
$E_{\alpha \beta }$
($\alpha \ne \beta $) also pass through the level crossing.
Correspondingly, in Fig.\ref{fig:L12-levelc-q3}(a), the ground state is given by the
state of the sector $N_\alpha:(0,0,0)$ and the else $(-1,1,0)$ is passing through.
The behavior of the ground state energy is cusp like. However, it should not be
a rigorous level crossing 
due to the the mixture of the edge states at both ends. 
It induces tiny (exponentially small as a function of the system size, $L$) level repulsion.
This level repulsion vanishes by taking $L\to\infty$.
This is negligibly small for the present parameter in Fig.\ref{fig:L12-levelc-q3}.
The emergent symmetry $Z^{\rm left}_\QN\times Z^{\rm right}_\QN$ ($L\to \infty$) protects
this (asymptotic) level crossing.
Just before the level crossing ($t=t_3-)$, the edge state due to $\alpha=1 $ near $j=1$ and
$\alpha=1 $ near $j=L$ are the ground state. The edge states at both ends
contribute to $P_1$ by $-1$ in pair (See Table \ref{table:lowEs}).
Note that
there is another contribution due to bulk as well.
After the crossing ($t=t_3+)$, the edge state changes
to the one due to $\alpha=3 $ near $j=1$ and
$\alpha=3 $ near $j=L$.
As for the CoM,  $P_1$, contribution
from  the edge state with energy  $E_{11}$ vanishes at $t=t_3$.
The bulk contribution remains the same (since it is continuous in time $t$) at $t=t_3$.
Then it implies $\Delta P_1=+1$. The similar consideration implies
$\Delta P_3=-1$.
In the present case, there is no further level crossing in
Fig.\ref{fig:L12-levelc-q3} (a).
Then according to Eq.(\ref{eq:CanI}), $I^1=+1$, $I^2=0,$ and $I^3=-1$.
Assuming this emergent symmetry and the level crossings, each jump
of the CoM $\Delta P_\alpha $ is identified for each level crossing
as shown in the caption of Fig.\ref{fig:L12-levelc-q3} and
Fig.\ref{fig:L10-levelc-q5} supplemented with the sum of the discontinuities
$I^\alpha $.
In the next section,
CoM's for $\QN=3$ case is directly calculated by
using the DMRG calculation.
Also direct calculation of the Chern number $C^\alpha $ are compared
in the following section.
It enables us to
confirm the bulk-edge correspondence Eq.(\ref{eq:charge-by-disc}) and Eq.(\ref{eq:bec-topo}).

\subsection{Explicit Chern numbers and Diophantine equation} 

To be simple let us first consider a system at $t_0=+0$ (see Fig.\ref{fig:lowspecB}).
In the pump cycle, one may see a series of the jumps in $P_\alpha $.
As for the sum of the discontinuities $I^\alpha$, a pair of the jumps
except the first and the last ones is cancelled
(See Fig.\ref{fig:L12-levelc-q3} and Fig.\ref{fig:L10-levelc-q5}).
The last jump is due to the level crossing
from some state with energy $E_{\alpha \alpha }$ to the state with $E_{\QN\QN}$.
It results in
$I^\QN=-1$.
Similarly  the first jump is a level crossing
from the state with energy
$E_{\alpha \alpha }$ ($\alpha =\tau_{\frac{\QN-1}2 }$)
to the state with energy  $E_{\beta \beta }$ ($\beta =\tau_{\frac{\QN-1}2 -1}$) for $\QN$:odd
and
from
$E_{\alpha \alpha }$ ($\alpha =\tau_{\frac{\QN}2 }$)
to $E_{\beta \beta }$ ($\beta =\tau_{\frac{\QN}2 -1}$) for $\QN$:even.
It results in
{$I^{\tau_{\frac{\QN-1}2}}=+1$ for $\QN$ (odd)}
and
  { $I^{\tau_{\frac\QN2 }}=+1$ for $\QN$ (even)}.

In a similar way as for the time dependence Eq.(\ref{eq:proto}), the edge states
appear at
$t/T=t_{\text{ini}}/T\equiv t_0/T+1/4 $
and vanishes
$t/T=t_{\text{fin}}/T\equiv t_0/T+ {3}/{4} $ for each $t_0$.
Within the period, $[t_{\text{ini}},t_{\text{fin}}]$, the level crossings due to
the edge state
$\beta =\tau_{j+1}$
to
$\alpha =\tau_j $
occur
at $t=t_j$, Eq.(\ref{eq:crs}),
that cause the jumps in
$\Delta \bar P^{\tau_j}=-1$ and
$\Delta \bar P^{\tau_{j+1}}=+1$
for all $j$'s that satisfy $t_{\text{ini}}<t_j <t_{\text{fin}}$ (See Fig.\ref{fig:lowspecB}). 
Since the paired jumps inside the period cancel with each other,
the first one $t_{j_{\text{ini}}}$, $t_{\text{ini}}<t_{j_{\text{ini}}}< t_{\text{ini}}+ 1/Q$
gives the sum of the discontinuity, $I^{\tau_{j_{\text{ini}+1 }}}=+1$. 
Similarly the last one
 $t_{j_{\text{fin}}}$, $t_{\text{fin}}-1/Q<t_{j_{\text{fin}}}< t_{\text{fin}}$
gives to the sum of the discontinuity $I^{\tau_{j_{\text{fin}}}}=-1$.
Otherwise $I^\alpha =0$
($\alpha \notin \{\tau_{j_{\text{ini}}+1},\tau_{j_{\text{fin}}}\}$).
The conditions are written as
\footnote{
 The first condition is written as 
 \begin{align*}
  \frac{ t_0}{T}+\frac {1}{4} < \frac {3}{4} - \frac {j_{\text{ini}}+\frac 1 2 }{Q}
  < \frac{ t_0}{T}+\frac {1}{4} + \frac {1}{Q}
  \\
  \frac{ t_0}{T} < \frac {1}{2} - \frac {j_{\text{ini}}+\frac 1 2 }{Q}
  < \frac{ t_0}{T} + \frac {1}{Q}
  \\
 Q  \frac{ t_0}{T} < \frac {Q}{2} - ({j_{\text{ini}}+\frac 1 2 })
 < Q\frac{ t_0}{T} +1
 \\
 Q(\frac 1 2 - \frac {t_0}{T}) -\frac {3}{2}
 < j_{\text{ini}}<
 Q(\frac 1 2 - \frac {t_0}{T}) -\frac {1}{2}
 \end{align*}
 Similarly the last one is written as follows.
 \begin{align*}
  \frac{ t_0}{T}+\frac {3}{4} - \frac {1}{Q}< \frac {3}{4} - \frac {j_{\text{fin}}+\frac 1 2 }{Q}
  < \frac{ t_0}{T}+\frac {3}{4} 
  \\
  \frac{ t_0}{T} - \frac {1}{Q}< - \frac {j_{\text{fin}}+\frac 1 2 }{Q}
  < \frac{ t_0}{T} 
  \\
 Q  \frac{ t_0}{T}-1 < - ({j_{\text{fin}}+\frac 1 2 })
 < Q\frac{ t_0}{T} 
 \\
- Q \frac {t_0}{T} -\frac 1 2 
 < j_{\text{fin}}<
- Q \frac {t_0}{T} +\frac 1 2 
 \end{align*}
Then it implies
 \begin{align*}
  j_{\text{ini}} +1&=\lfloor Q(\frac 1 2 - \frac {t_0}{T}) +\frac {1}{2}  \rfloor
  \\
  j_{\text{fin}} &=
  \lfloor - Q \frac {t_0}{T} +\frac 1 2 \rfloor.
 \end{align*}
}
 \begin{align}
  \frac{ t_0}{T}+\frac {1}{4} < \frac {3}{4} - \frac {j_{\text{ini}}+\frac 1 2 }{Q}
  < \frac{ t_0}{T}+\frac {1}{4} + \frac {1}{Q},
  \\
  \frac{ t_0}{T}+\frac {3}{4} - \frac {1}{Q}< \frac {3}{4} - \frac {j_{\text{fin}}+\frac 1 2 }{Q}
  < \frac{ t_0}{T}+\frac {3}{4} .
 \end{align}
It implies
 \begin{align}
  j_{\text{ini}} +1&=\lfloor Q(\frac 1 2 - \frac {t_0}{T}) +\frac {1}{2}  \rfloor,
  \\
  j_{\text{fin}} &=
\lfloor - Q \frac {t_0}{T} +\frac 1 2 \rfloor,
 \end{align}
 where $\lfloor x\rfloor$ is the largest integer less than $x$.
Finally we have with using the bulk-edge correspondence
 \begin{align}
  I^\alpha (t_0)&= C^\alpha (t_0)=
  \left\{
  \begin{array}{cl}
   {-1} & \alpha \equiv \tau_{\lfloor  Q(1- \frac {t_0}{T}) +\frac 1 2 \rfloor}\ \text{mod } Q
   \\
   {+1} & \alpha \equiv \tau_{\lfloor Q(\frac 1 2 - \frac {t_0}{T}) +\frac {1}{2}  \rfloor}\ \text{mod } Q
   \\
   0 &{\text{otherwise}}   
  \end{array}\right.,
  \label{eq:aChern}
 \end{align}
 It implies a series of topological transitions associated with
 the shift of the dimerization $t_0$.

  The Chern numbers of the generic path $\ell_{V_\alpha G V_\beta }$,
  ($1\le\alpha<\beta\le Q$), $C^{\ell_{V_\alpha G V_\beta }}$
  is simply given by
 \begin{align}
  C^{\ell_{V_\alpha G V_\beta }} &= 
  \sum_{\gamma =\alpha }^ {\beta-1} C^{\gamma }.
 \end{align}

 \section{Numerical evaluation of Topological numbers}
 \label{sec:num}
In this section, extensive data for the
numerical evaluation of
the low energy spectra, the CoMs, and the topological numbers
(the sum of the jumps and the Chern numbers) 
are shown.
They are given by the DMRG and the exact diagonalization.


\subsection{Low energy spectra of the finite size systems}
\label{sec:spec-finite}
Low energy spectra of the $\QN=3$, $L=12$ system
and the $\QN=5$, $L=10$ system
are shown in Fig.\ref{fig:L12-levelc-q3} and Fig.\ref{fig:L10-levelc-q5}.
The results for the $\QN=3$ case are 
consistently compared with the DMRG calculation shown in Sec.\ref{sec:DMRG-st-Com}.

\begin{widetext}

 \begin{figure}[t]
  \captionsetup[sub]{skip=0pt}
 \begin{minipage}[b]{0.48\linewidth}
  \centering
  \includegraphics[keepaspectratio, scale=0.23]{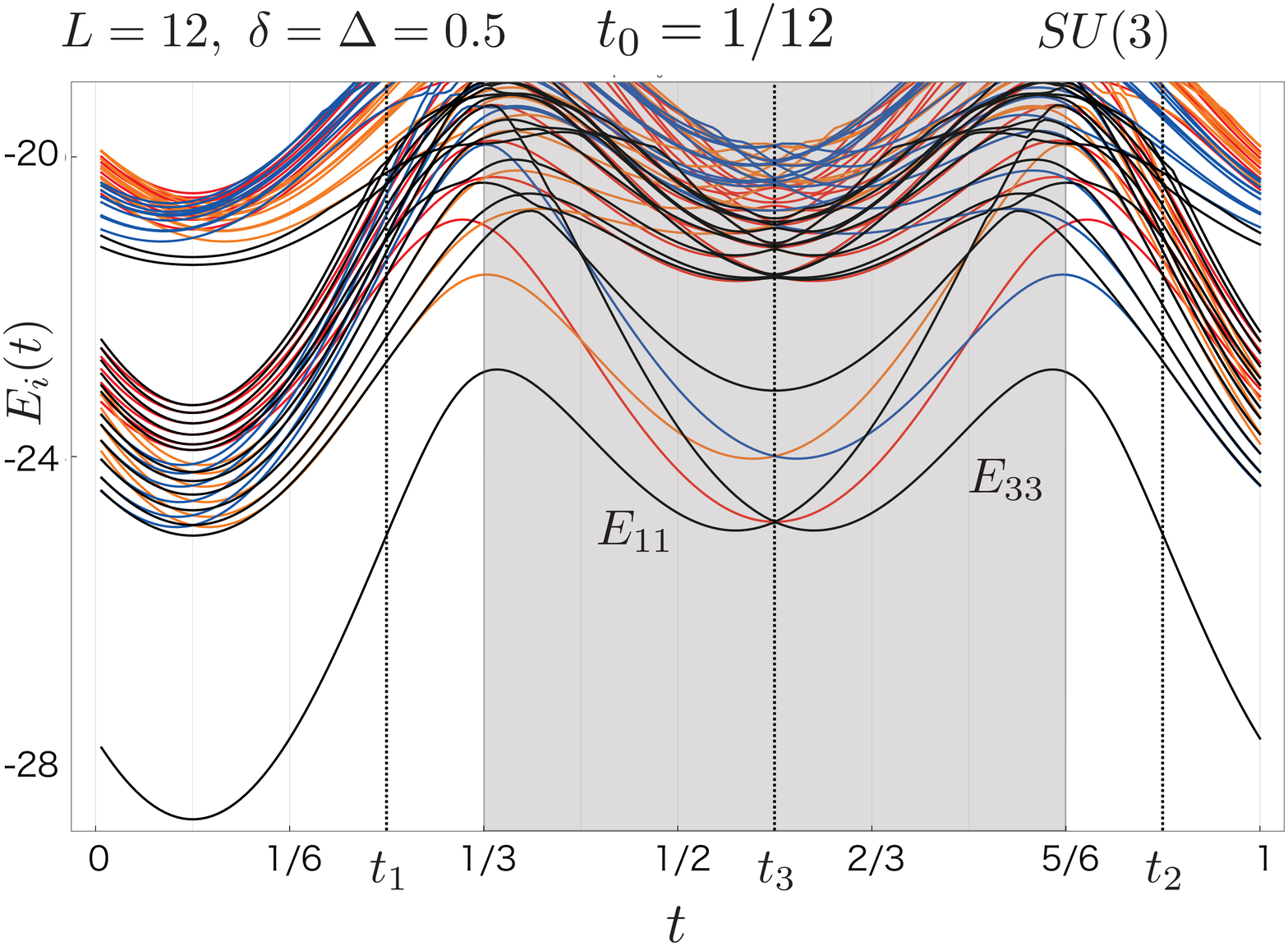}
  \subcaption{
   $\Delta P_1(t_3)=+1 $ and
   $\Delta P_3(t_3)=-1 $ implies
   $I^\alpha :(+1,0,-1)$.
  }\label{fig:q3a}
 \end{minipage}
 \begin{minipage}[b]{0.48\linewidth}
  \centering
  \includegraphics[keepaspectratio, scale=0.23]{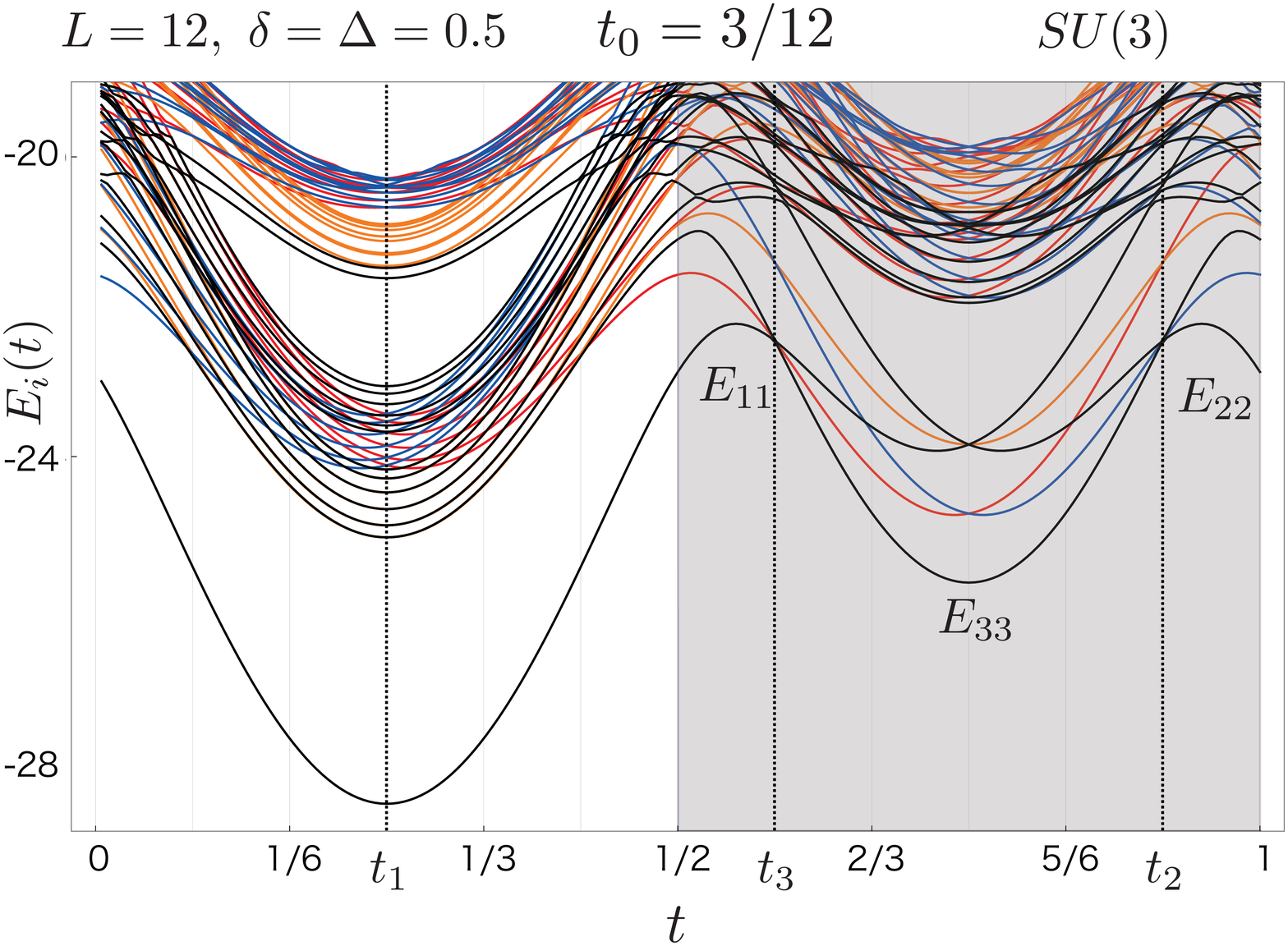}
  \subcaption{
   $\Delta P_1(t_3)=+1 $,
   $\Delta P_3(t_3)=-\Delta P_3(t_2)=-1 $
   and
   $\Delta P_2(t_2)=-1 $ implies
   $I^\alpha :(+1,-1,0)$.
  }\label{fig:q3b}
 \end{minipage}
 \\
 \begin{minipage}[b]{0.48\linewidth}
  \centering
  \includegraphics[keepaspectratio, scale=0.23]{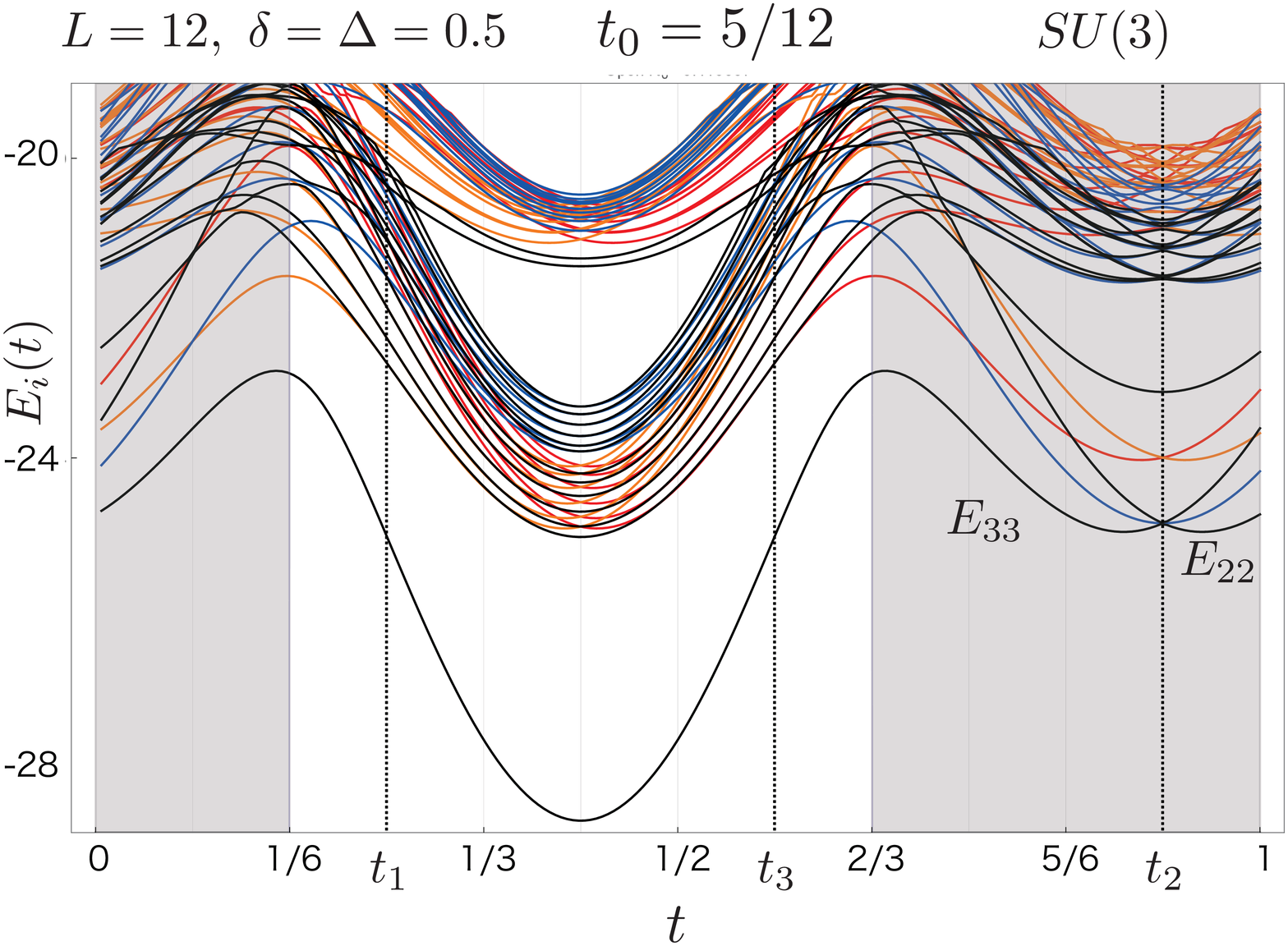}
  \subcaption{
   $\Delta P_3(t_2)=+1 $,
   and
   $\Delta P_2(t_2)=-1 $,
   implies
   $I^\alpha :(0,-1,+1)$.
  }\label{fig:q3fc}
 \end{minipage}
 \begin{minipage}[b]{0.48\linewidth}
  \centering
  \includegraphics[keepaspectratio, scale=0.23]{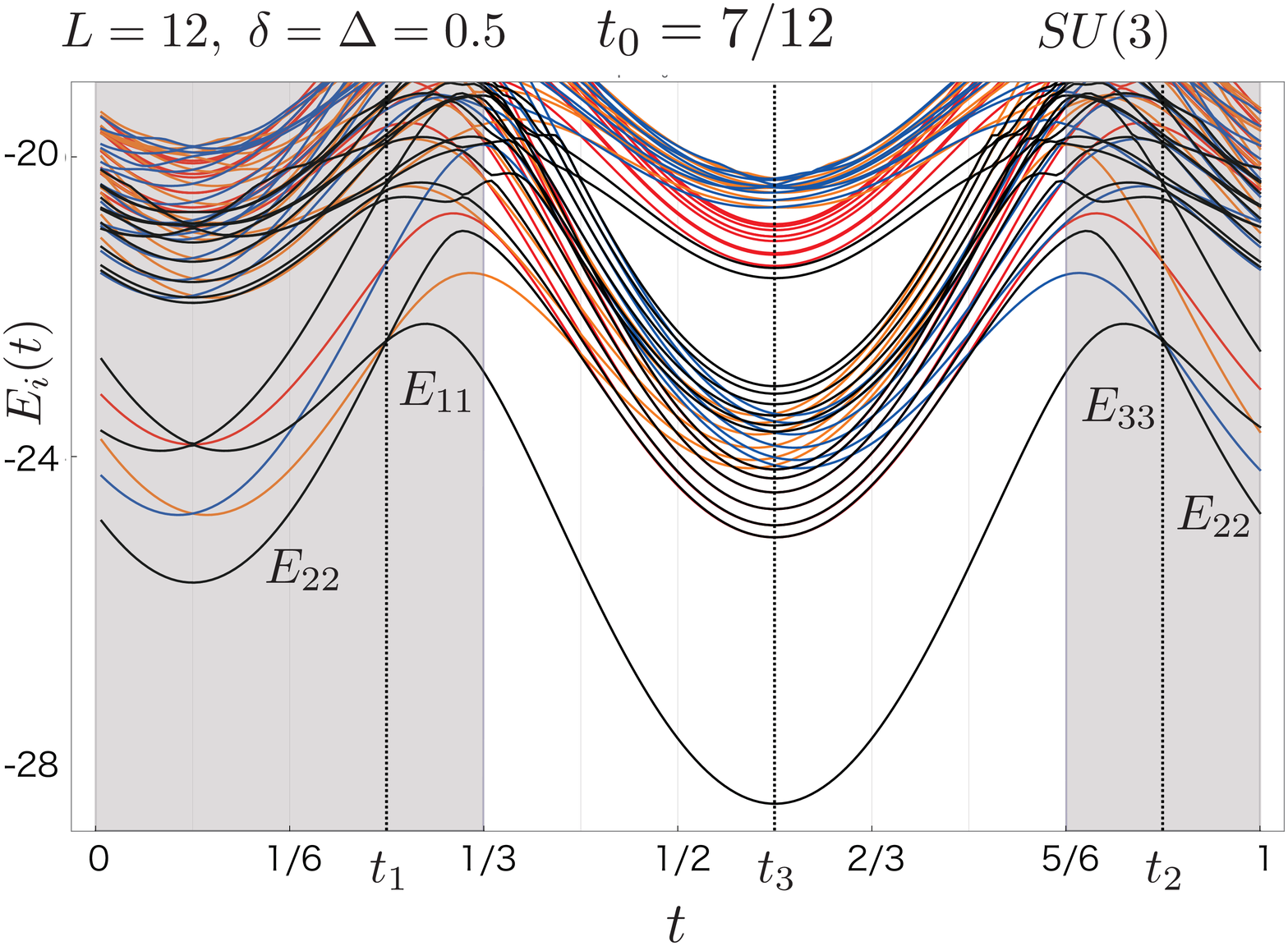}
  \subcaption{
   $\Delta P_3(t_2)=+1 $,
   $\Delta P_2(t_2)=-\Delta P_2(t_1)=-1 $
   and
   $\Delta P_1(t_1)=-1 $ implies
   $I^\alpha :(-1,0,+1)$. 
  }\label{fig:q3d}
 \end{minipage}
 \\
 \begin{minipage}[b]{0.48\linewidth}
  \centering
  \includegraphics[keepaspectratio, scale=0.23]{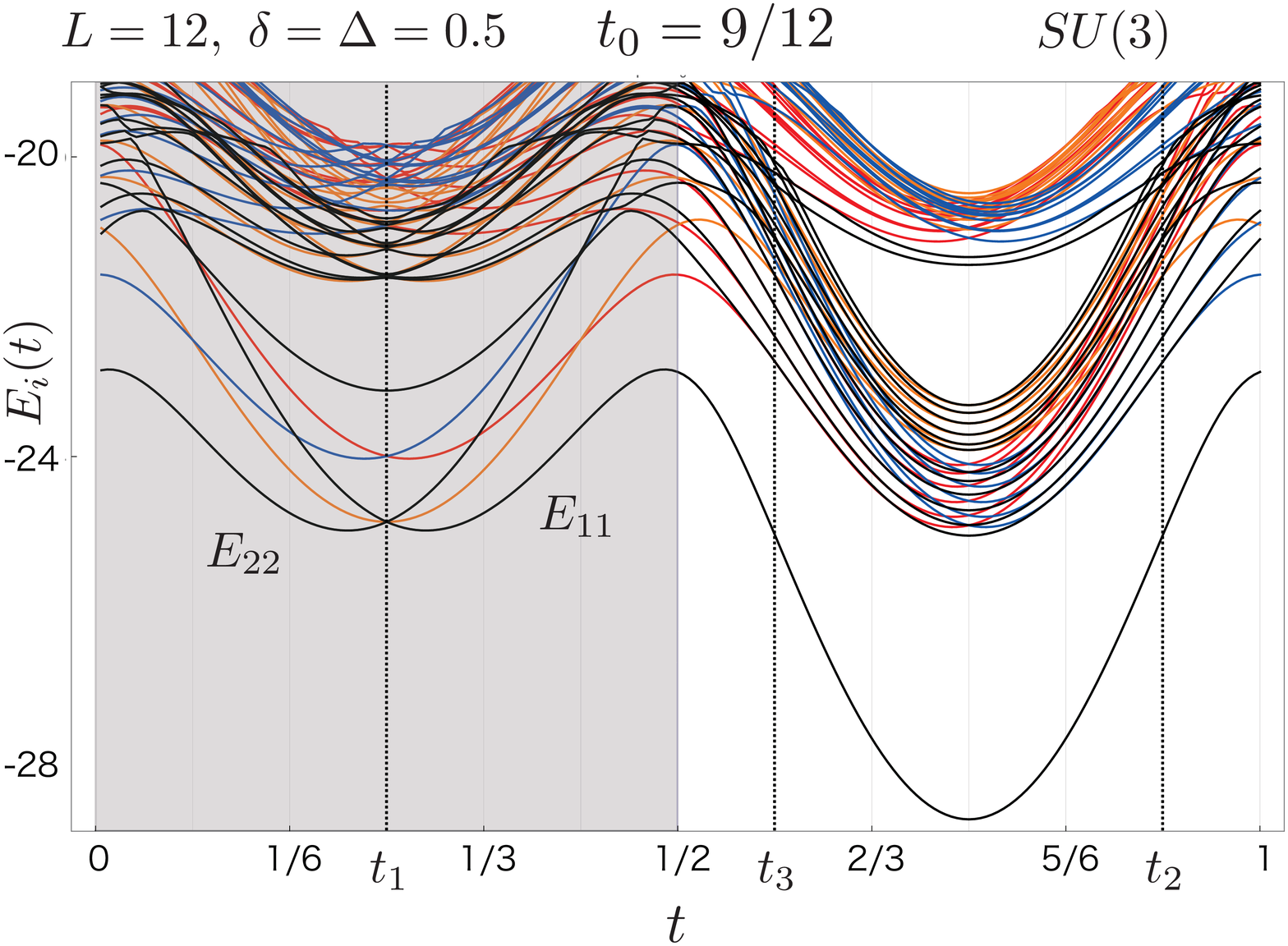}
  \subcaption{ 
   $\Delta P_2(t_1)=+1 $,
   $\Delta P_1(t_1)=-1$
   implies
   $I^\alpha :(-1,+1,0)$.
  }\label{fig:q3e}
 \end{minipage}
 \begin{minipage}[b]{0.48\linewidth}
  \centering
  \includegraphics[keepaspectratio, scale=0.23]{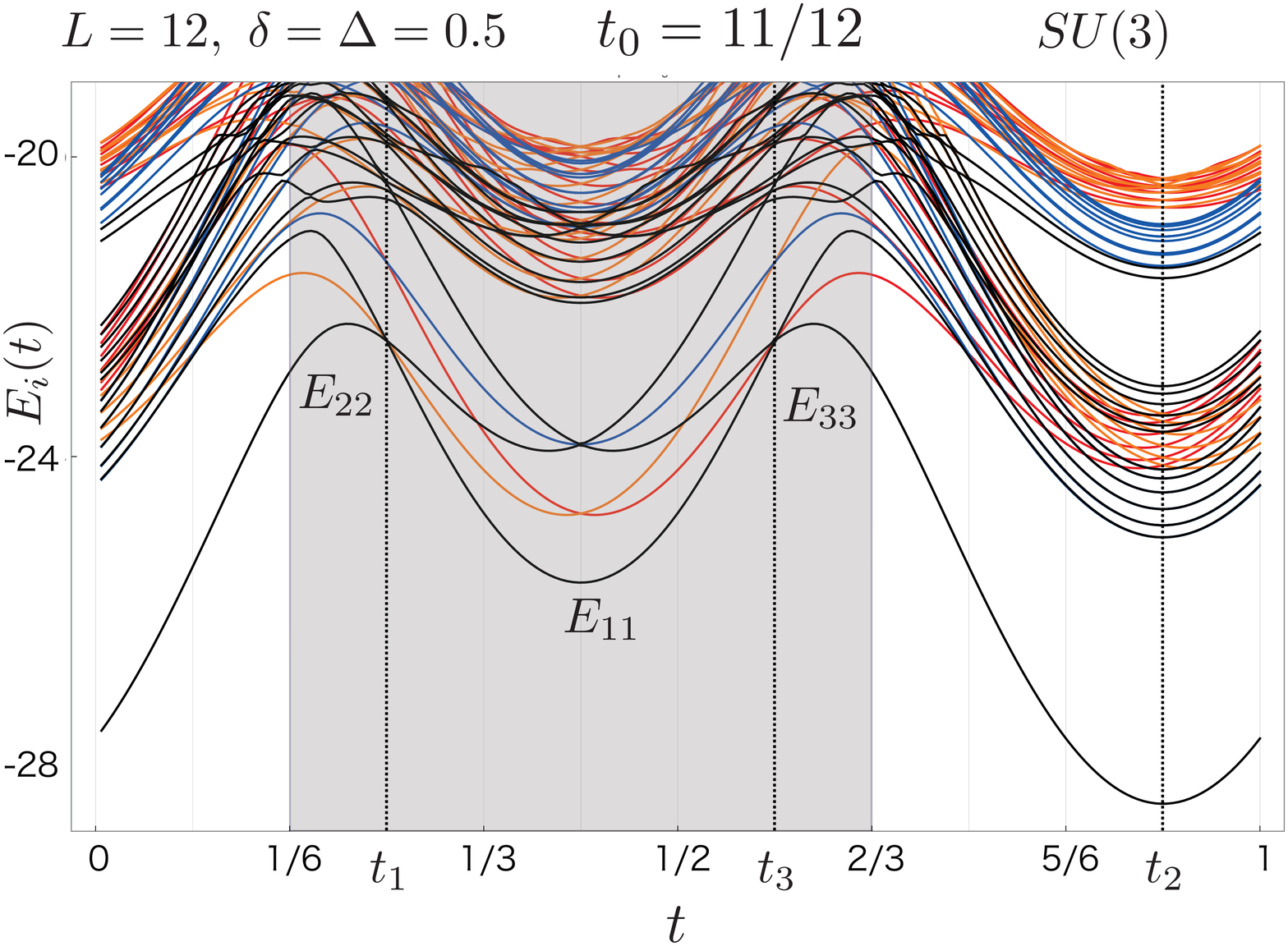}
  \subcaption{
   $\Delta P_2(t_1)=+1 $,
   $\Delta P_1(t_1)=-\Delta P_1(t_3)=+1 $
   and
   $\Delta P_3(t_3)=-1 $ implies
   $I^\alpha :(0,+1,-1)$.
  }\label{fig:q3f}
 \end{minipage}
 \caption{\label{fig:L12-levelc-q3}
$t_0$ dependence of
  the lowest 15 energy levels of the $SU(3)$ symmetric Hamiltonian
  of the 12 site system with open boundary condition within $\bar N_\alpha $
  sectors, where $\delta =0.5, \Delta =0.5 $ { and $\Phi=1/3$} ($\QN=3$).
  The black lines are for $\bar N_\alpha :(0,0,0)$,
  the red ones are  for $(1,-1,0)$ and $(-1,1,0)$,
  the orange ones are  for $(0,1,-1)$ and $(0,-1,1)$,
  and
  the blue ones are  for $(-1,0,1)$ and $(1,0,-1)$.
  The gray region is for $|J_0(t)|<|J_e(t)|$ where 
  the low energy spectrum is composed of
  the multiplet of the edge states of the dimension
  $\QN^2$.
  One expects $\QN\times\QN$ emergent symmetry, in the $L\to\infty$ limit,
  that is responsible for the level crossings within the multiplet.
  The lowest eigen state is identified by $E_{\alpha \alpha }$ assuming the
  emergent $\QN\times\QN$ symmetry for the infinite system.
  The dimensions of the Hilbert spaces are $35169$ for $(0,0,0)$ and  $27888$ for the else   
  (Compare with the Fig.\ref{fig:lowspecB}).
  Due to Eq.(\ref{eq:CanI}), $I^\alpha =\sum_i \Delta P^\alpha (t_i)$.
  }
\end{figure}

\begin{figure}[t]
 \begin{minipage}[b]{0.48\linewidth}
   \centering
  \includegraphics[keepaspectratio, scale=0.230]{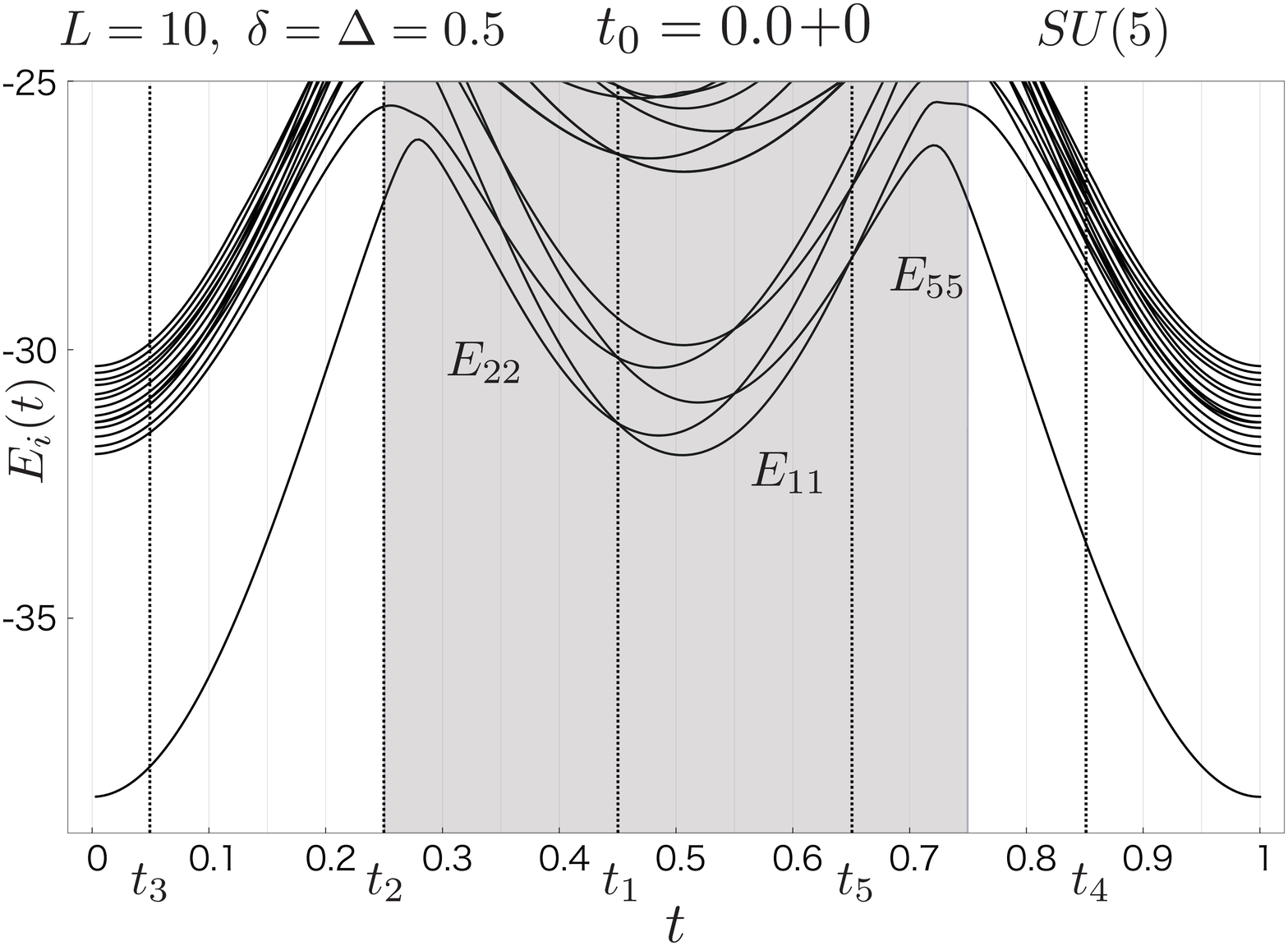}
  \subcaption{ ($t_0=0.0+0$)
   $\Delta P_2(t_1)=+1 $,
   $\Delta P_1(t_1)=-\Delta P_1(t_5)= -1$
   and
   $\Delta P_5(t_5)=-1 $,   
   implies
   $I^\alpha :(0,+1,0,0,-1)$.
  }\label{fig:q5a}
 \end{minipage}
 \begin{minipage}[b]{0.48\linewidth}
  \centering
  \includegraphics[keepaspectratio, scale=0.230]{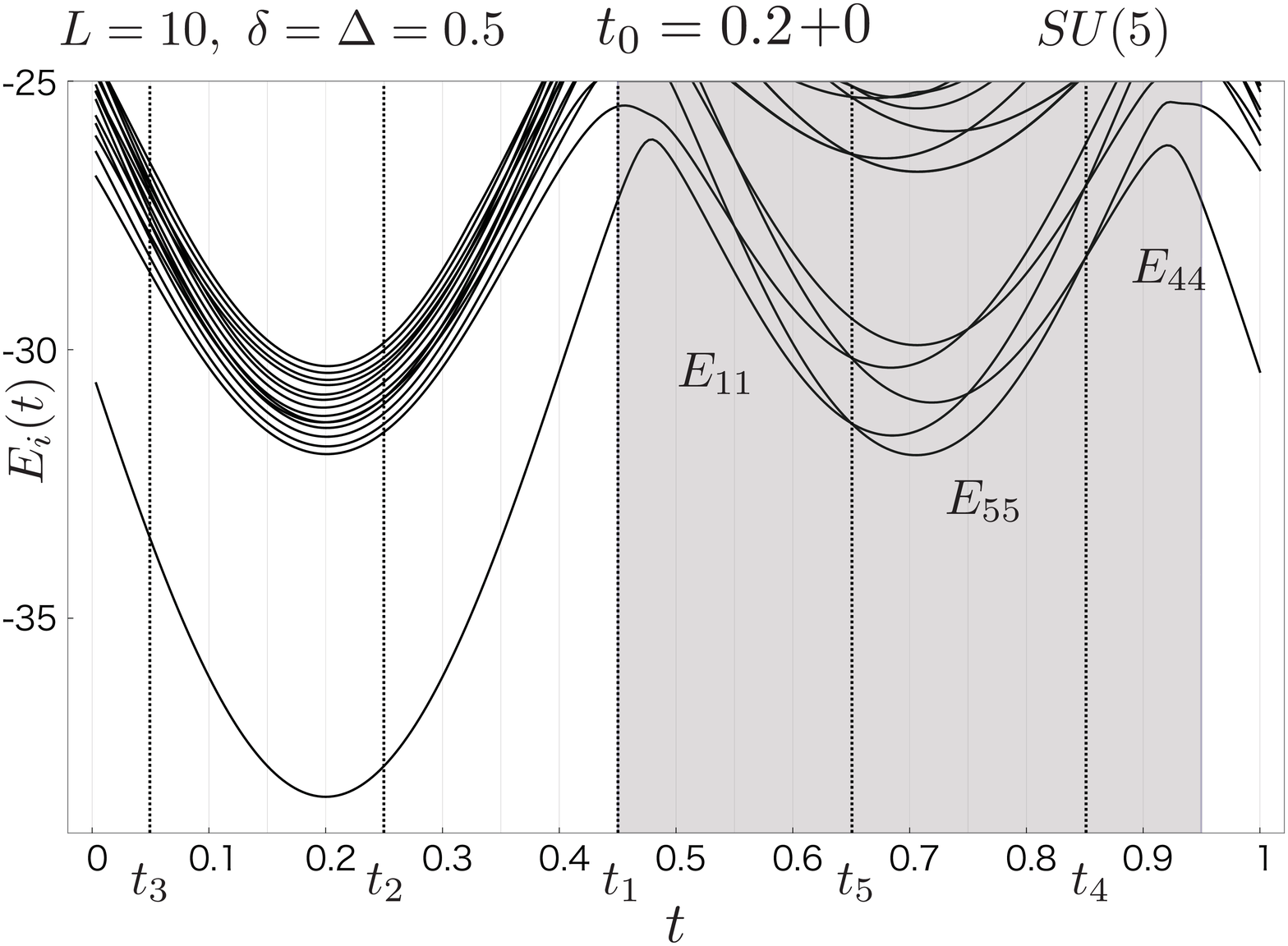}
  \subcaption{ ($t_0=0.2+0$)
   $\Delta P_1(t_5)=+1 $,
   $\Delta P_5(t_5)=-\Delta P_5(t_4)= -1$
   and
   $\Delta P_4(t_4)=-1 $,   
   implies
   $I^\alpha :(+1,0,0,-1,0)$.
  }\label{fig:q5b}
 \end{minipage}
\\ 
 \begin{minipage}[b]{0.48\linewidth}
   \centering
  \includegraphics[keepaspectratio, scale=0.230]{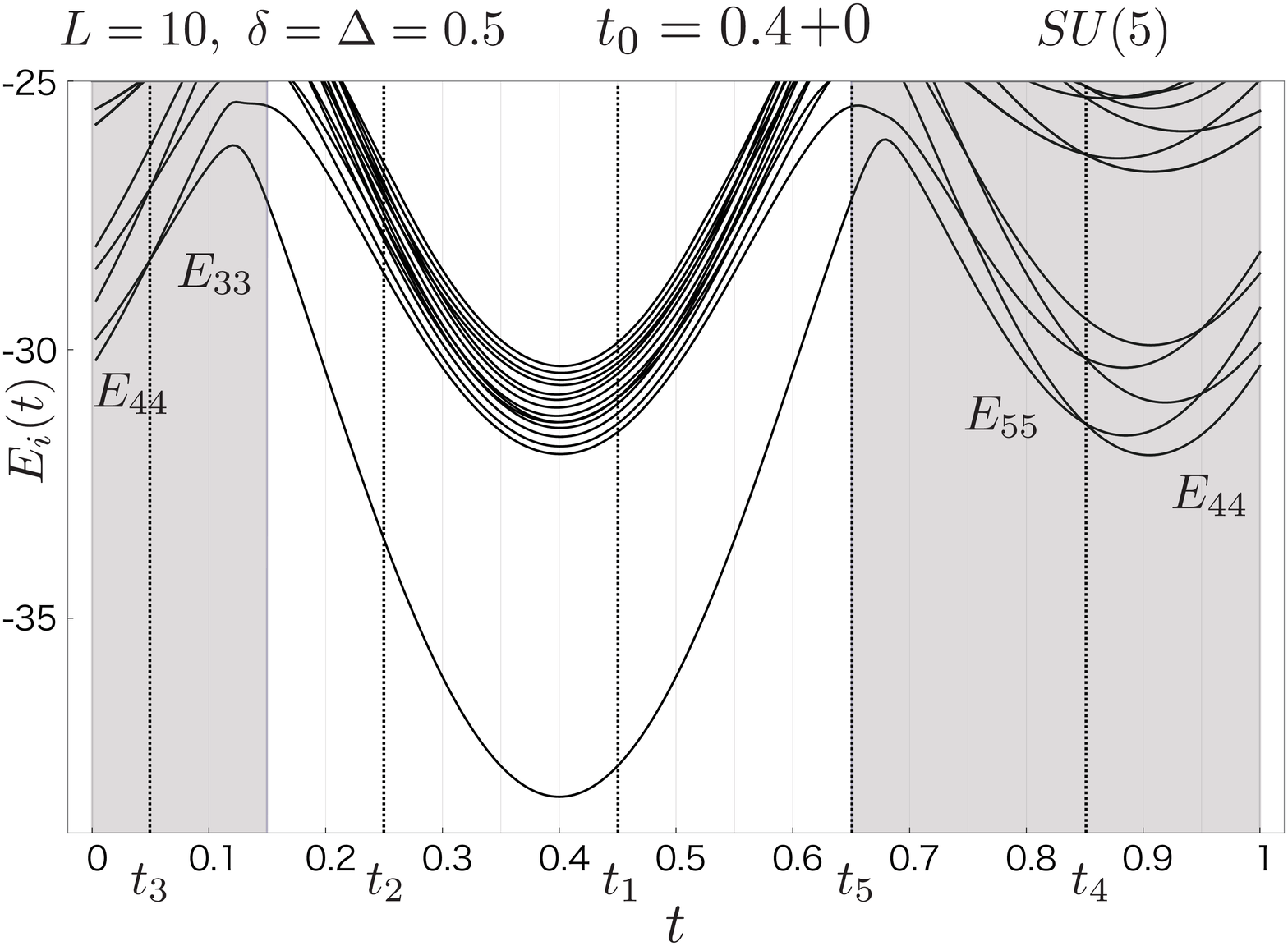}
  \subcaption{ ($t_0=0.4+0$)
   $\Delta P_5(t_4)=+1 $,
   $\Delta P_4(t_4)=-\Delta P_4(t_3)= -1$
   and
   $\Delta P_3(t_3)=-1 $,   
   implies
   $I^\alpha :(0,0,-1,0,+1)$.
  }\label{fig:q5c}
 \end{minipage}
 \begin{minipage}[b]{0.48\linewidth}
  \centering
  \includegraphics[keepaspectratio, scale=0.230]{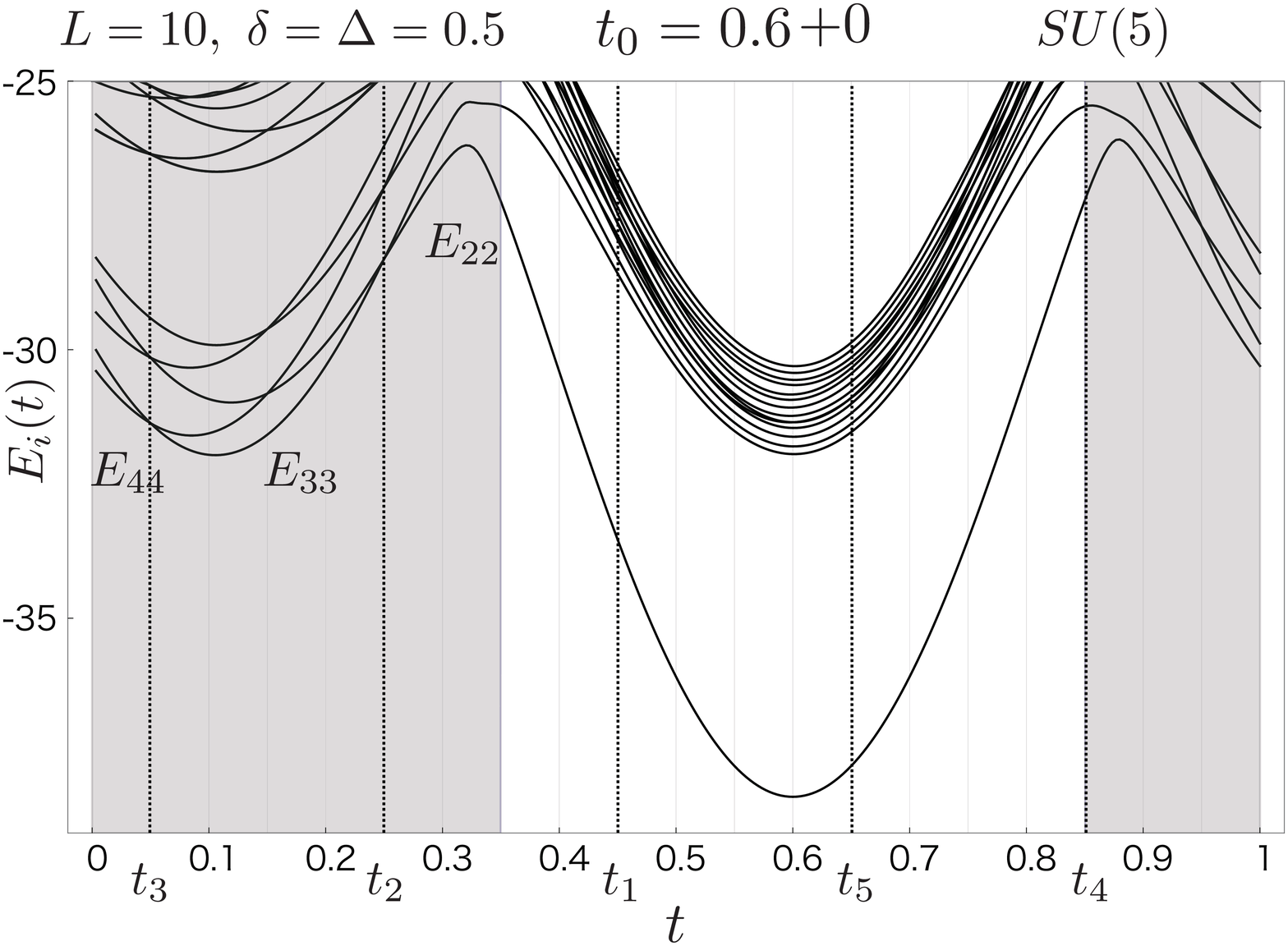}
  \subcaption{ ($t_0=0.6+0$)
   $\Delta P_4(t_3)=+1 $,
   $\Delta P_3(t_3)=-\Delta P_3(t_2)= -1$
   and
   $\Delta P_2(t_2)=-1 $,   
   implies
   $I^\alpha :(0,-1,0,+1,0)$.
  }\label{fig:q5d}
 \end{minipage}
 \\
 \begin{minipage}[b]{0.48\linewidth}
   \centering
  \includegraphics[keepaspectratio, scale=0.230]{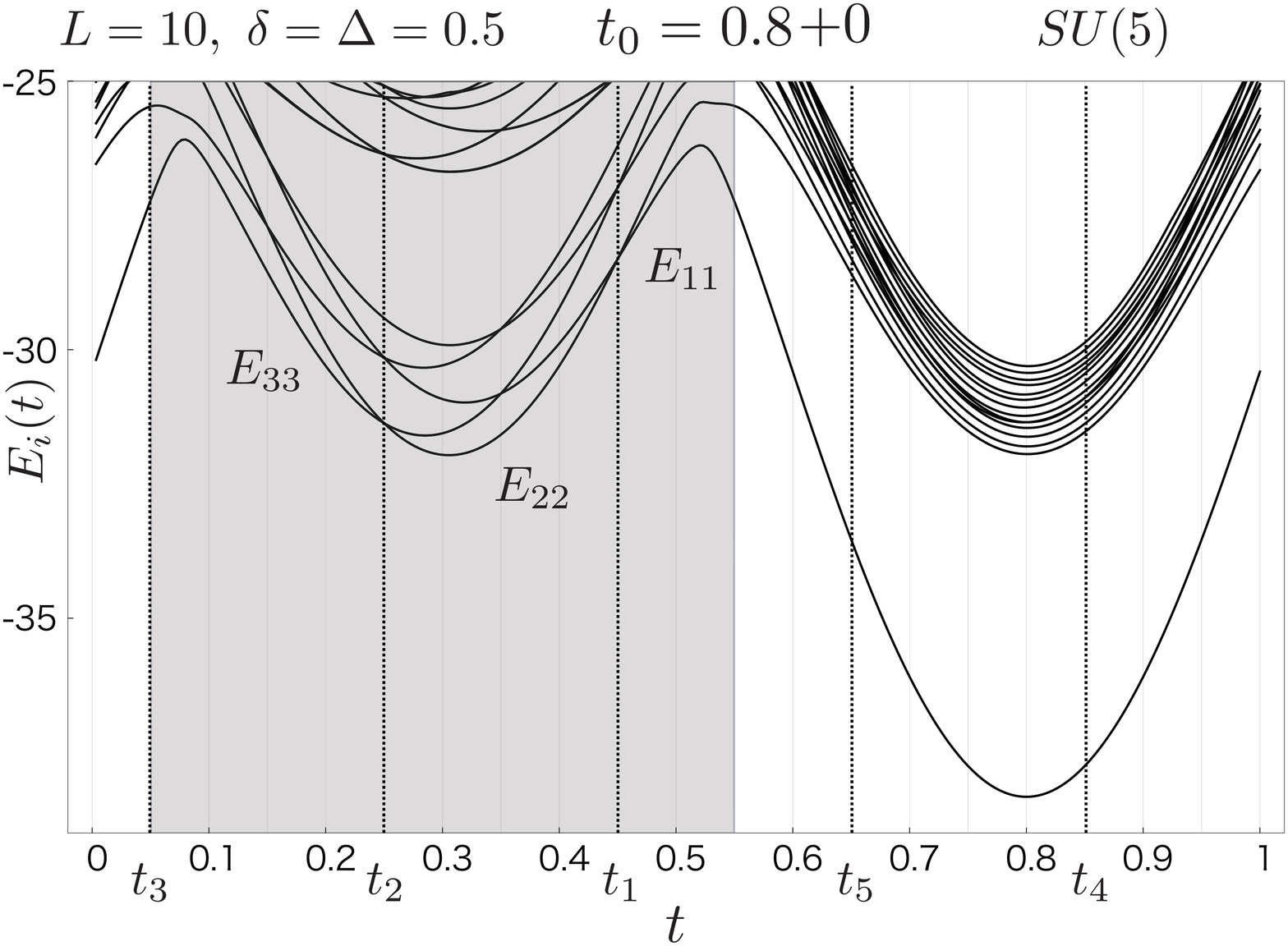}
  \subcaption{ ($t_0=0.8+0$)
   $\Delta P_3(t_2)=+1 $,
   $\Delta P_2(t_2)=-\Delta P_2(t_1)= -1$
   and
   $\Delta P_(t_1)=-1 $,   
   implies
   $I^\alpha :(-1,0,+1,0,0)$.
  }\label{fig:q5e}
 \end{minipage}
 
 \caption{\label{fig:L10-levelc-q5}
  $t_0$ dependence of
  the lowest 15 energy levels of the $SU(5)$ symmetric Hamiltonian
  of the 10 site system with open boundary condition within $\bar N_\alpha $
  sectors, where $\delta =0.5, \Delta =0.5$ { and $\Phi=1/5$} ($\QN=5$).
  Only the date for
  $\bar N_\alpha :(0,0,0,0,0)$ are shown, which give ground state energies.
  The gray region is for $|J_0(t)|<|J_e(t)|$ where 
  the low energy spectrum is composed of
  the multiplet of the edge states of the dimension
  $\QN^2$.
  One expects $\QN\times\QN$ emergent symmetry, in the $L\to\infty$ limit,
  that is responsible for the level crossings within the multiplet.
  The lowest eigen state is identified by $E_{\alpha \alpha }$ assuming the
  emergent $\QN\times\QN$ symmetry for the infinite system.
  The dimensions of the Hilbert spaces are $127905$
  (Compare with the Fig.\ref{fig:lowspecB}).
   Due to Eq.(\ref{eq:CanI}), $I^\alpha =\sum_i \Delta P^\alpha (t_i)$.} 
 \end{figure}
 
\end{widetext}

\subsection{Discontinuities of CoM by DMRG}
\label{sec:DMRG-st-Com}
In this section, the CoM's as discussed are directly calculated by using the
DMRG method.
The 
$Q=3$ fermion system can be simulated by mapping to
the following biquadratic spin model  with open boundary condition
(See appendix \ref{sec:color}, Eq.(\ref{eq:fermion-spin})),  
\begin{eqnarray}
&&H_{S}=H^{(2)}_{S}+H^S_{B},
\end{eqnarray}
where 
\begin{eqnarray}
&&H^{(2)}_S=\sum^{L/2}_{\ell=1}\biggl[J_o(t)(\vec{S}_{2\ell-1}\cdot \vec{S}_{2\ell})^2 +J_e(t)(\vec{S}_{2\ell}\cdot \vec{S}_{2\ell+1})^2\biggr],\nonumber\\
&&H^S_{B}(t)=-\sum_{j,\alpha}\Delta_{\alpha}(S^{\alpha}_j)^2 .
\end{eqnarray} 
Here $J_o=J_{i\in \text{odd}}$, $J_e=J_{i\in \text{even}}$ and 
$\Delta_\alpha$ is given by Eq.(\ref{eq:proto}).

We calculate the staggered quadratic spin center of mass (CoM), given by
\begin{eqnarray}
P^s_{1}(t)&=&\sum^{L}_{j=1}(-1)^{j}x_j\langle S^{x2}_j\rangle{={\delta_L} + P_1},\nonumber\\
P^s_{2}(t)&=&\sum^{L}_{j=1}(-1)^{j}x_j\langle S^{y2}_j\rangle{={\delta_L}  +P_2},\nonumber\\
P^s_{3}(t)&=&\sum^{L}_{j=1}(-1)^{j}x_j\langle S^{z2}_j\rangle{={\delta _L} +P_3},
\label{sqsCoM}
\end{eqnarray}
where
$\delta _L=0\, 
(L:\text{odd})$,
$ 1/2\, (L:\text{even})$,
$j_0=(L+1)/2$ and $\langle \cdot\rangle$ means taking the expectation value for the groundstate at the time $t$.
The CoM of the spins $P^s_\alpha  $, $(\alpha =1,2,3)$ is {directly related to that of fermions, $P_\alpha  $.
Note that the factor $(-1)^{j-1}$ in  Eq.(\ref{eq:CoM}) and the summation over $j$. 
}
In the following, we fix the parameters as
{$J_0=-1$, $\delta J=0.5$} and $\Delta=0.5$, same to the parameter set of the results in Fig.8-10.

To calculate the behavior of the CoM, we employ DMRG algorithm in TeNPy package\cite{tenpy}. 
The numerical results for various $t_0$ are shown in Fig.~\ref{FigC1}.  
We observe the behavior of the CoM in all data divides into two parts, continuous part and jump part. The time evolution of the CoM in the continuous part indicates the presence of the bulk current. 
We verify that each jump point are identical to the level crossing points ($t_1$, $t_2$ and $t_3$), expected from Fig.\ref{fig:lowspecB} and Fig.\ref{fig:L12-levelc-q3}. 
{
The jumps of the CoM, $P_\alpha $, and the sum them, $I^\alpha  $,
are summarized in Table \ref{table:DMRG}.
This is  consistent to the low energy spectra shown in
Fig.\ref{fig:L12-levelc-q3} (see its caption).
}

As for a finite size effect mentioned in the appendix \ref{sec:disc},
we have shown a concrete example, in Fig.~\ref{FigC2}.
The jump  of $\Delta P^{s}_{1}$ and $\Delta P^{s}_{3}$ approach to 1 and -1
for $L\to \infty$. This agrees to the exponential localization of the edge states as discussed. 

\begin{widetext}
\centering
\begin{figure*}[t]
\includegraphics[width=18cm]{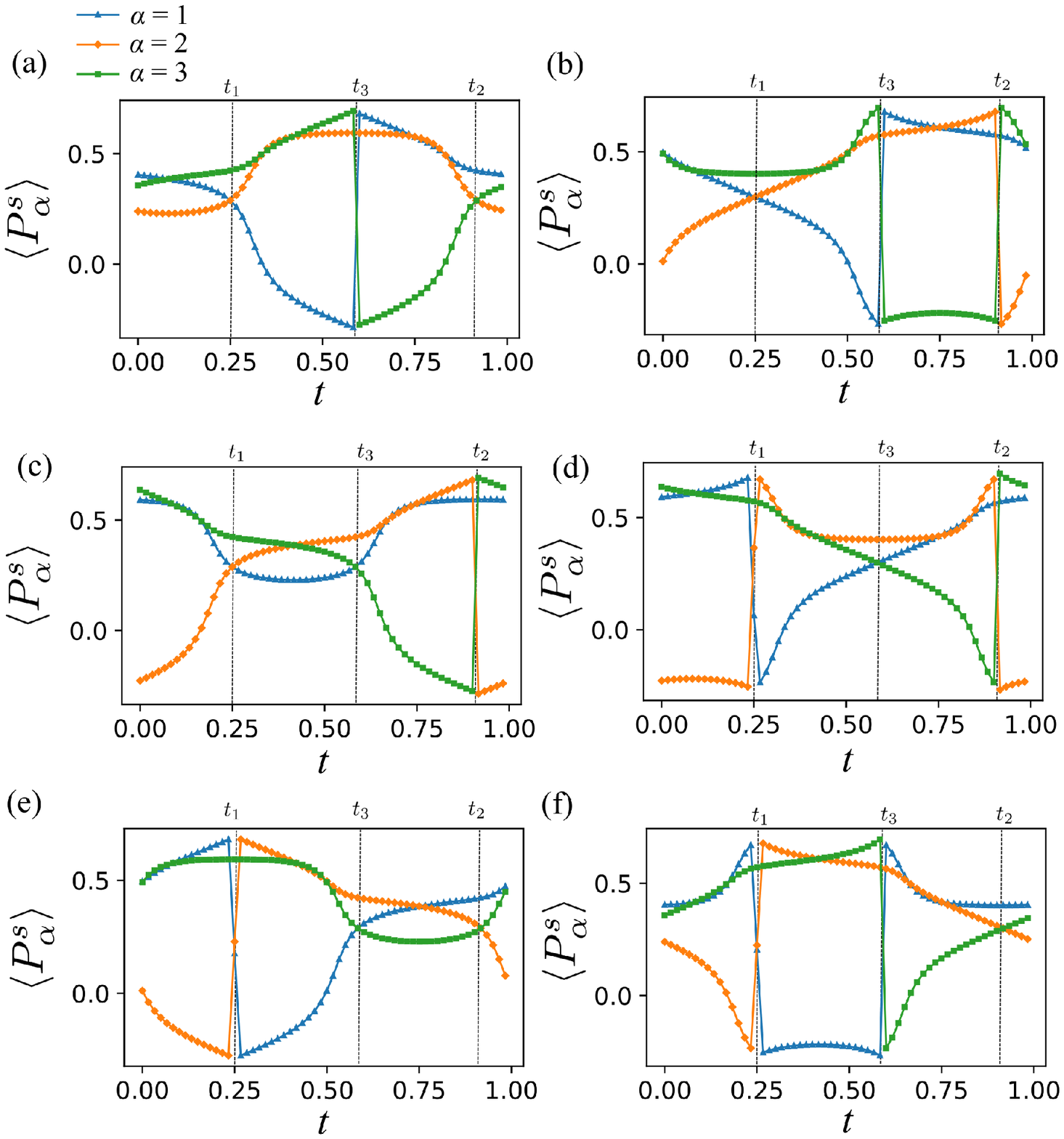}
\caption{The behavior of the sq-CoM's $P^{s}_{\alpha}$ for various $t_0$ and components. The system size is $L=64$ with open boundary condition. 
The data shows $t_0/T=1/12$[(a)], $3/12$[(b)], 
$5/12$[(c)], $7/12$[(d)], $9/12$[(e)] and $11/12$[(f)].}
\label{FigC1}
\end{figure*}



\begin{center}
\renewcommand{\arraystretch}{1.2}
\begin{table}[t]
  \caption{
  Jumps of the CoM, the sum of them in Fig. \ref{FigC1}
  and
  values
  of
  the numerically obtained Chern numbers picked up from  Fig.\ref{fig:ChernQ3}(a),(b) and (c).}
  \label{table:DMRG}
  \centering
  \begin{tabular}{|c|c|c|c|c|c|}
    \hline
    $t_0$
    &    $(\Delta P_1,\Delta P_2,\Delta P_3)|_{t=t_1} $
    &    $(\Delta P_1,\Delta P_2,\Delta P_3)|_{t=t_2} $
    &  $(\Delta P_1,\Delta P_2,\Delta P_3)|_{t=t_3} $
    &  $(I^1,I^2,I^3) $        &  $(C^1,C^2,C^3=C^0) $ 
    \\
    \hline  \hline
    $\frac {1}{12}T $  & $ (0,0,0) $ &$ (0,0,0)$&$(+1,0,-1)$&$(+1,0,-1)$ &$(+1,0,-1)$ 
\\  \hline
    $\frac {3}{12}T $  & $ (0,0,0) $ &$ (0,-1,+1)$&$(+1,0,-1)$&$(+1,-1,0)$&$(+1,-1,0)$
\\  \hline
    $\frac {5}{12}T $  & $ (0,0,0) $ &$ (0,-1,+1)$&$(0,0,0)$&$(0,-1,+1)$&$(0,-1,+1)$ 
\\  \hline
    $\frac {7}{12}T $  & $ (-1,+1,0) $ &$ (0,-1,+1)$&$(0,0,0)$&$(-1,0,+1)$&$(-1,0,+1)$ 
\\  \hline
    $\frac {9}{12}T $  & $ (-1,+1,0) $ &$ (0,0,0)$&$(0,0,0)$&$(-1,+1,0)$&$(-1,+1,0)$ 
\\  \hline
    $\frac {11}{12}T $  & $ (-1,+1,0) $ &$ (0,0,0)$&$(+1,0,-1)$&$(0,+1,-1)$&$(0,+1,-1)$ 
\\  \hline
\end{tabular}
\end{table}
\end{center}

\centering
\begin{figure}[t]
\includegraphics[width=10cm]{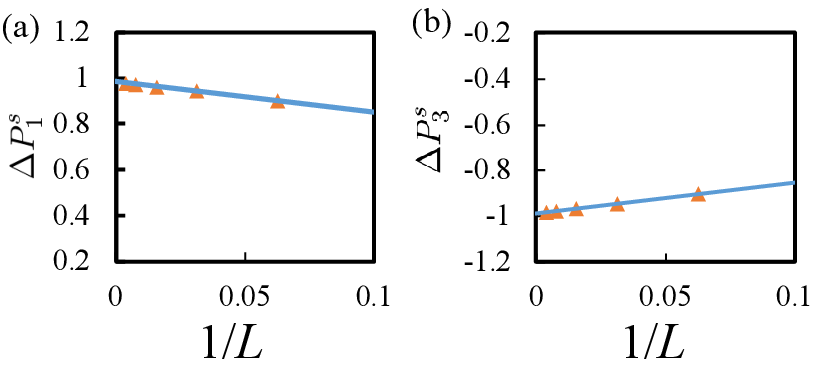}
\caption{System size dependence of the jumps of CoM's at $t=t_3$ and $t_0/T=1/12$. 
Results for (a) $\Delta P^{s}_{1}$ and (b) $\Delta P^{s}_{3}$.}
\label{FigC2}
\end{figure}


\end{widetext}


\subsection{Direct evaluation of Chern numbers}
\label{sec:ChernBEC}
Using the bulk-edge correspondence of the topological pump,
the Chern number, that is a total pumped charge of the bulk
should be the same to the discontinuity of the CoMs, Eq.(\ref{eq:aChern}).
We have confirmed it by a direct evaluation of the Chern number
Eq.(\ref{eq:bec-qandC})
obtained by the integral of the field strength $B_\tw$.
This is done
by using the Fukui-Hatsugai-Suzuki formula \cite{Fukui05}
for $\QN=3$, $\Phi=1/3,2/3$,
$\QN=4$, $\Phi=1/4,3/4$
and 
$\QN=5$, $\Phi=1/5,2/5,3/5,4/5$. We have plotted the Chern numbers
as a function of  $t_0$.
The results are shown in Figs.\ref{fig:ChernQ3},
\ref{fig:ChernQ4} and \ref{fig:ChernQ5}.
The analytic formula Eq.(\ref{eq:aChern}) is plotted by the solid lines
and the numerical values
obtained by discretized integration formula by the Fukui-Hatsugai-Suzuki
formula
are shown in the red circles. They agree with each other almost completely
except a few points for $\QN=5$, $\Phi=2/5$ and $\Phi=3/5$. They are near the
topological phase transitions where any numerical calculation can be unstable.

\vfill

\begin{widetext}
 \begin{center}
  \begin{figure}[t]
  \captionsetup[sub]{skip=0pt}
 \begin{minipage}[t]{0.48\linewidth}
  \centering
  \includegraphics[keepaspectratio, scale=0.33]{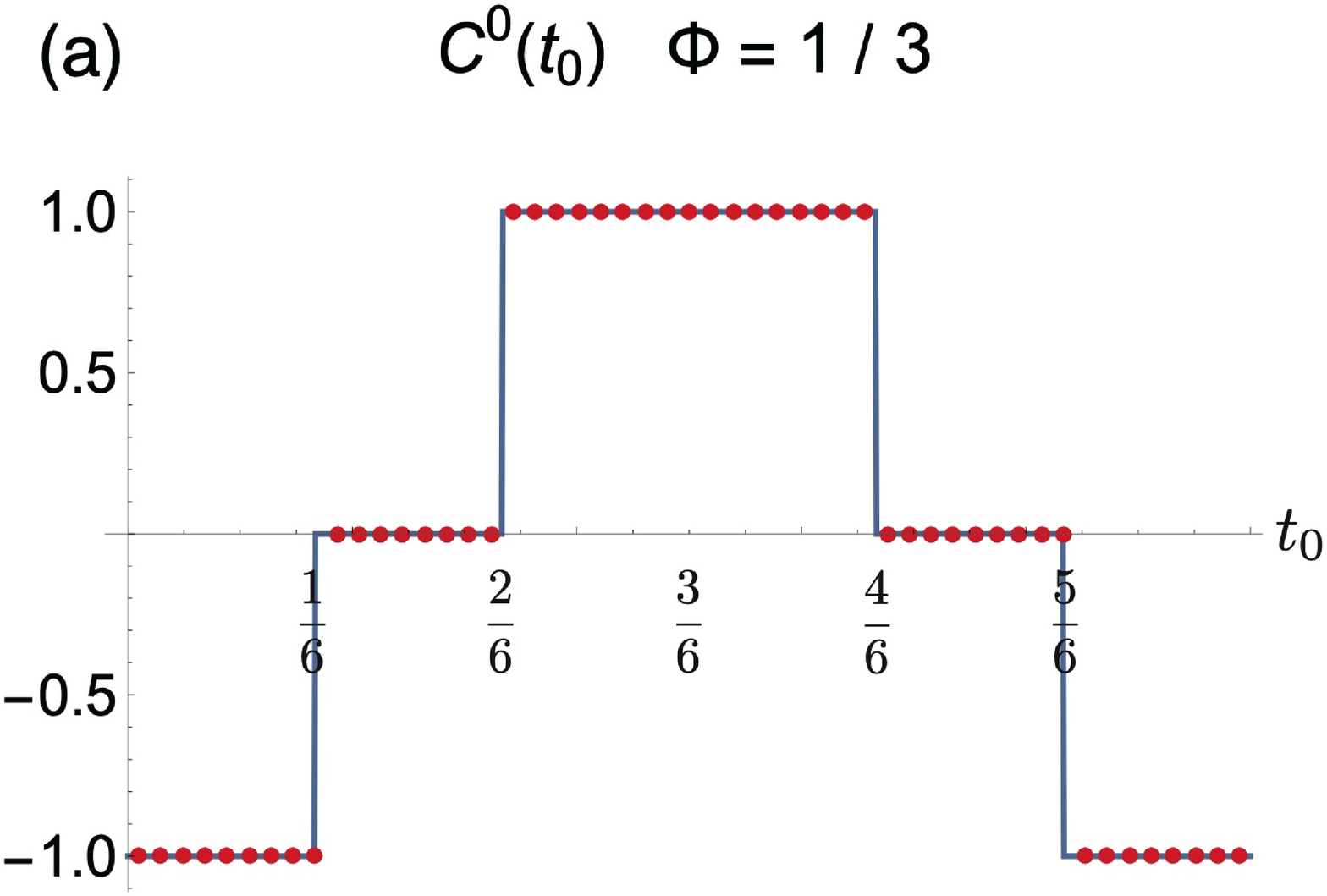}
 \end{minipage}
 \begin{minipage}[t]{0.48\linewidth}
  \centering
  \includegraphics[keepaspectratio, scale=0.33]{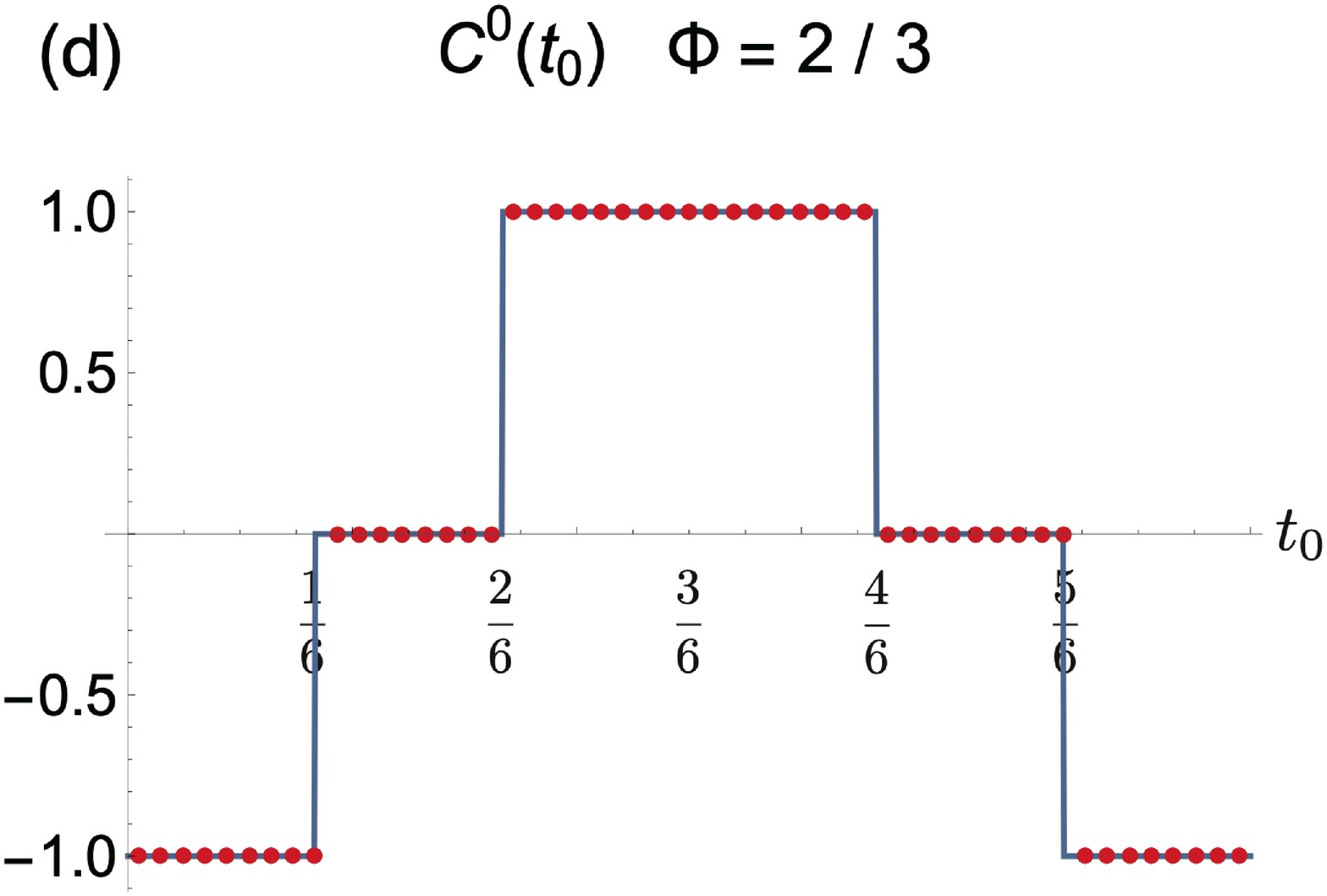}
 \end{minipage}
 \\
 \begin{minipage}[t]{0.48\linewidth}
  \centering
  \includegraphics[keepaspectratio, scale=0.33]{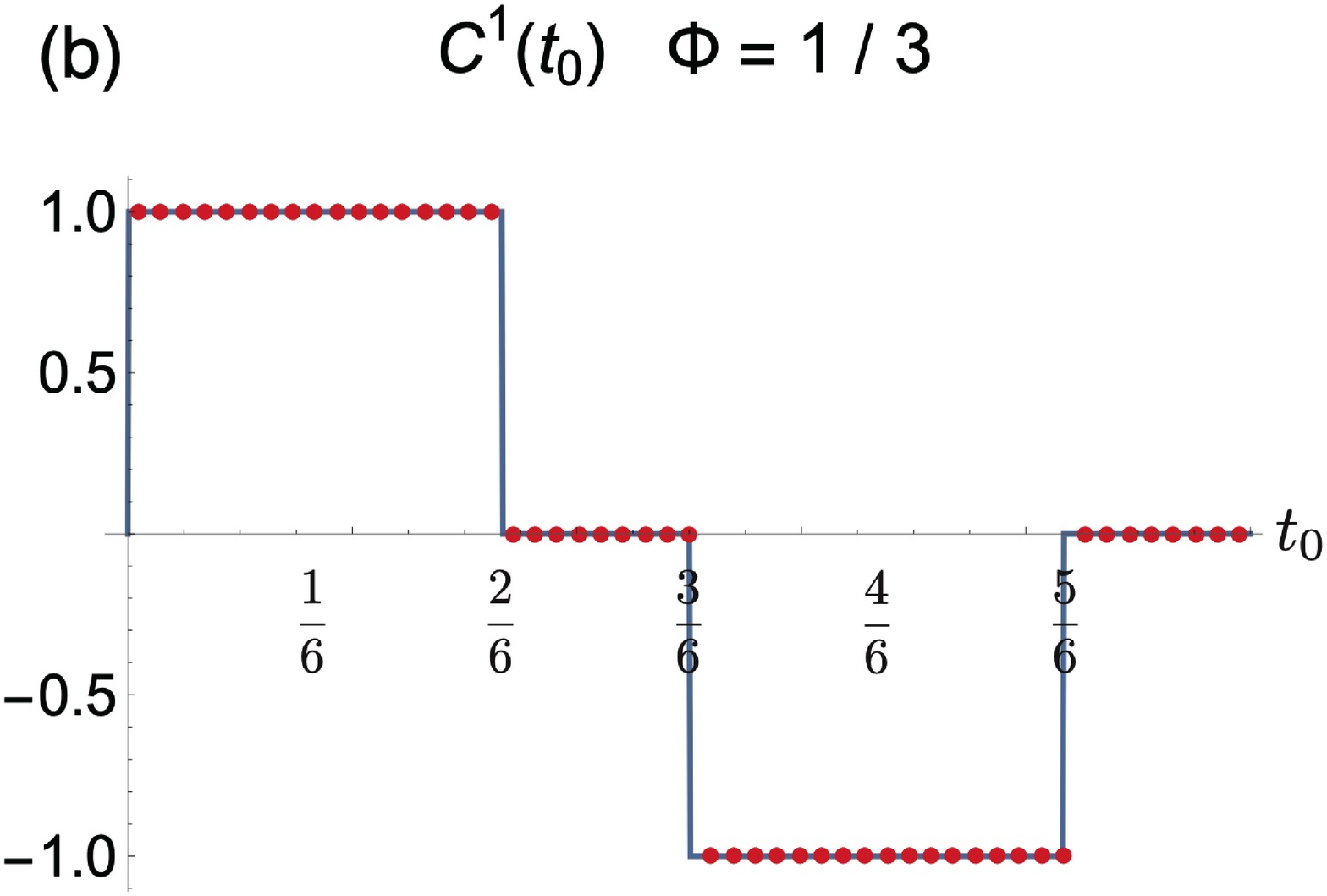}
 \end{minipage}
 \begin{minipage}[t]{0.48\linewidth}
  \centering
  \includegraphics[keepaspectratio, scale=0.33]{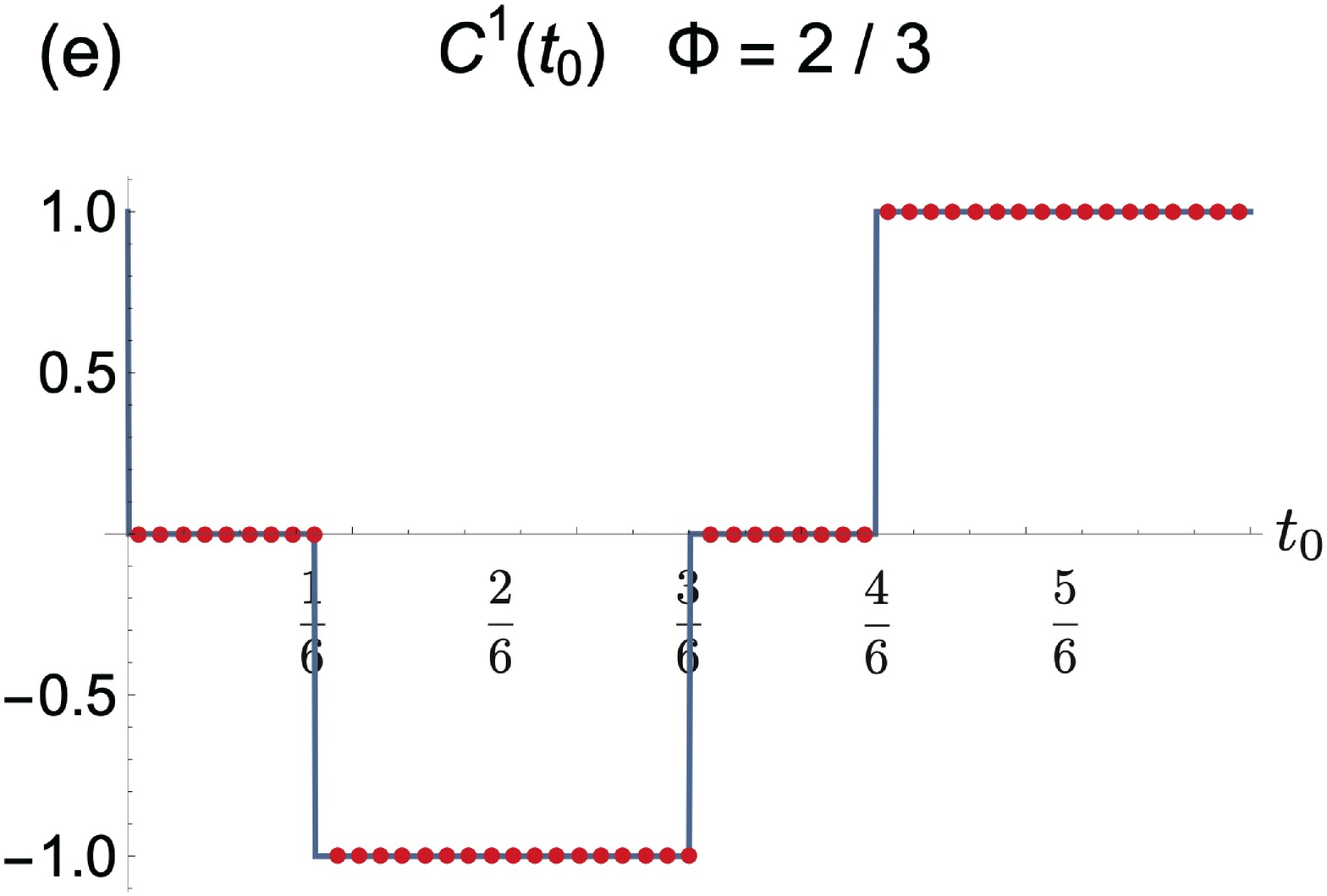}
 \end{minipage}
 \\
 \begin{minipage}[t]{0.48\linewidth}
  \centering
  \includegraphics[keepaspectratio, scale=0.33]{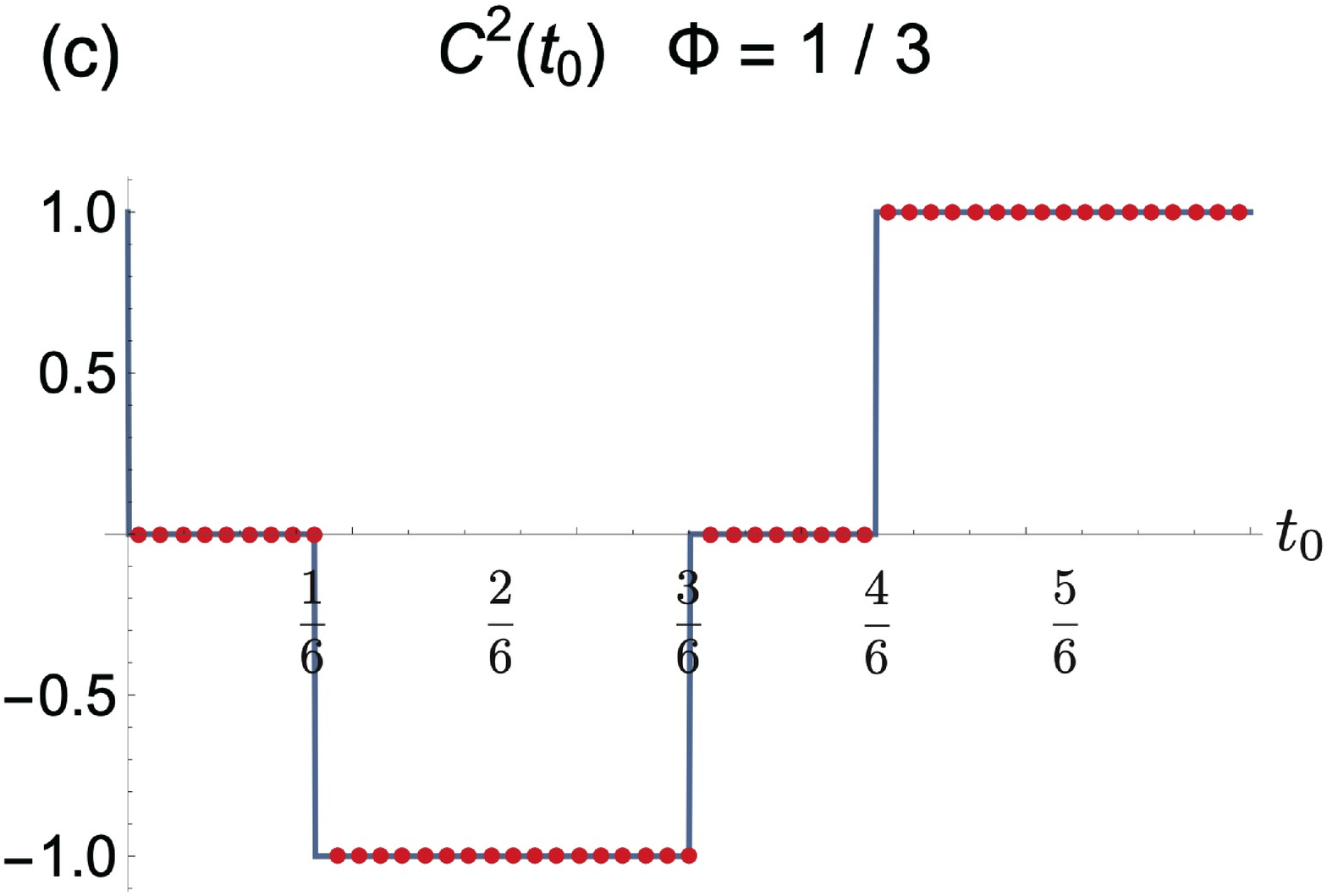}
 \end{minipage}
 \begin{minipage}[t]{0.48\linewidth}
  \centering
  \includegraphics[keepaspectratio, scale=0.33]{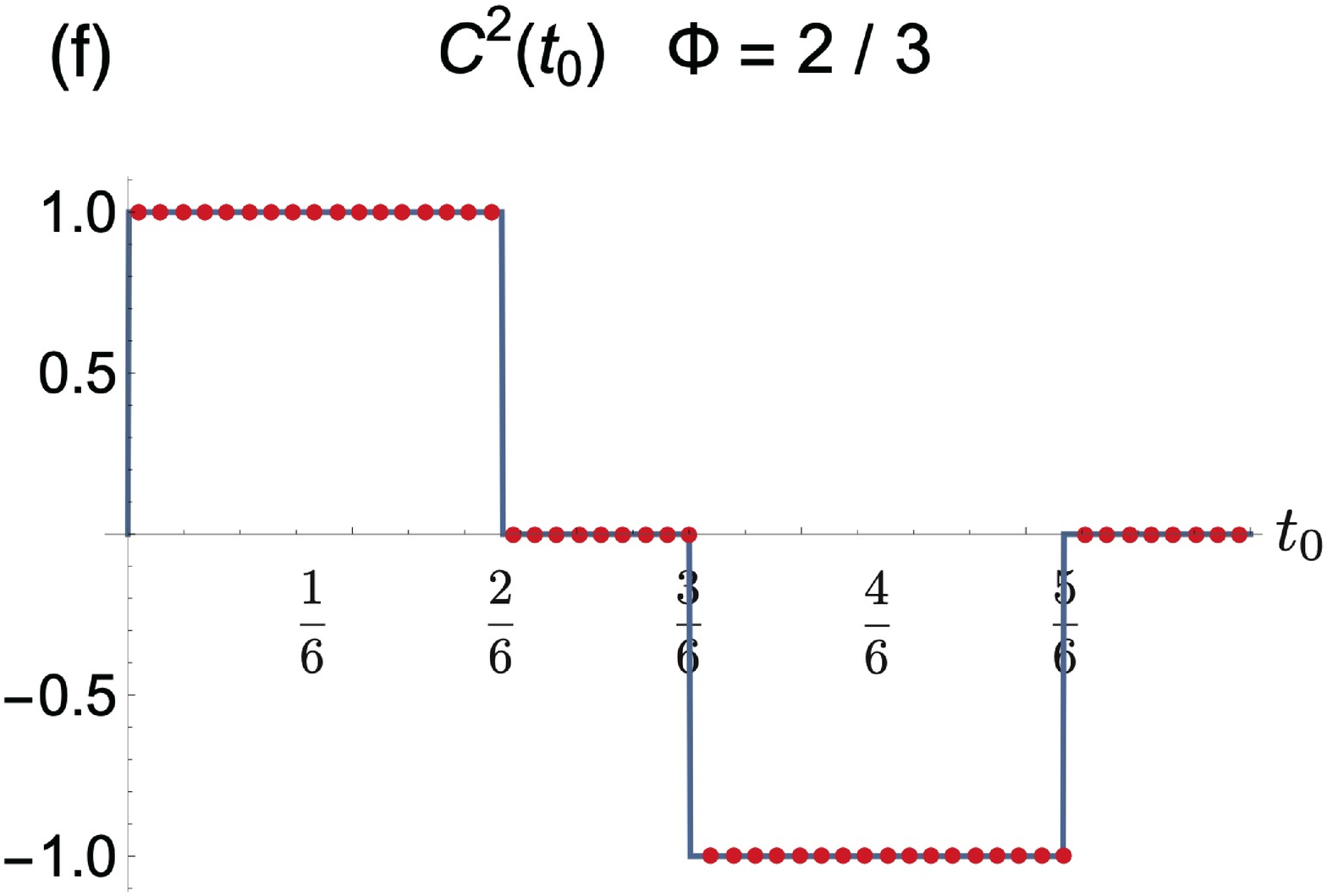}
 \end{minipage}
 \\
 \caption{\label{fig:ChernQ3}
  $t_0$ dependence of
  the Chern numbers. $\QN=3,\Phi=\frac {1}{3} $ and $\frac {2}{3} $.
  $J_0=-1.0, \delta =0.5$ and $\Delta =0.5$. The dimension of
  the Hilbert space is 4653 ($L=10$).
  The data points are for $t_0=(i-0.5)/51, i=1,\cdots,51$.
  Red points are numerical evaluation of Eq.(\ref{eq:bec-qandC}) by
  Fukui-Hatsugai-Suzuki formula \cite{Fukui05}
  and the lines are analytical results, Eq.(\ref{eq:aChern}).
  }
 \end{figure}
\end{center} 



 \begin{figure}[h]
  \captionsetup[sub]{skip=0pt}
 \begin{minipage}[b]{0.48\linewidth}
  \centering
  \includegraphics[keepaspectratio, scale=0.3]{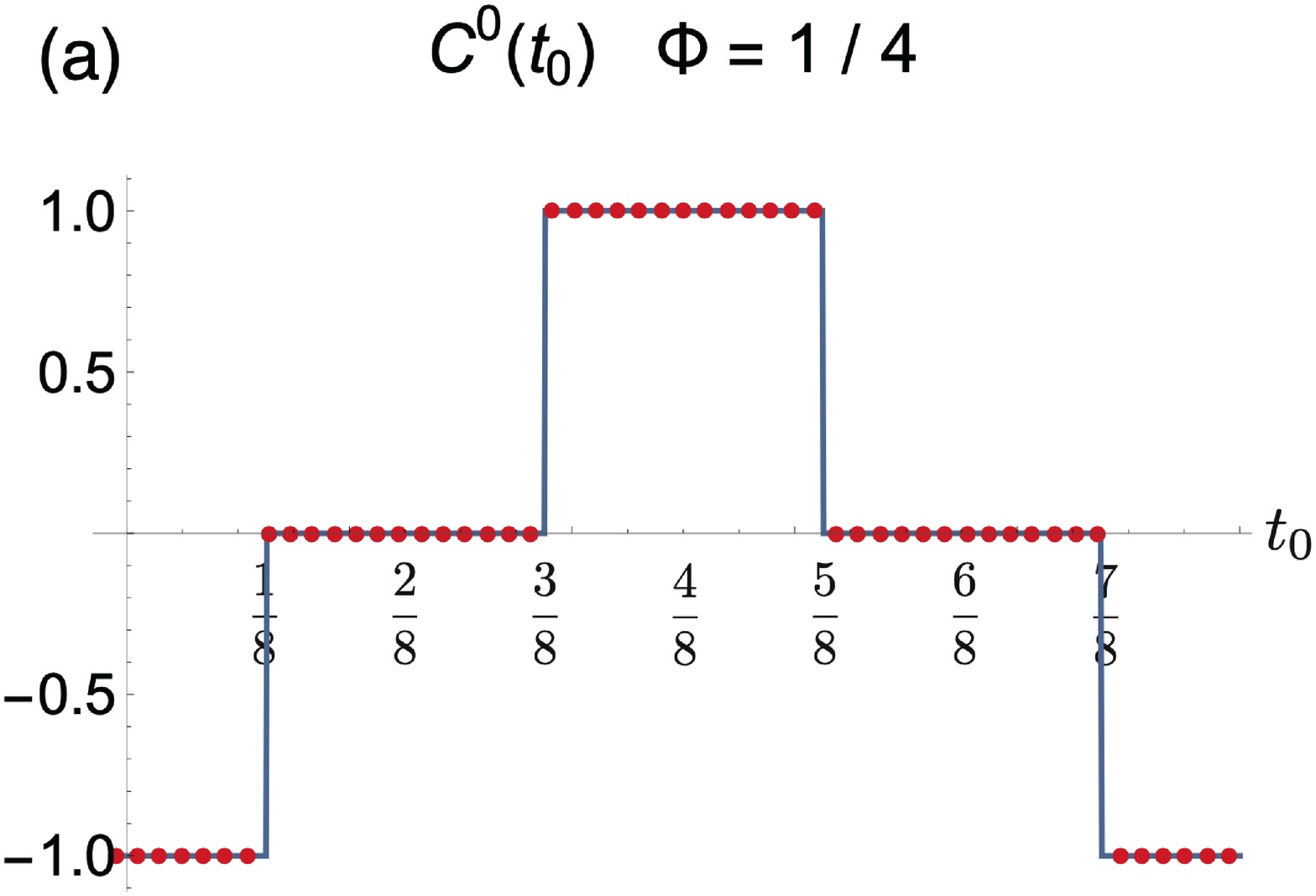}
 \end{minipage}
 \begin{minipage}[b]{0.48\linewidth}
  \centering
  \includegraphics[keepaspectratio, scale=0.3]{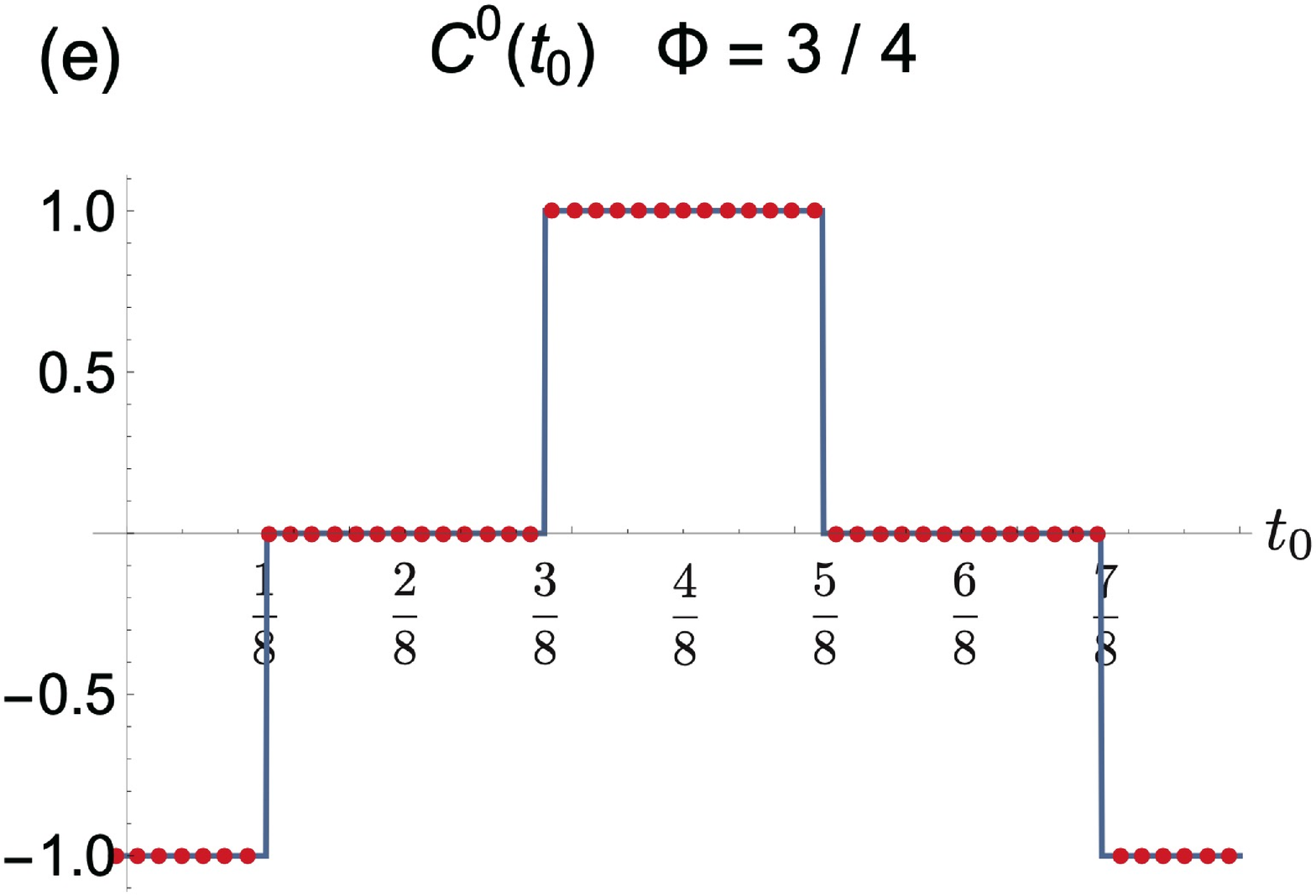}
 \end{minipage}
 \\
 \begin{minipage}[b]{0.48\linewidth}
  \centering
  \includegraphics[keepaspectratio, scale=0.30]{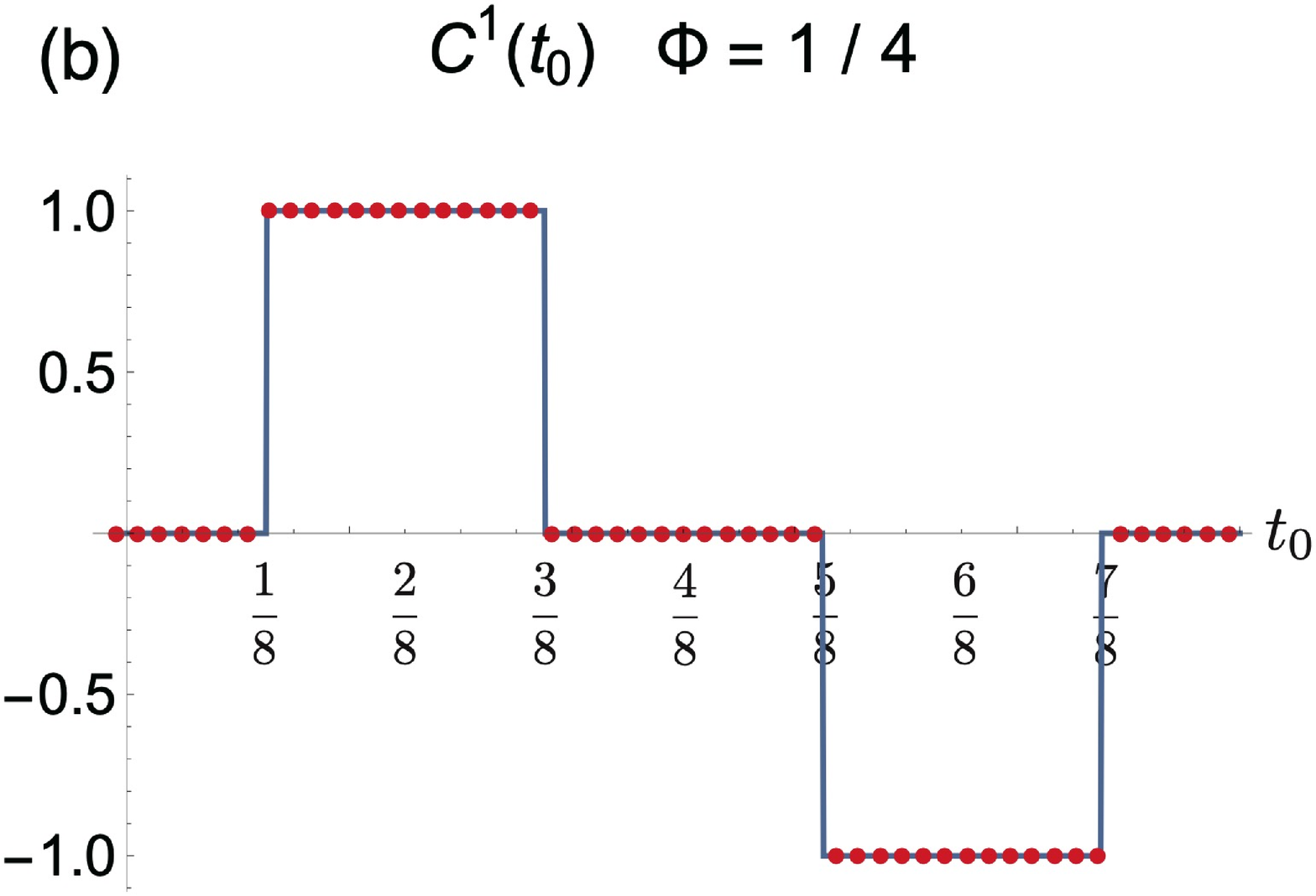}
 \end{minipage}
 \begin{minipage}[b]{0.48\linewidth}
  \centering
  \includegraphics[keepaspectratio, scale=0.30]{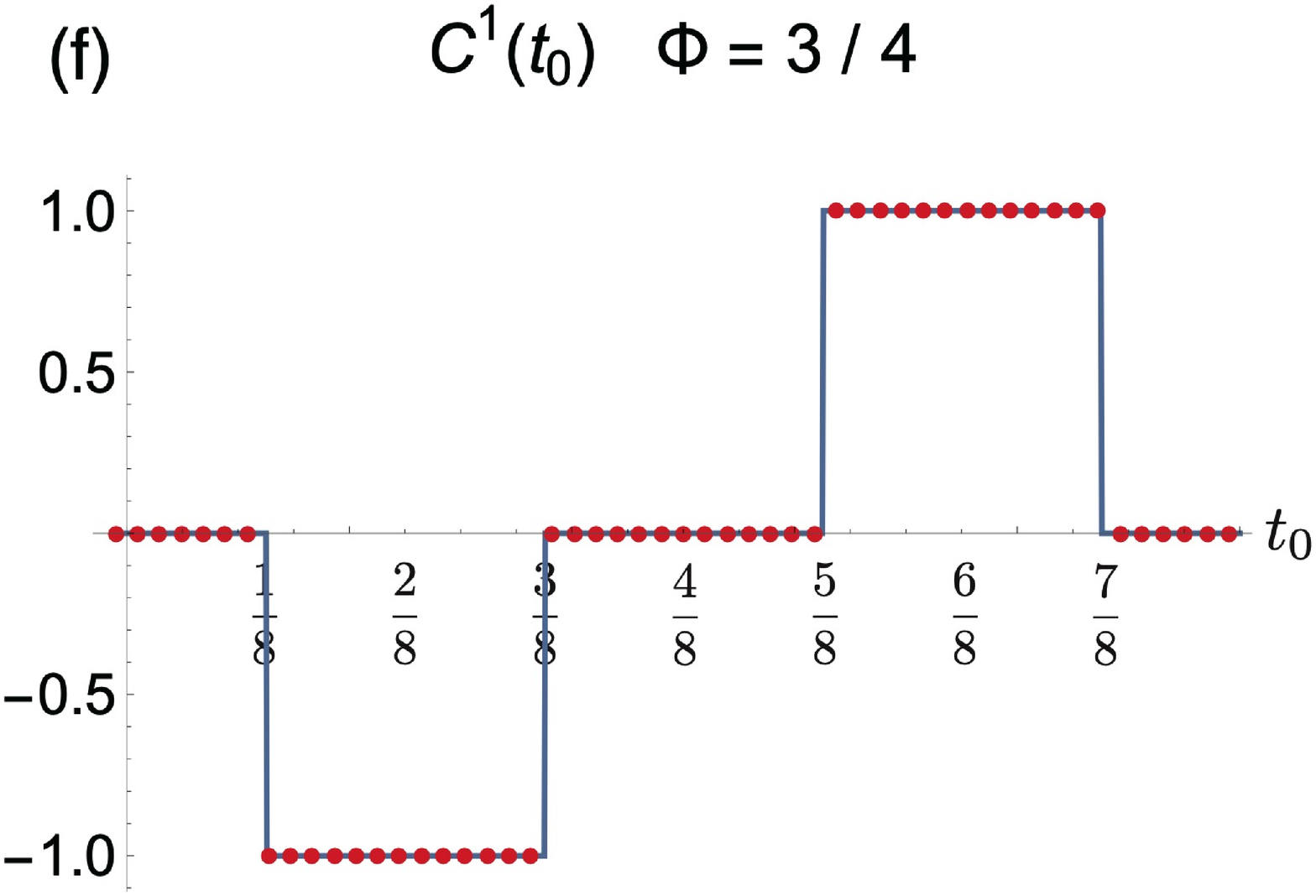}
 \end{minipage}
 \\
 \begin{minipage}[b]{0.48\linewidth}
  \centering
  \includegraphics[keepaspectratio, scale=0.30]{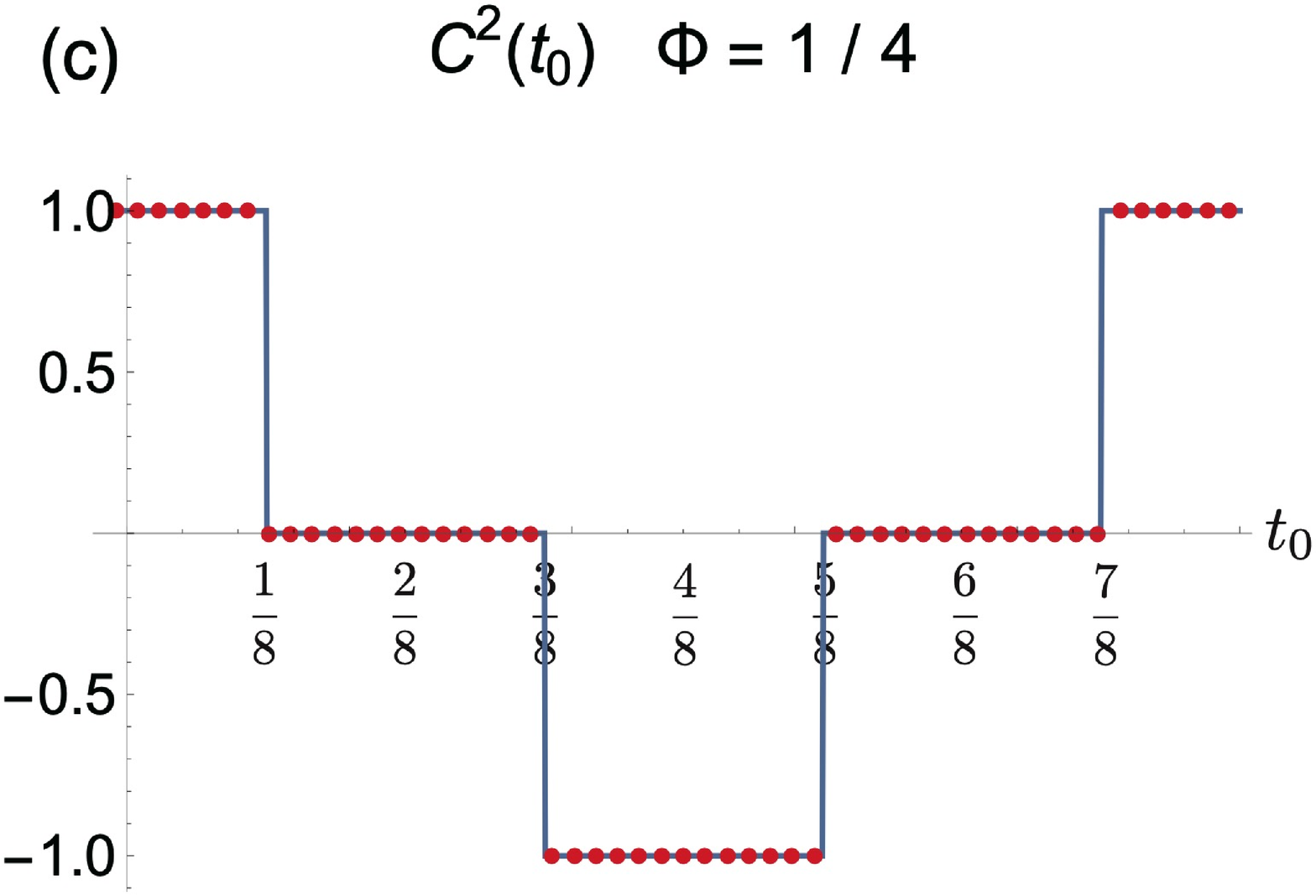}
 \end{minipage}
 \begin{minipage}[b]{0.48\linewidth}
  \centering
  \includegraphics[keepaspectratio, scale=0.30]{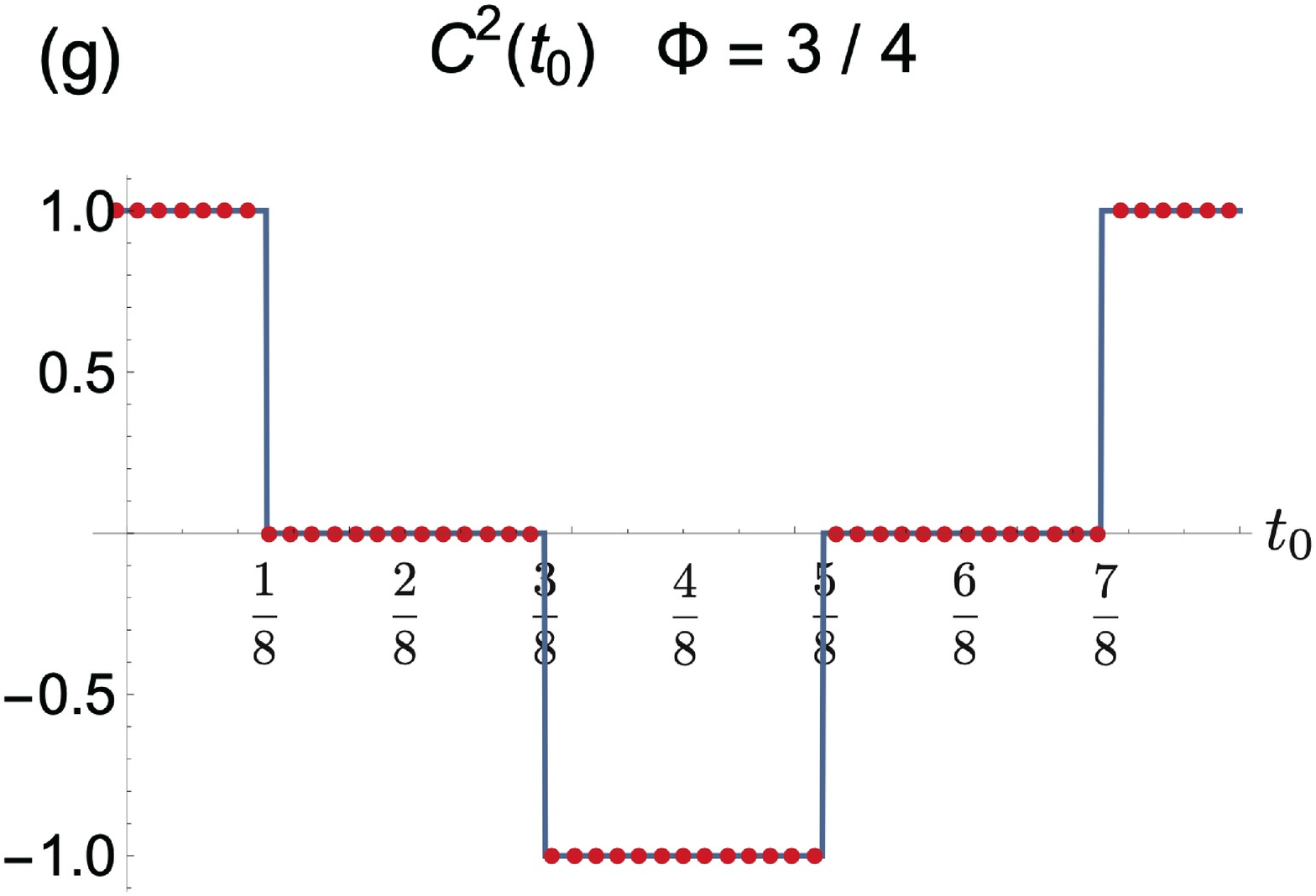}
 \end{minipage}
 \\
 \begin{minipage}[b]{0.48\linewidth}
  \centering
  \includegraphics[keepaspectratio, scale=0.30]{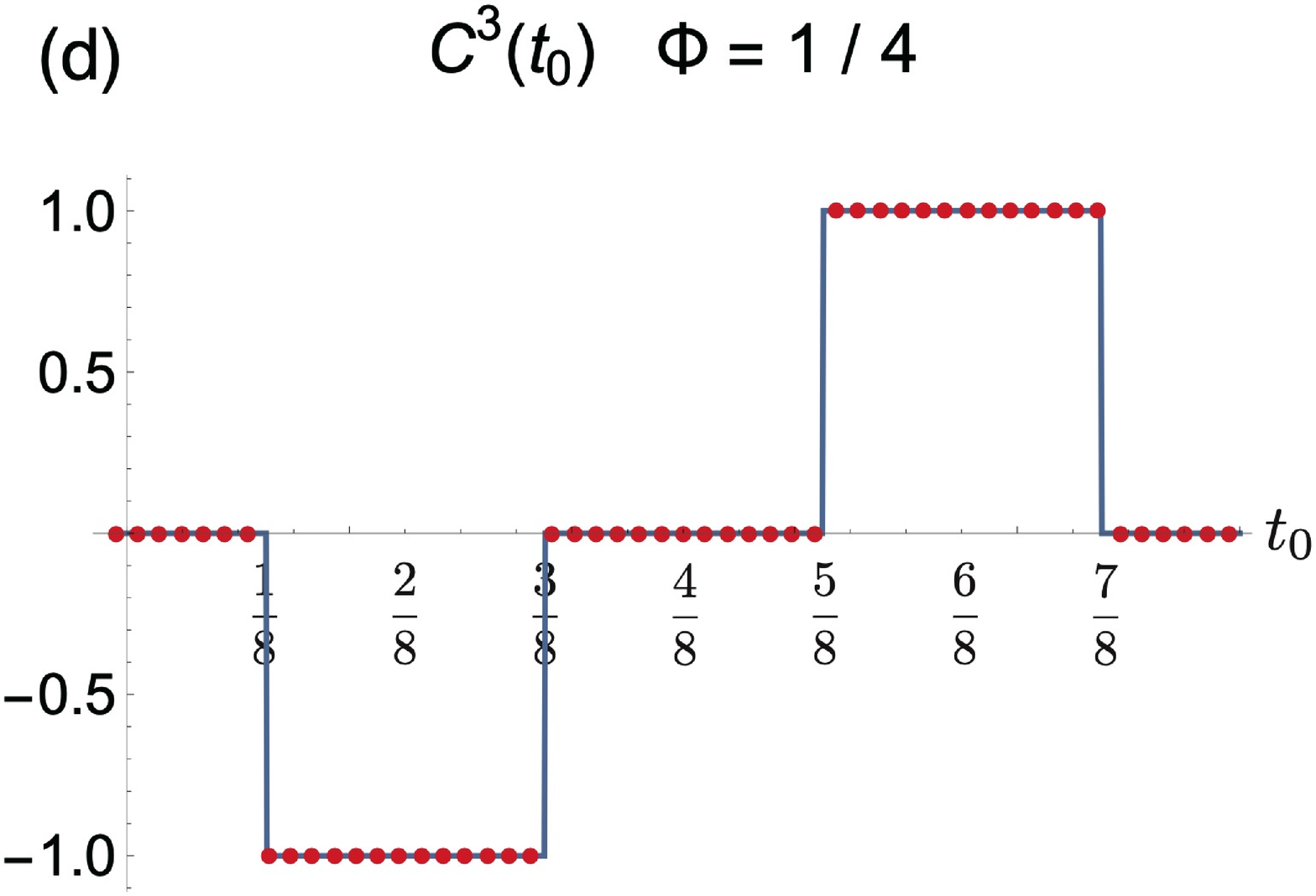}
 \end{minipage}
 \begin{minipage}[b]{0.48\linewidth}
  \centering
  \includegraphics[keepaspectratio, scale=0.30]{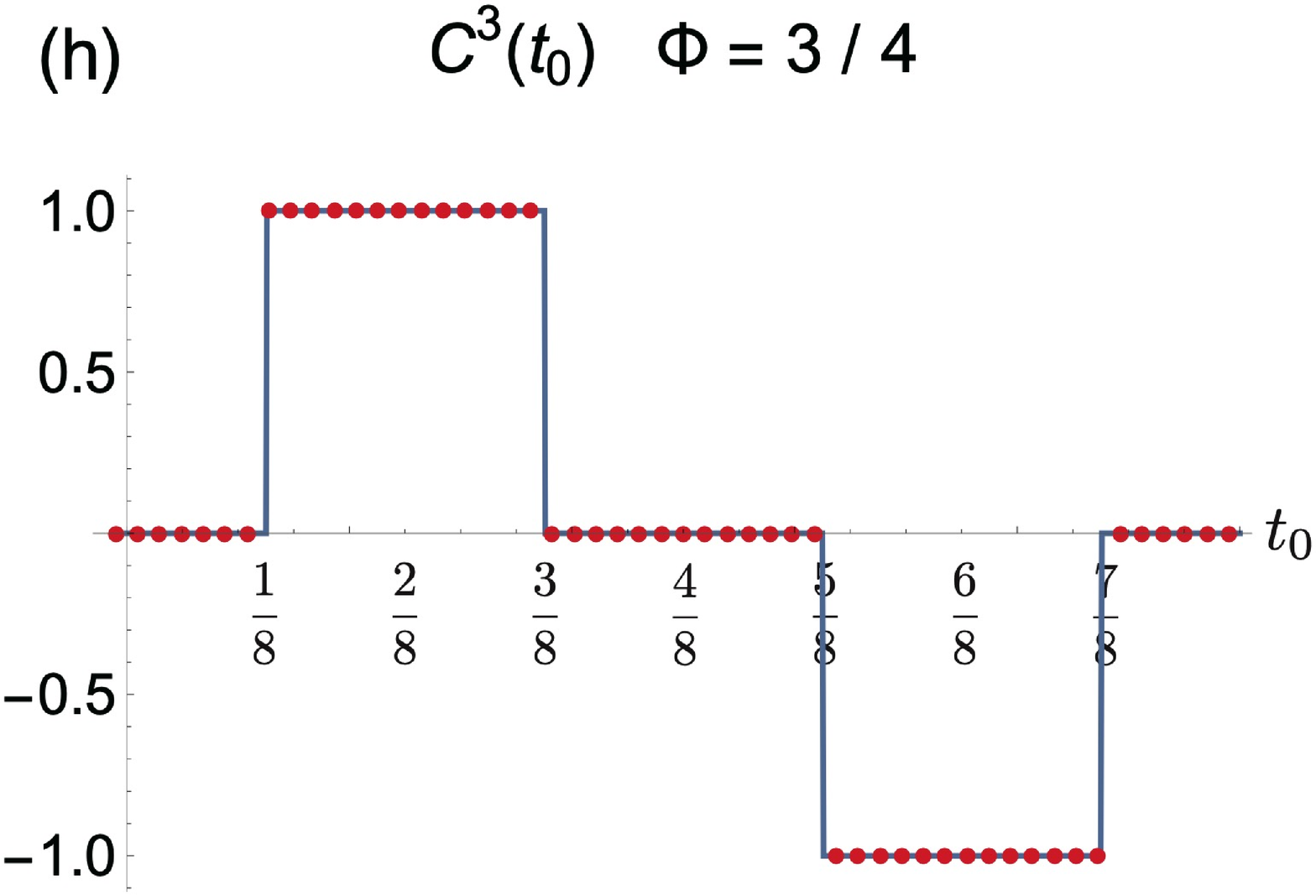}
 \end{minipage}
 \\
 \caption{\label{fig:ChernQ4}
  $t_0$ dependence of
  the Chern numbers. $\QN=4,\Phi=\frac {1}{4} $ and $\frac {3}{4} $.
  $J_0=-1.0, \delta =0.5$ and $\Delta =0.5$. The dimension of
  the Hilbert space is 2716 ($L=8$). The data points are for
  $t_0=(i-0.5)/51, i=1,\cdots,51$.
  Red points are numerical evaluation of Eq.(\ref{eq:bec-qandC}) by
  Fukui-Hatsugai-Suzuki formula \cite{Fukui05}
  and the lines are analytical results, Eq.(\ref{eq:aChern}).
  }
 \end{figure}



 \begin{figure}[t]
  \captionsetup[sub]{skip=0pt}
 \begin{minipage}[b]{0.48\linewidth}
  \centering
  \includegraphics[keepaspectratio, scale=0.24]{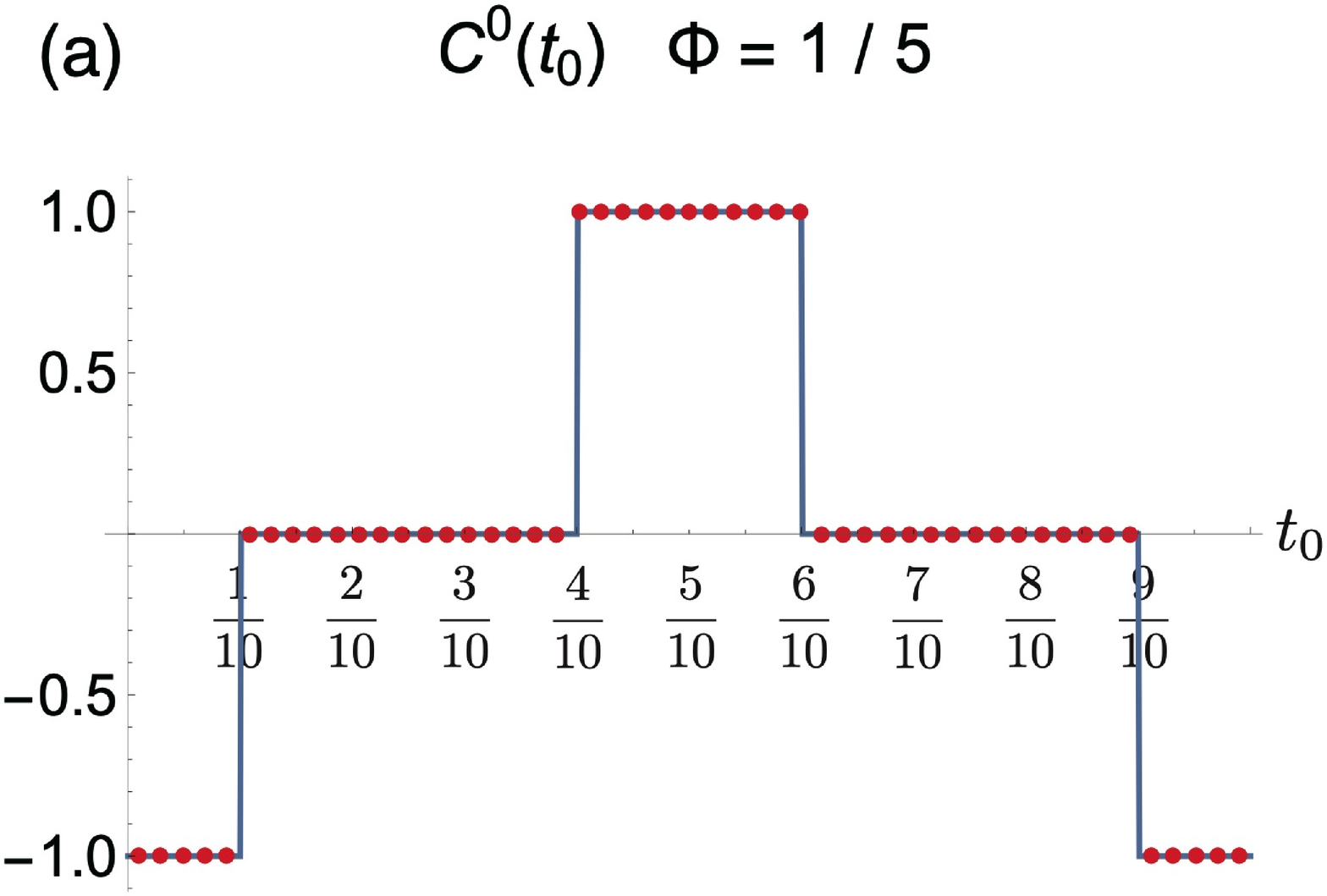}
 \end{minipage}
 \begin{minipage}[b]{0.48\linewidth}
  \centering
  \includegraphics[keepaspectratio, scale=0.24]{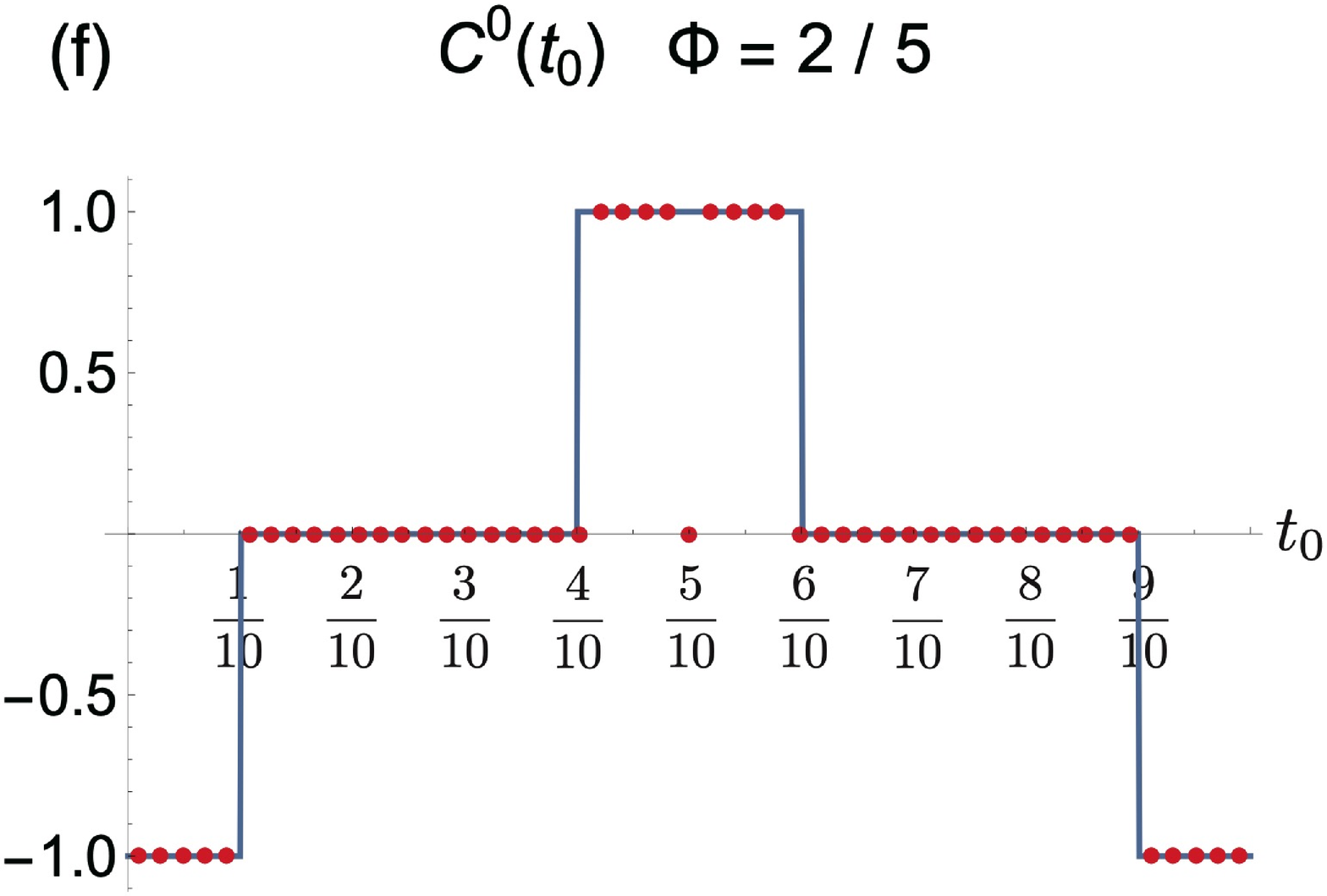}
 \end{minipage}
 \\
 \begin{minipage}[b]{0.48\linewidth}
  \centering
  \includegraphics[keepaspectratio, scale=0.24]{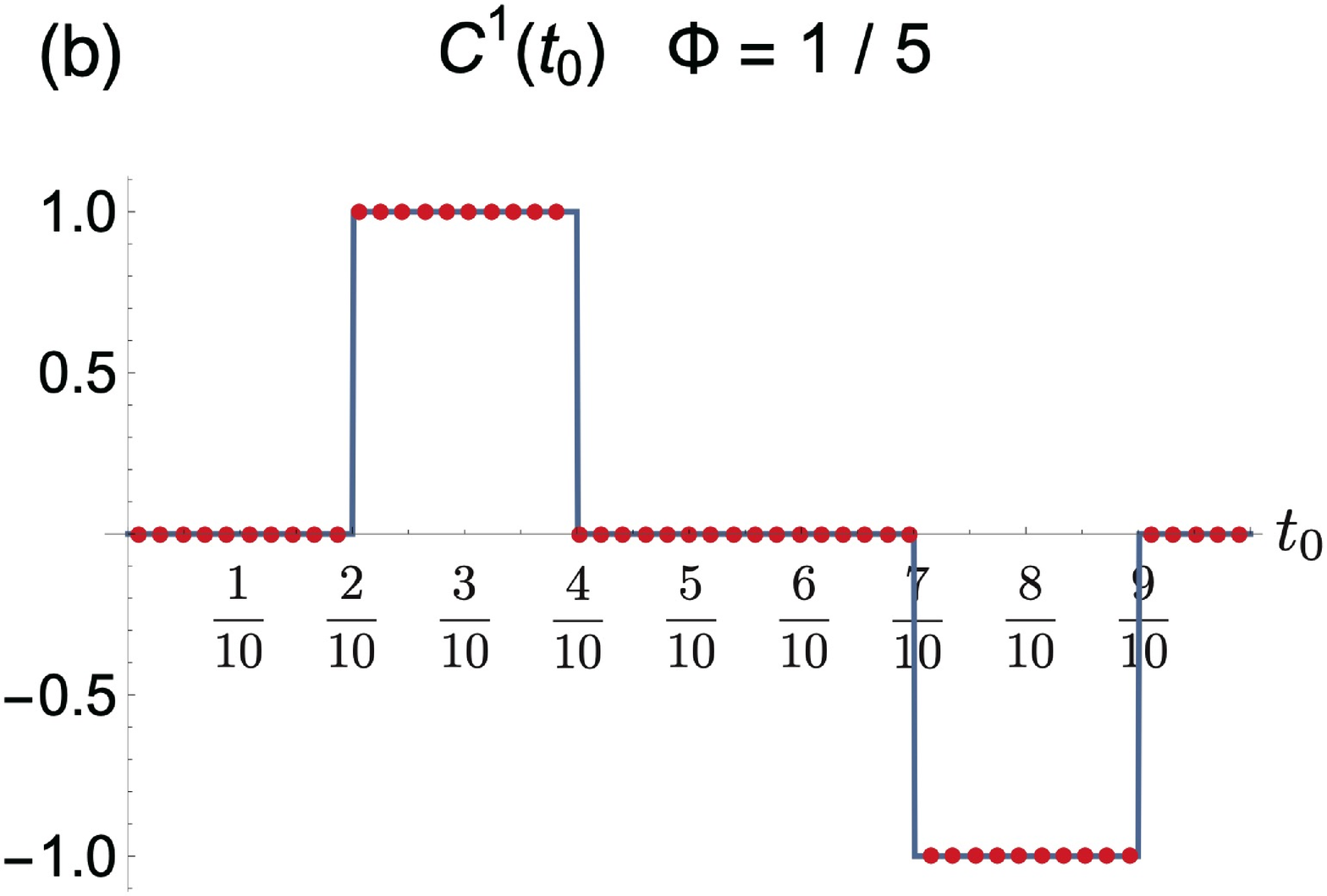}
 \end{minipage}
 \begin{minipage}[b]{0.48\linewidth}
  \centering
  \includegraphics[keepaspectratio, scale=0.24]{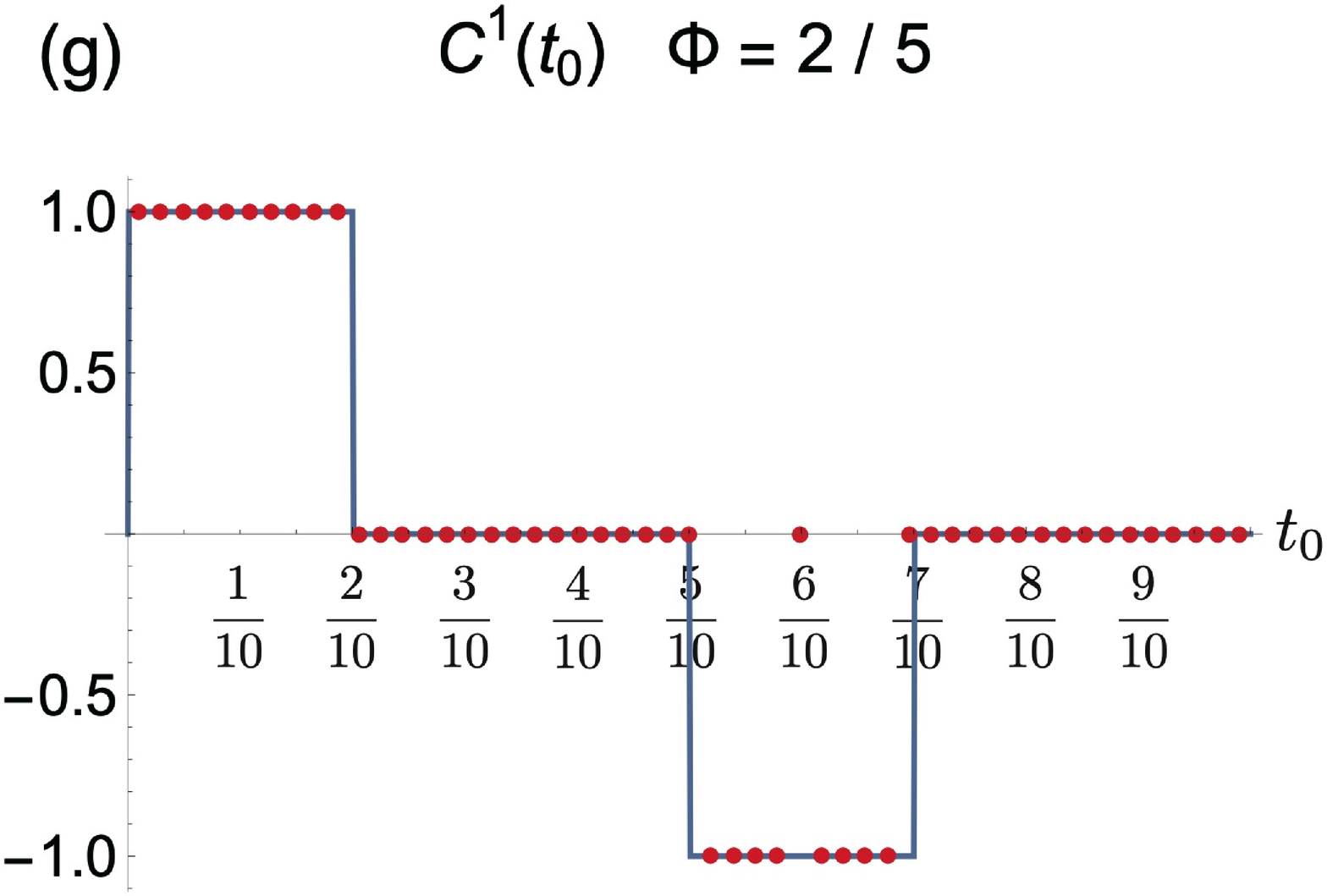}
 \end{minipage}
 \\
 \begin{minipage}[b]{0.48\linewidth}
  \centering
  \includegraphics[keepaspectratio, scale=0.24]{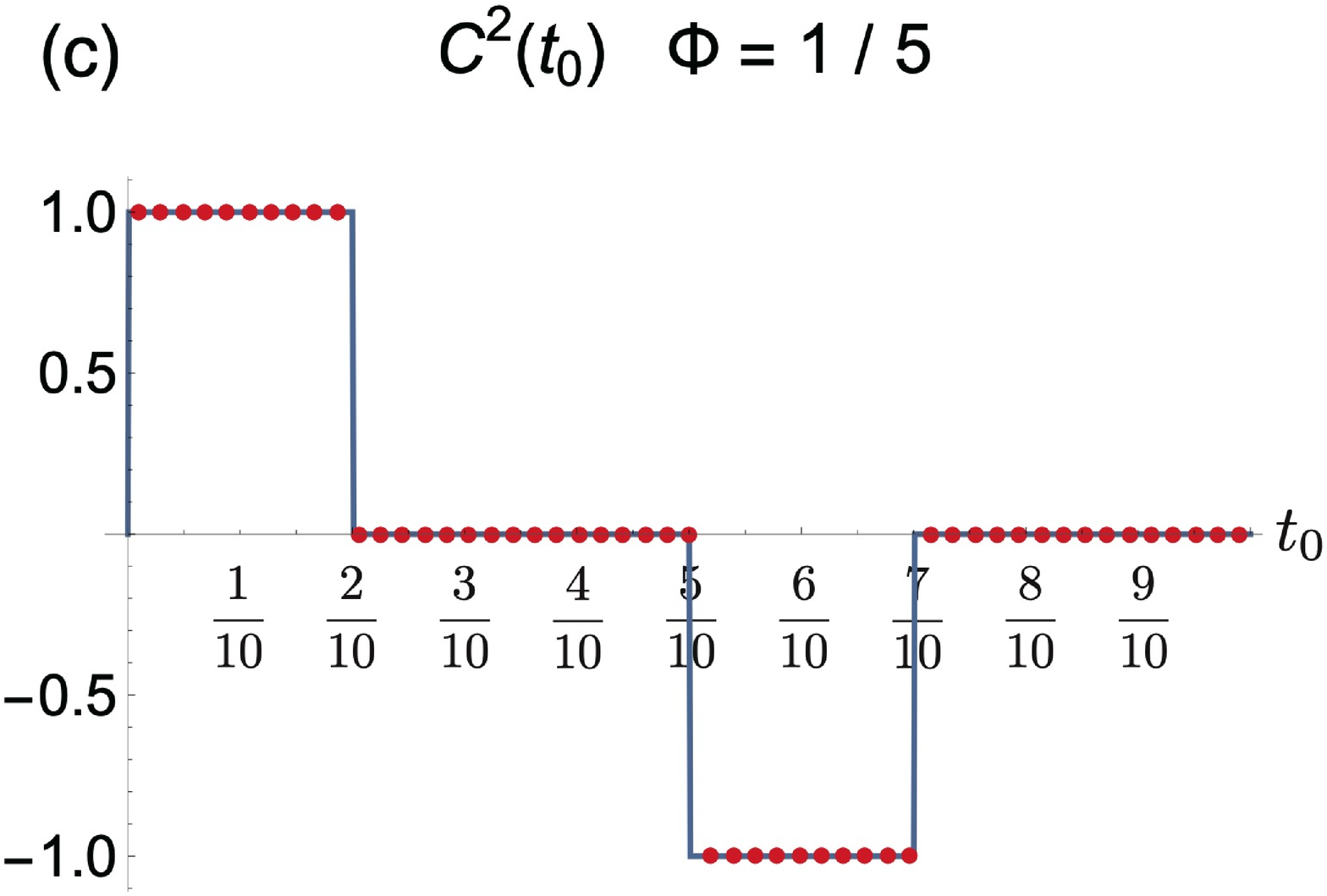}
 \end{minipage}
 \begin{minipage}[b]{0.48\linewidth}
  \centering
  \includegraphics[keepaspectratio, scale=0.24]{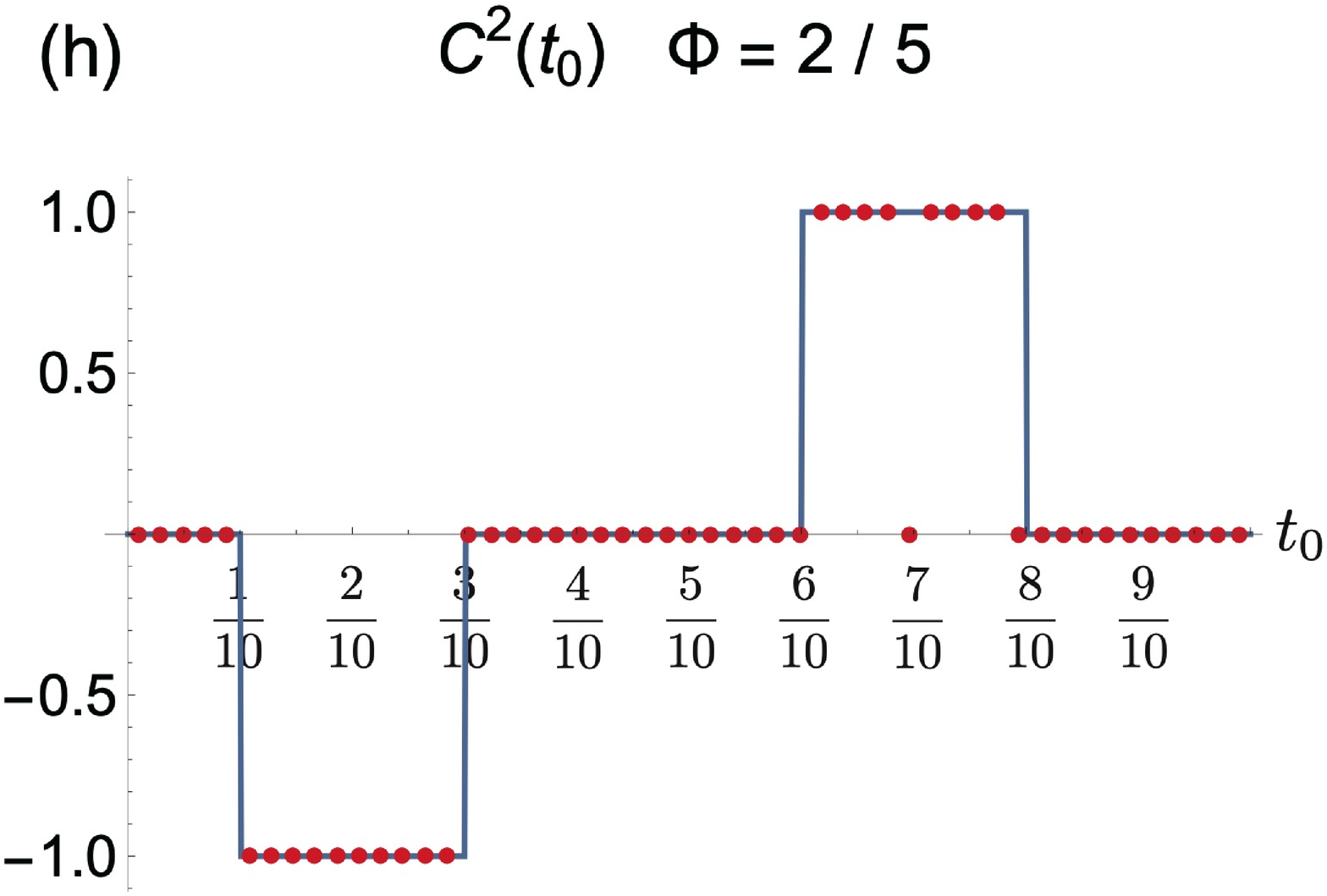}
 \end{minipage}
 \\
 \begin{minipage}[b]{0.48\linewidth}
  \centering
  \includegraphics[keepaspectratio, scale=0.24]{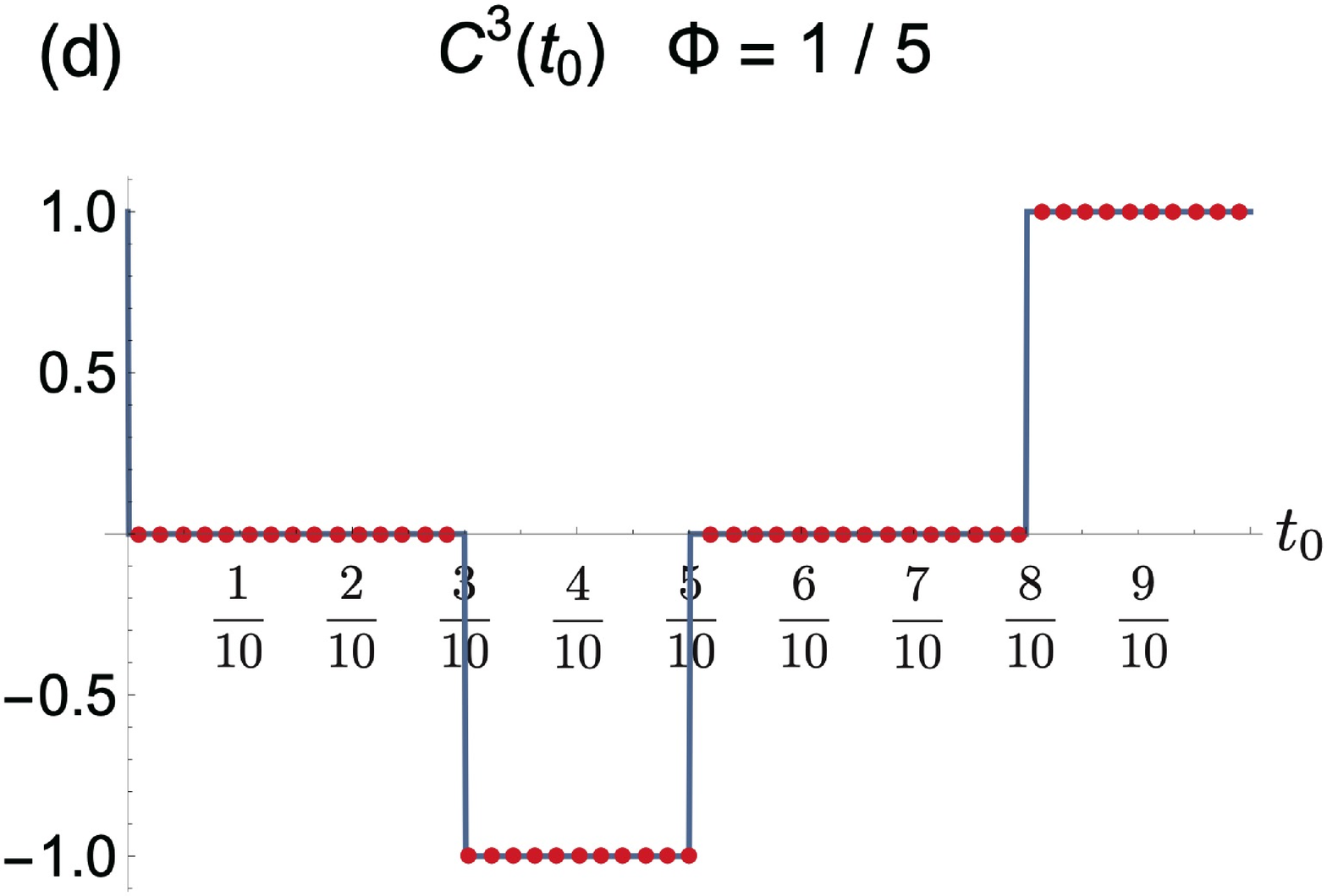}
 \end{minipage}
 \begin{minipage}[b]{0.48\linewidth}
  \centering
  \includegraphics[keepaspectratio, scale=0.24]{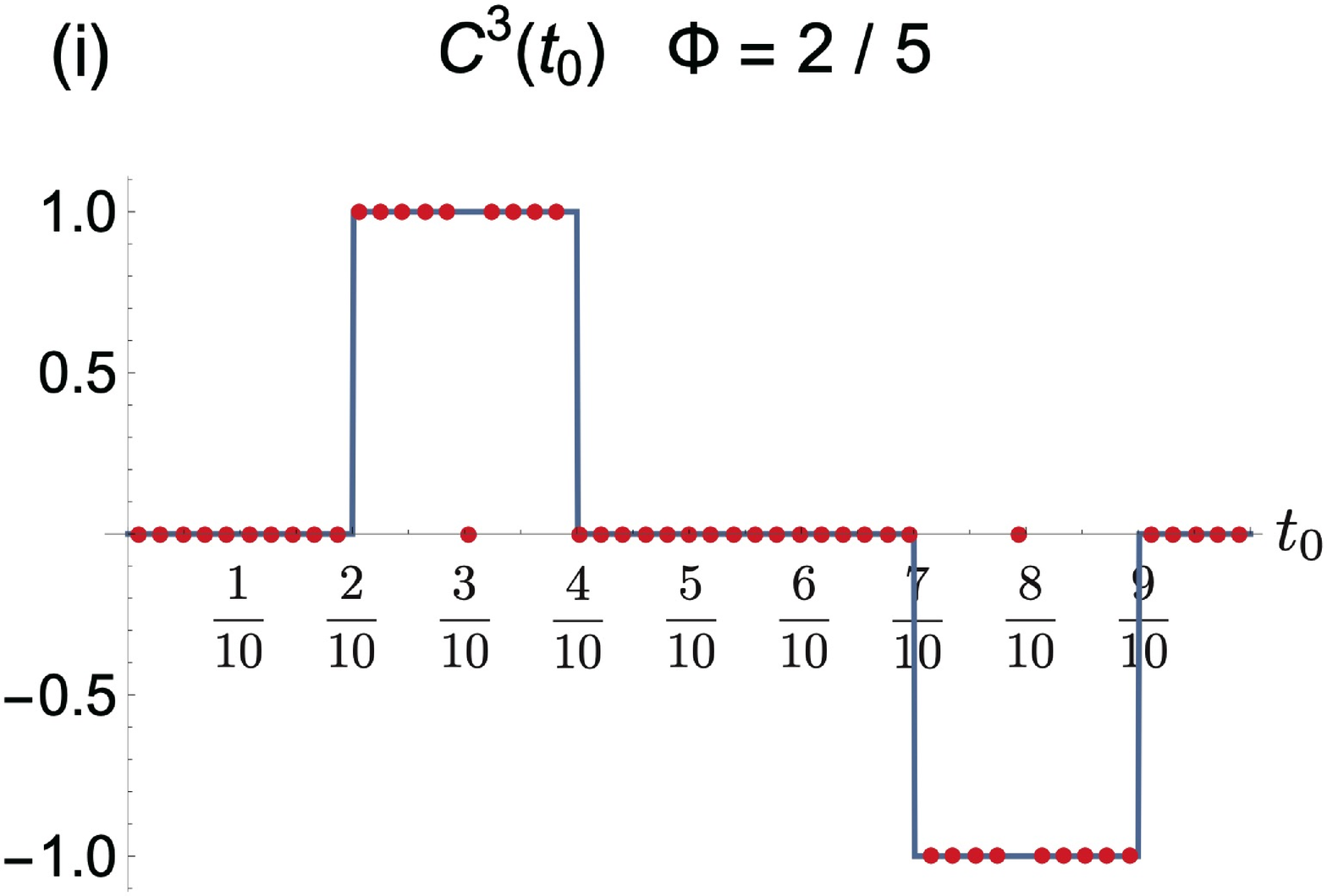}
 \end{minipage}
 \\
 \begin{minipage}[b]{0.48\linewidth}
  \centering
  \includegraphics[keepaspectratio, scale=0.24]{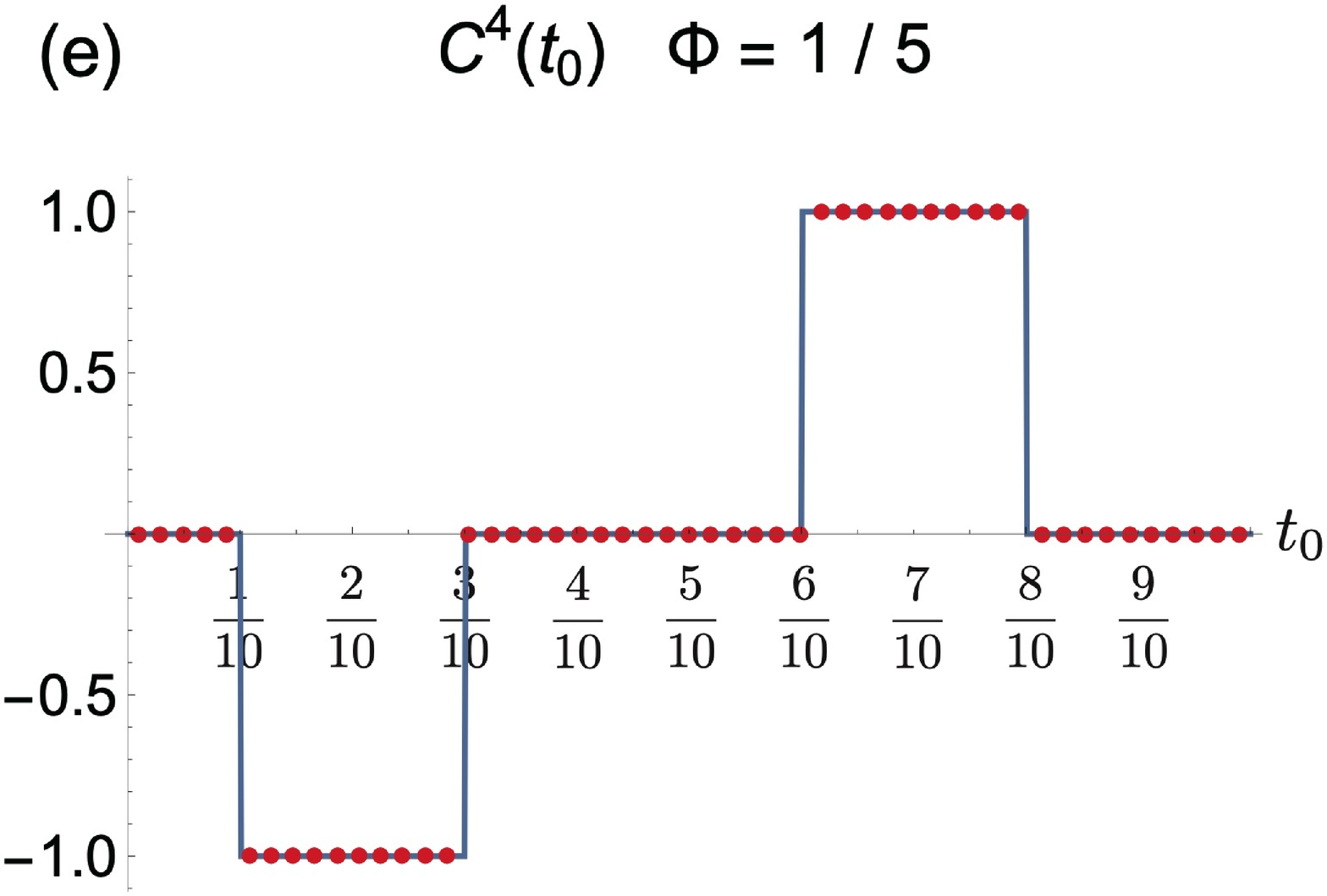}
 \end{minipage}
 \begin{minipage}[b]{0.48\linewidth}
  \centering
  \includegraphics[keepaspectratio, scale=0.24]{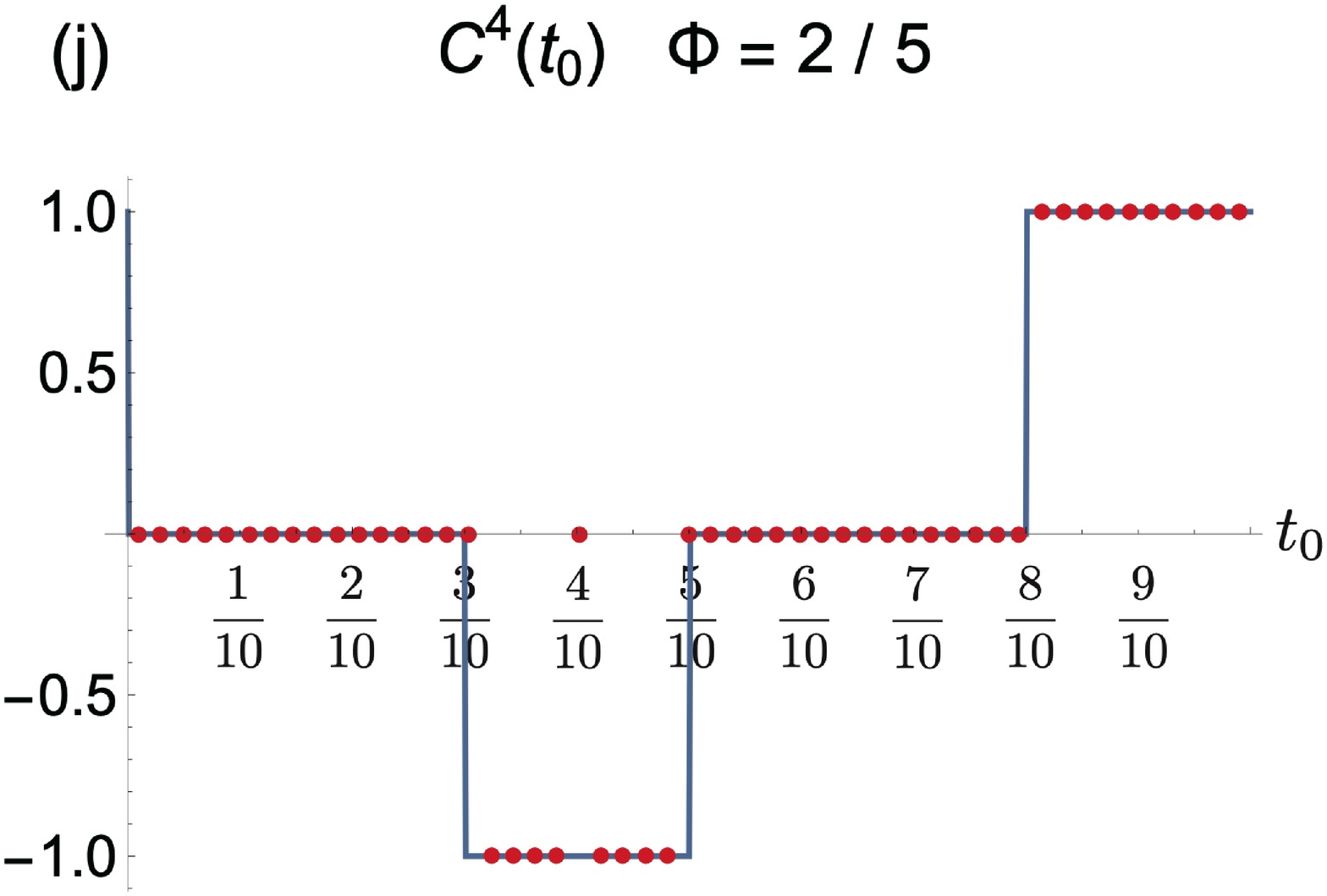}
 \end{minipage}
 \\
 \caption{\label{fig:ChernQ5}
  $t_0$ dependence of
  the Chern numbers. $\QN=5,\Phi=\frac {1}{5} $ and $\frac {2}{5} $.
  $J_0=-1.0, \delta =0.5$ and $\Delta =0.5$. The dimension of
  the Hilbert space is 545 ($L=6$). The data points are for
  $t_0=(i-0.5)/51, i=1,\cdots,51$.
  Red points are numerical evaluation of Eq.(\ref{eq:bec-qandC}) by
  Fukui-Hatsugai-Suzuki formula \cite{Fukui05}
  and the lines are analytical results, Eq.(\ref{eq:aChern}).
 }
 \end{figure}



 \begin{figure}[t]
  \captionsetup[sub]{skip=0pt}
 \begin{minipage}[b]{0.48\linewidth}
  \centering
  \includegraphics[keepaspectratio, scale=0.24]{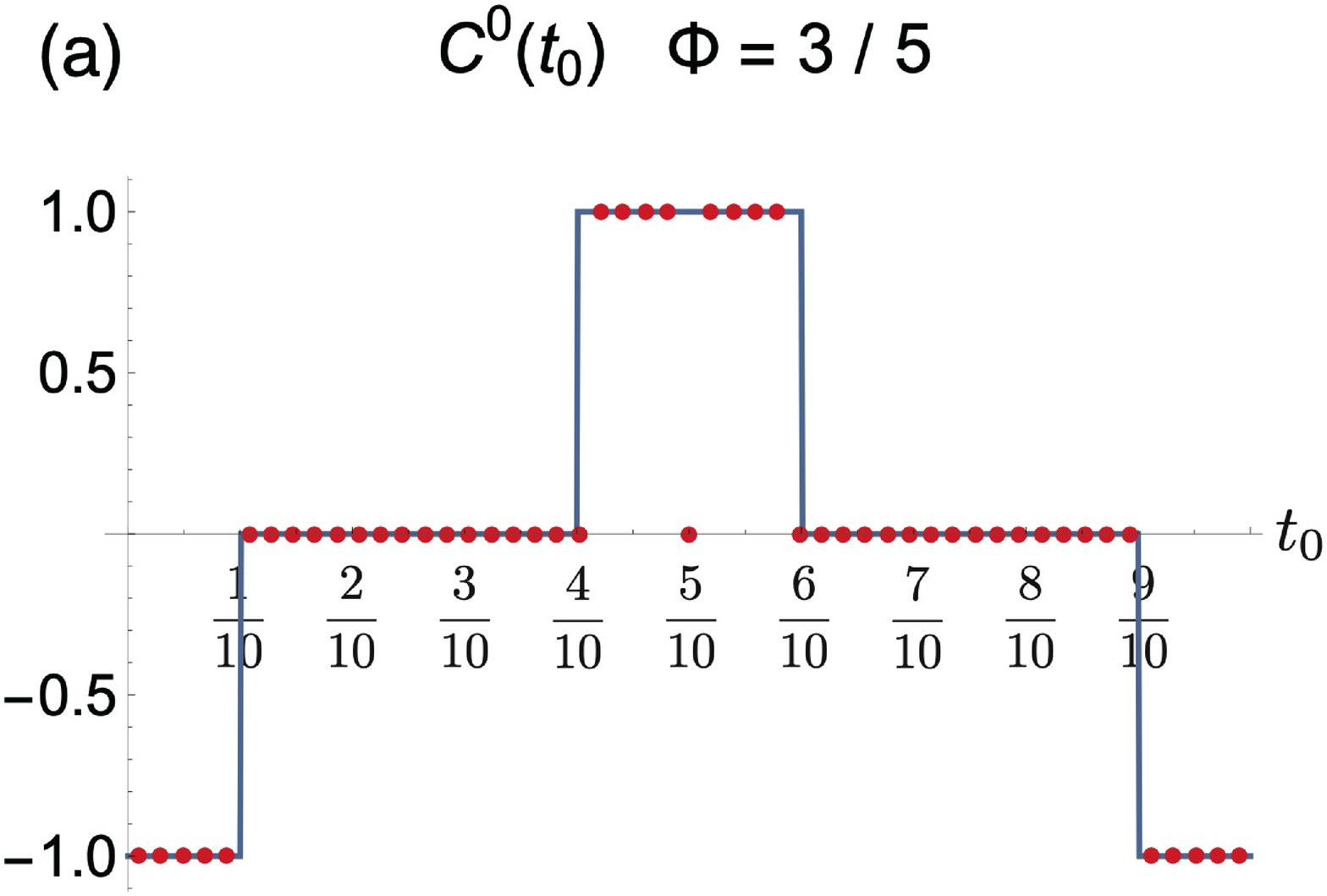}
 \end{minipage}
 \begin{minipage}[b]{0.48\linewidth}
  \centering
  \includegraphics[keepaspectratio, scale=0.24]{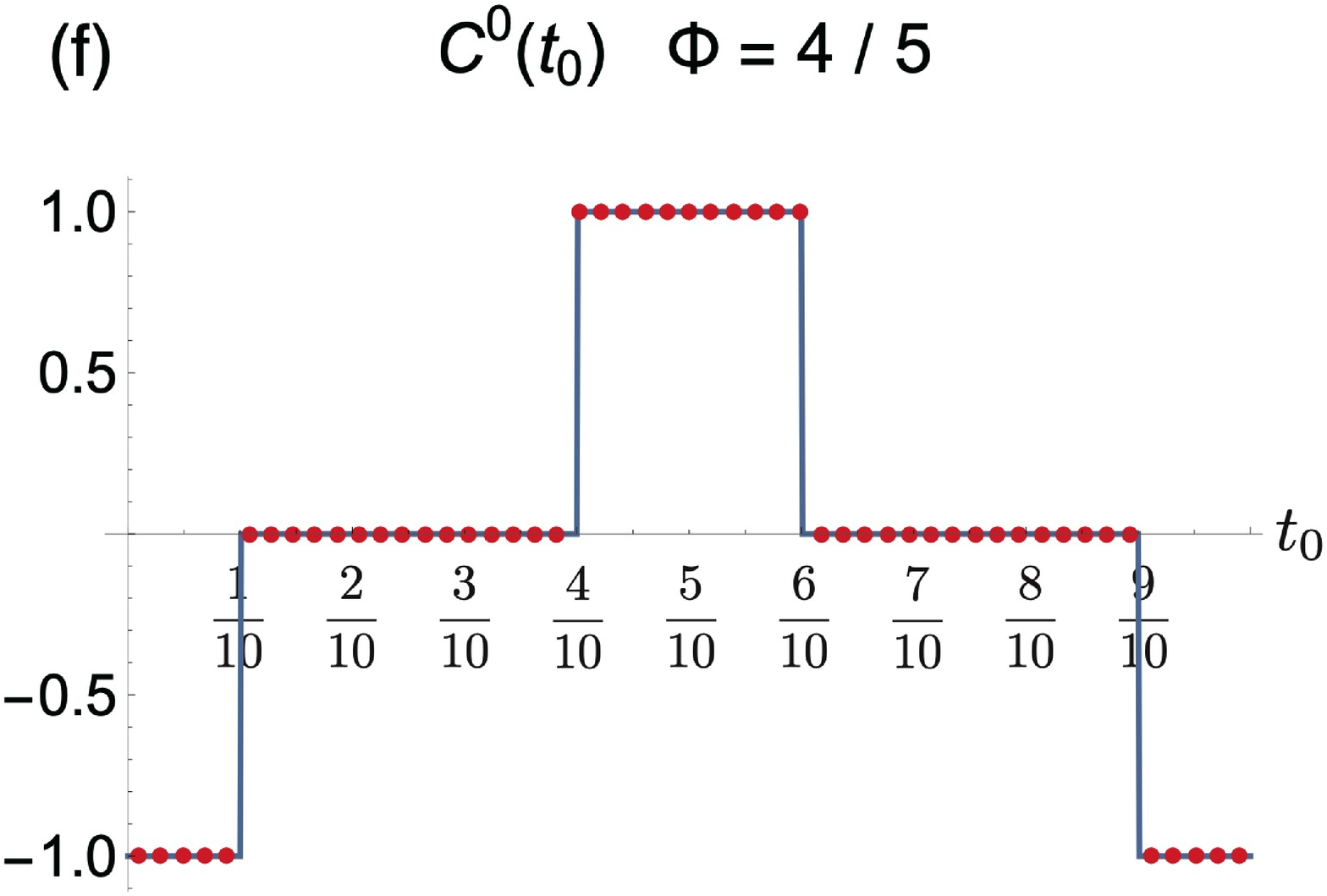}
 \end{minipage}
 \\
 \begin{minipage}[b]{0.48\linewidth}
  \centering
  \includegraphics[keepaspectratio, scale=0.24]{chern-fig-f/fig_24.eps}
 \end{minipage}
 \begin{minipage}[b]{0.48\linewidth}
  \centering
  \includegraphics[keepaspectratio, scale=0.24]{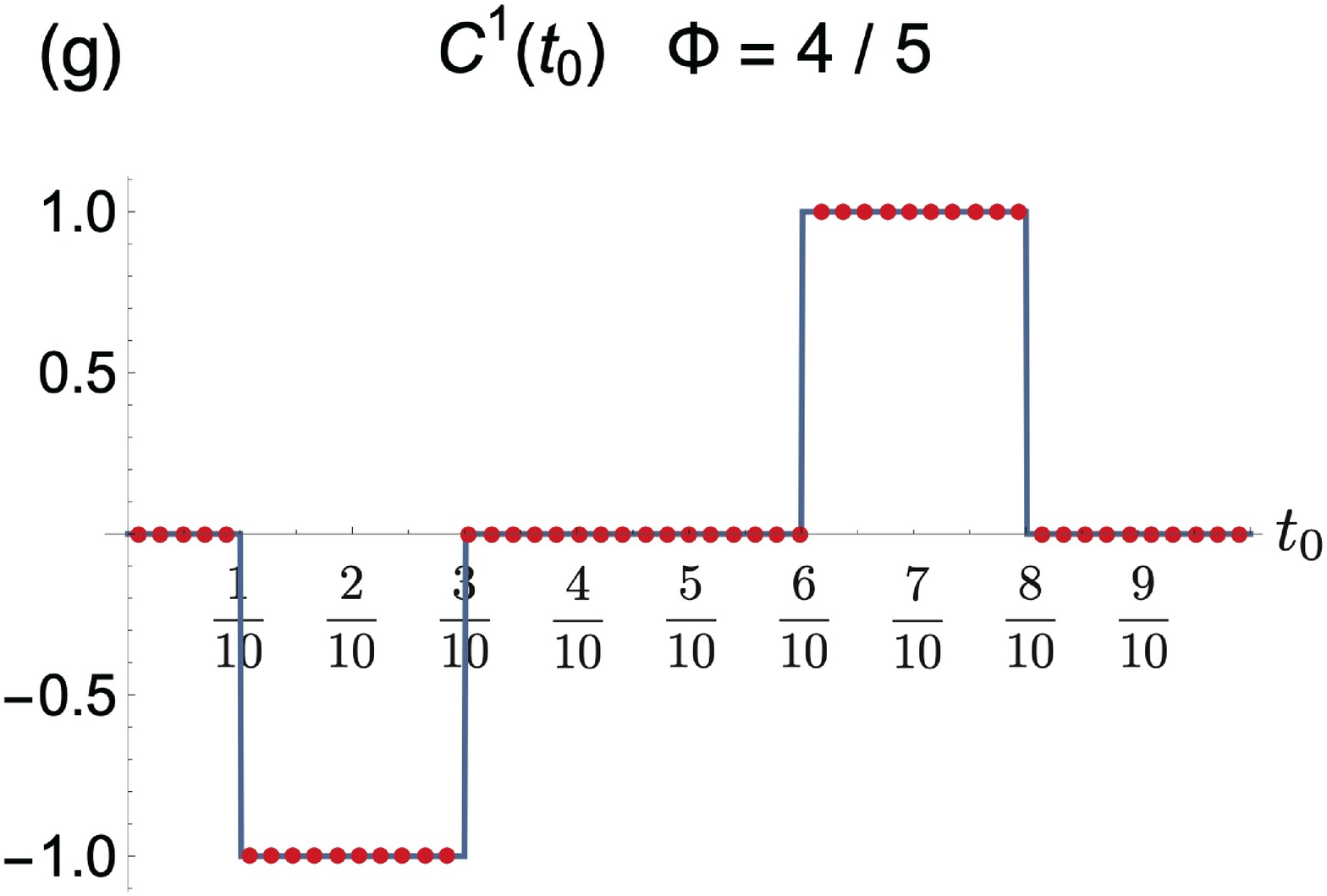}
 \end{minipage}
 \\
 \begin{minipage}[b]{0.48\linewidth}
  \centering
  \includegraphics[keepaspectratio, scale=0.24]{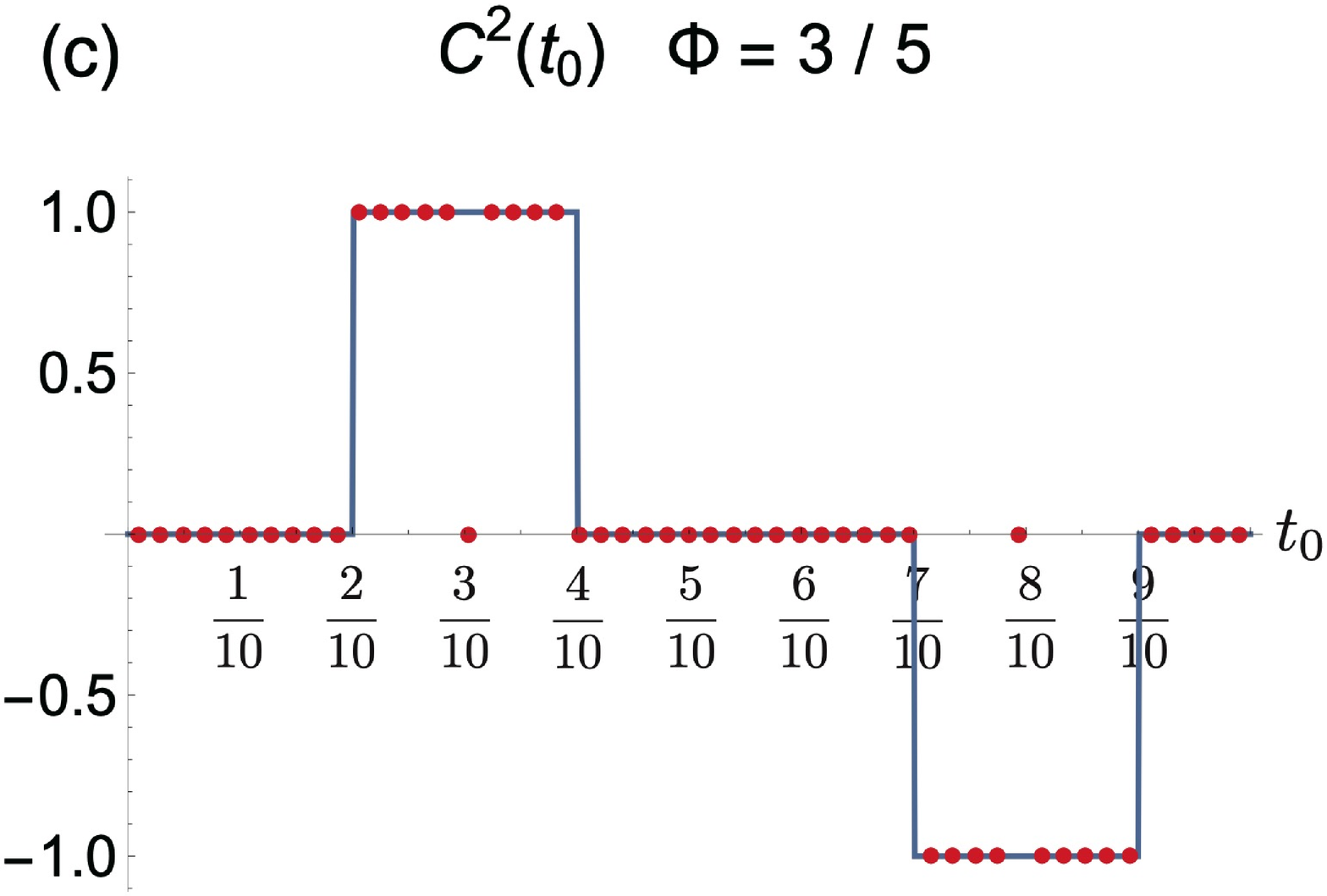}
 \end{minipage}
 \begin{minipage}[b]{0.48\linewidth}
  \centering
  \includegraphics[keepaspectratio, scale=0.24]{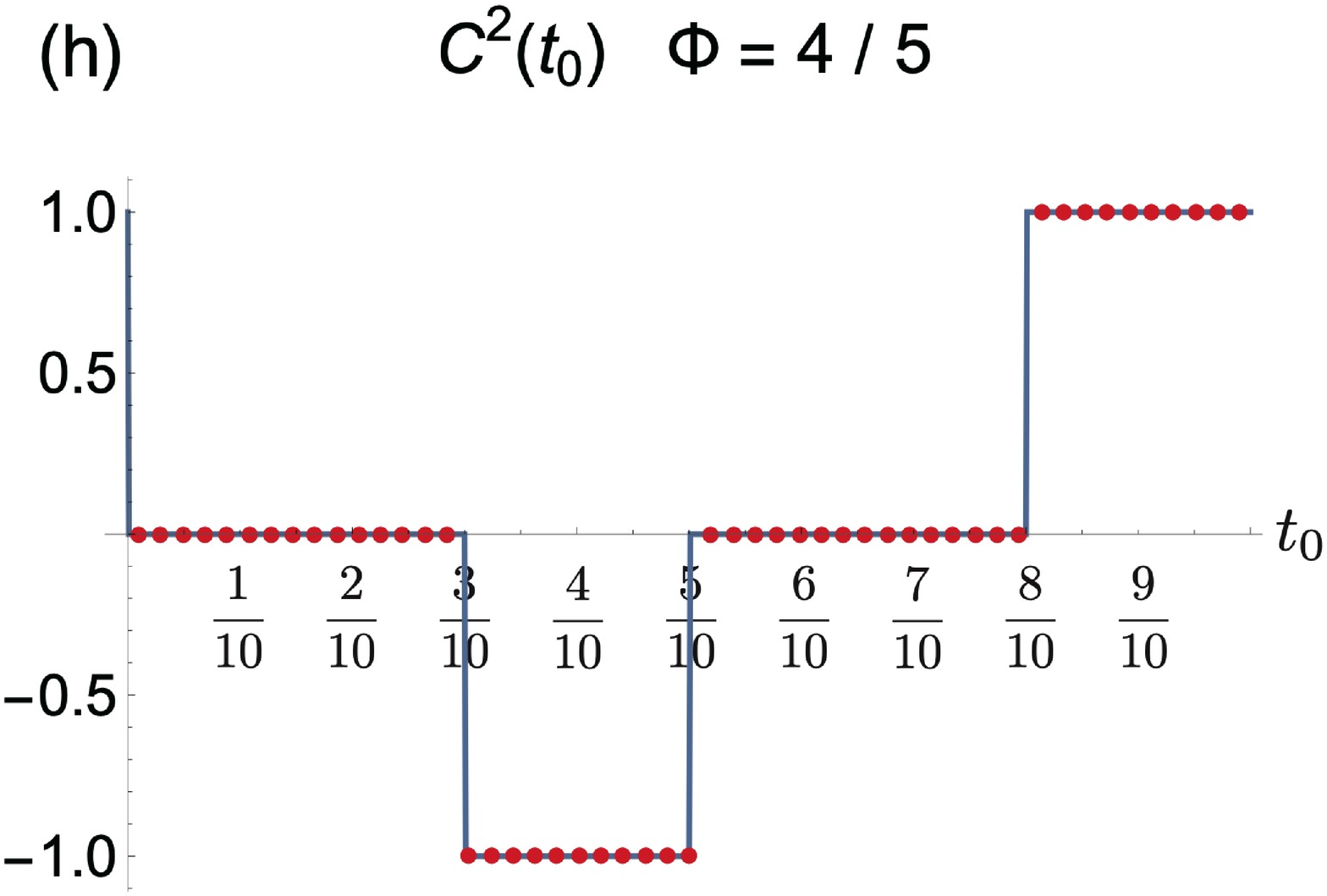}
 \end{minipage}
 \\
 \begin{minipage}[b]{0.48\linewidth}
  \centering
  \includegraphics[keepaspectratio, scale=0.24]{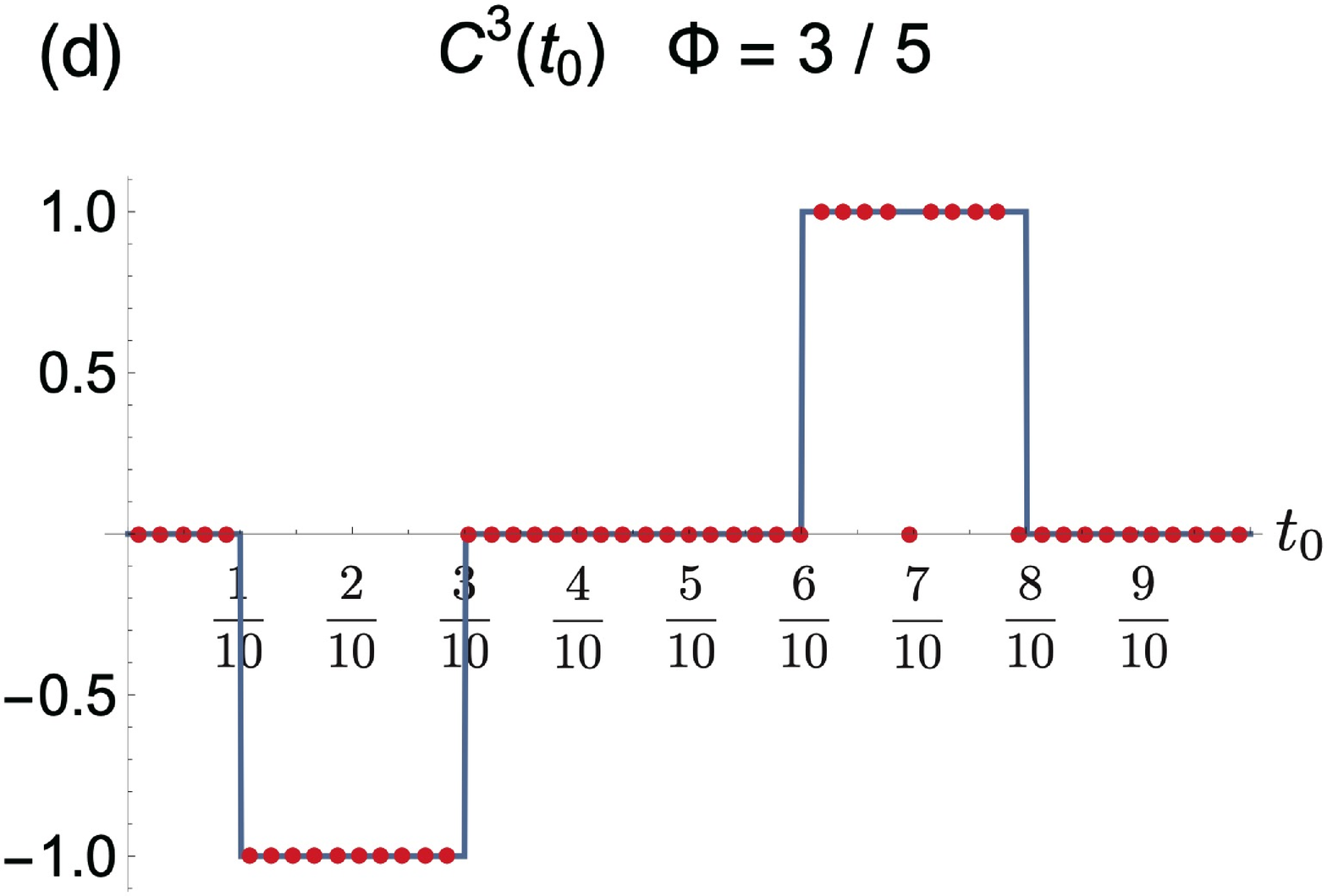}
 \end{minipage}
 \begin{minipage}[b]{0.48\linewidth}
  \centering
  \includegraphics[keepaspectratio, scale=0.24]{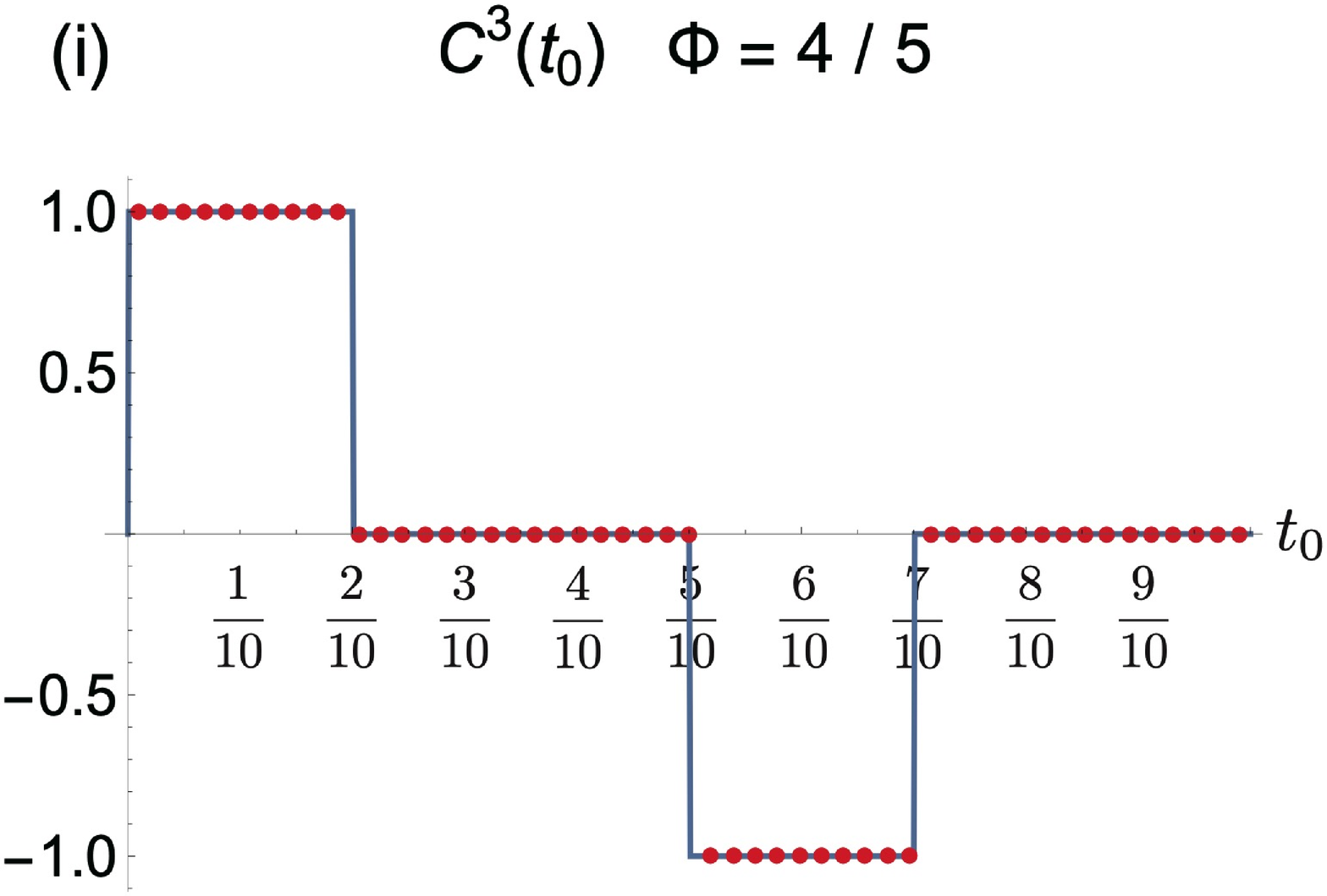}
 \end{minipage}
 \\
 \begin{minipage}[b]{0.48\linewidth}
  \centering
  \includegraphics[keepaspectratio, scale=0.24]{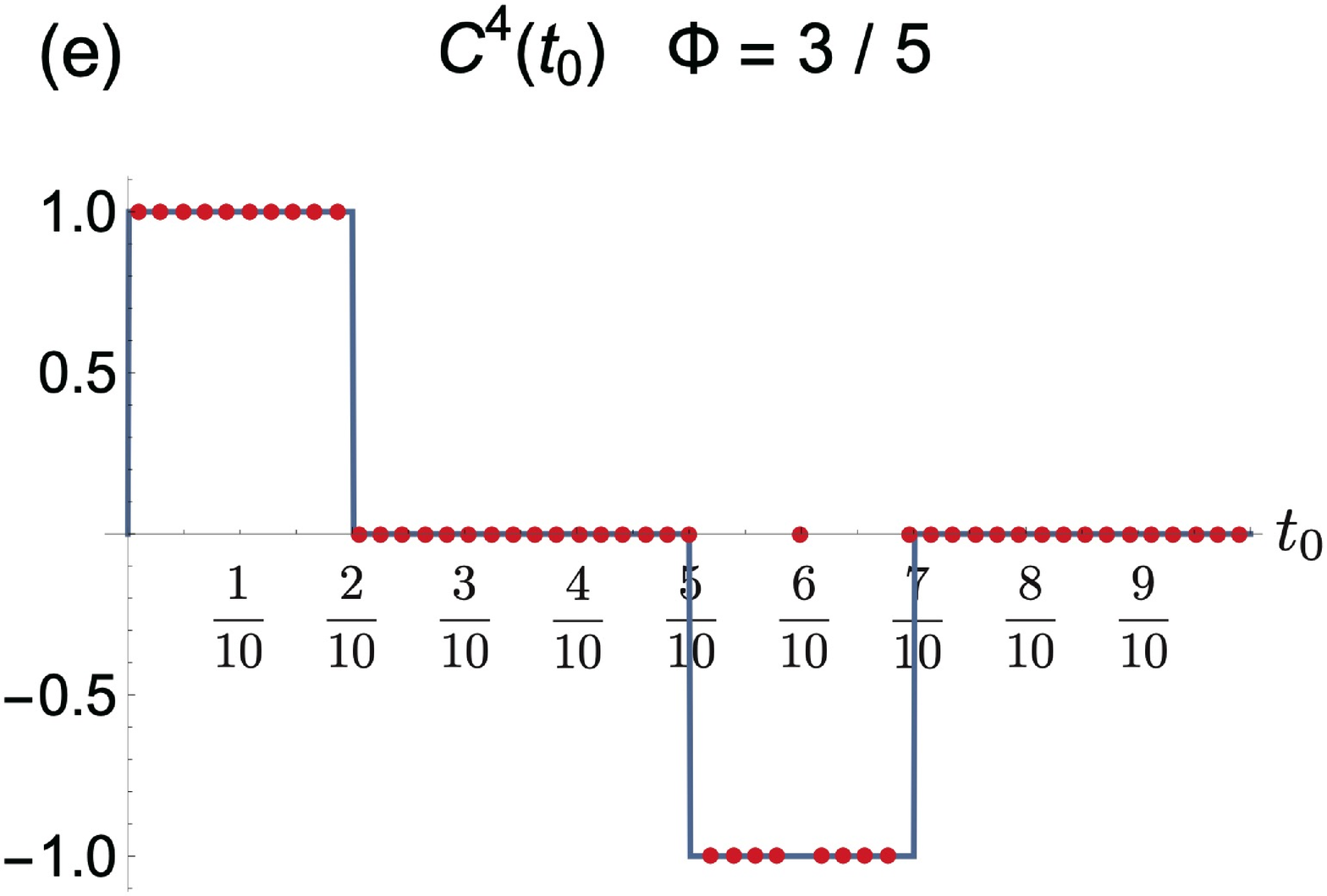}
 \end{minipage}
 \begin{minipage}[b]{0.48\linewidth}
  \centering
  \includegraphics[keepaspectratio, scale=0.24]{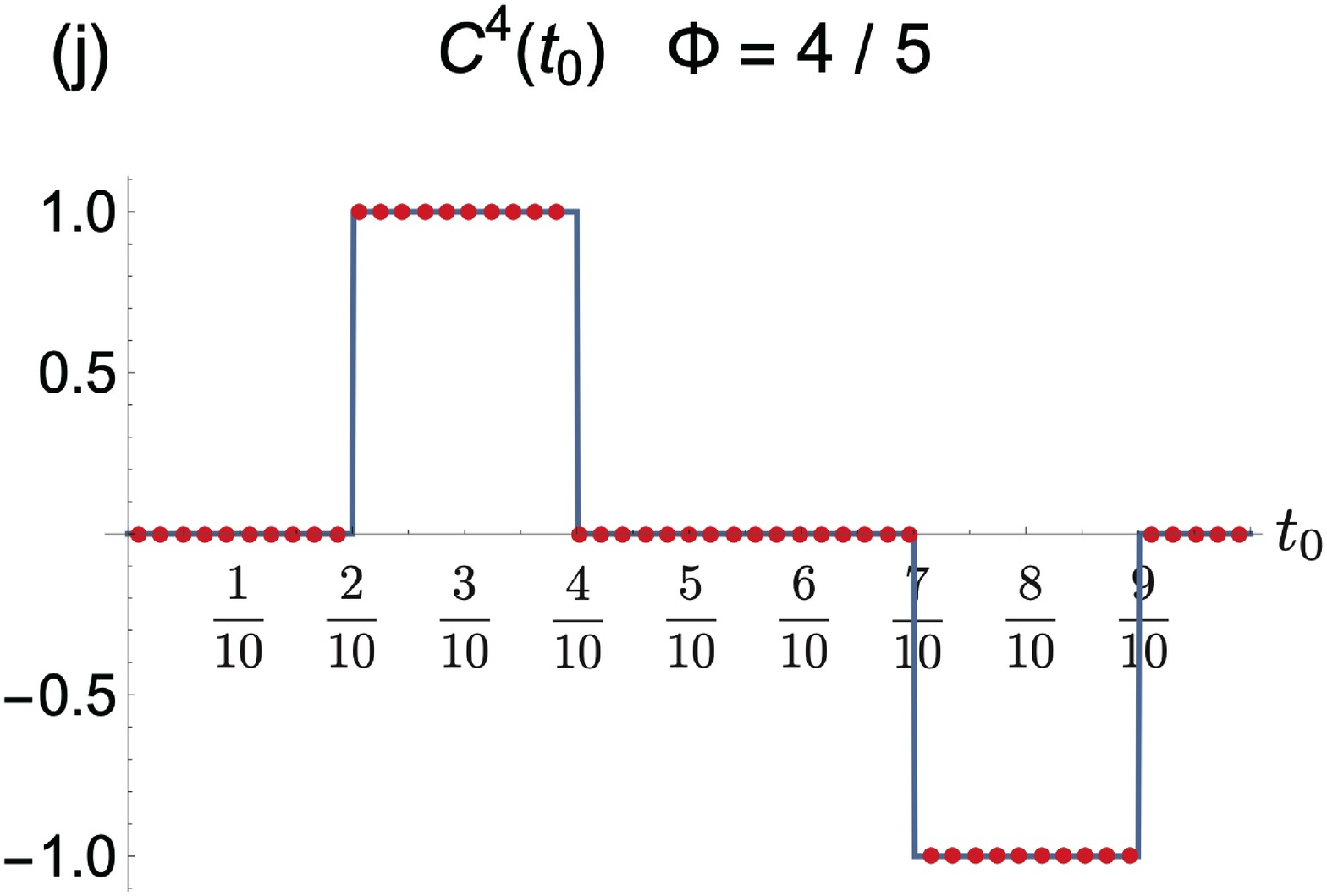}
 \end{minipage}
 \\
 \caption{\label{fig:ChernQ5-21}
  $t_0$ dependence of
  the Chern numbers. $\QN=5,\Phi=\frac {3}{5} $ and $\frac {4}{5} $
  $J_0=-1.0, \delta =0.5$ and $\Delta =0.5$. The dimension of
  the Hilbert space is 545 ($L=6$). The data points are for
  $t_0=(i-0.5)/51, i=1,\cdots,51$.
  Red points are numerical evaluation of Eq.(\ref{eq:bec-qandC}) by
  Fukui-Hatsugai-Suzuki formula \cite{Fukui05}
  and the lines are analytical results, Eq.(\ref{eq:aChern}).
  }
 \end{figure}

\end{widetext}



\section{Summary}
\label{sec:summary}
A topological pump
of the $\SUQ$ invariant
{quantum} chain  is proposed where the gauge invariance
of the colored fermions plays a central role.
Introducing a symmetry breaking perturbation,
$\QN$ Chern numbers, which are given by the integral over
the torus defined by the symmetric path in
the synthetic Brillouin zone and the time cycle,
characterize the bulk pump topologically.
As for an open boundary condition, the sums of the discontinuities in
$\QN$ different center of masses, which generate the large gauge transformation,
give topological numbers of the pump as well.
These discontinuities are topological numbers due to the edge states.
Relations among 
the open/periodic/twisted boundary conditions are
discussed in details,
that justifies the bulk-edge correspondence.
Using this bulk-edge correspondence,
an explicit analytic formula for the $\QN$ Chern numbers is
given associated with the Diophantine equation due to TKNN.
The low energy spectra and
the topological quantities are numerically evaluated, that justify
the consistency of the whole discussion.
}

\begin{acknowledgments}
 The work is supported in part by KAKENHI from JSPS
 Grants No.17H06138 (Y.H.),
 Grants No.21K13849 (Y.K.)
and CREST JPMJCR19T1 (Y.H.).
\end{acknowledgments}
  
  

\appendix

\section{Color description of $S=1$ spins \cite{AFFLECK1986409,AFFLECK1988582}}
\label{sec:color}

Let us first summarize a vector representation of $S=1$ angular momentum
in relation to the rotations in 3D as
\begin{align*}
 R^x(\alpha ) &=
 \matthree
   {1}{}{}
   {}{\cos \alpha }{- \sin \alpha }
   {}{\sin \alpha }{\cos \alpha }=1\oplus(\cos \alpha -i \sigma _y \sin \alpha )
   \\
   &= 1\oplus e^{-i \alpha \sigma _y}\equiv e^{-i \alpha T^1},
   \\
   T^1 &= 
-i \matthree
   {0}{}{}
   {}{0}{1}
   {}{-1}{0},\ T^1_{ij}=-i \epsilon_{1ij} .
\end{align*}
Similarly
\begin{align*}
 R_y(\beta) &= e^{-i \beta T^2},
 \\
 R_z(\gamma ) &= e^{-i \gamma T^3} ,
\end{align*}
 where
\begin{align*}
 T^a_{ij} &= -i \epsilon_{aij} .
\end{align*}
To summarize, they are explicitly defined by
\begin{align*}
   T^1 &= 
 \matthree
   {0}{0}{0}
   {0}{0}{-i}
   {0}{i}{0},
   T^2 = 
 \matthree
   {0}{0}{i}
   {0}{0}{0}
   {-i}{0}{0},
   T^3 = 
 \matthree
   {0}{-i}{0}
   {i}{0}{0}
   {0}{0}{0}.
\end{align*}

They are spins as
\begin{align*}
 ( \cmt{T^a}{T^b})_{ij} &= - \epsilon_{aik} \epsilon_{bkj} +\epsilon_{bik} \epsilon_{akj}
 = \epsilon_{aik} \epsilon_{bjk} -\epsilon_{bik} \epsilon_{ajk}
 \\
 &= \delta _{ab}\delta _{ij}-\delta _{aj}\delta _{ib}
 -\delta _{ba}\delta _{ij}
+\delta _{bj}\delta _{ai} 
 \\
 &=\delta _{ai}\delta _{bj}
 -\delta _{aj}\delta _{ib},
\end{align*}
\begin{align*}
 i \epsilon_{abc} T^c_{ij} &= \epsilon_{abc} \epsilon_{cij} =
 \epsilon_{abc} \epsilon_{ijc}
 \\
 &= \delta _{ai}\delta _{bj}-\delta _{aj}\delta _{bi}.
\end{align*}
Thus
\begin{align*}
 \cmt{T^a}{T^b} &= i \epsilon_{abc} T^c.
\end{align*}

Further
\begin{align*}
 ( T^a T^a)_{ij} &=  T^a_{ik} T^a_{kj}=-\epsilon_{aik} \epsilon_{akj} =\epsilon_{aki} \epsilon_{akj} =2 \delta _{ij},
 \\
 T^a T^a &= 2 E_3.
\end{align*}
This implies $S=1$.

Let us consider a bilinear-biquadratic Hamiltonian of $S=1$ quantum spin chain as
\begin{align*}
 H &= \sum_n \big[\cos \omega_S (\bm{S} _n\cdot \bm{S} _{n+1})
+ \sin \omega (\bm{S} _n\cdot \bm{S} _{n+1})^2\big].
\end{align*}

The spin 1 operators are written by color fermions, $c ^\dagger = (c_{+ 1} ^\dagger ,c_0 ^\dagger ,c_{- 1} ^\dagger )$, due to Affleck as
\begin{align*}
S^a &= c ^\dagger T^a c=c_\alpha ^\dagger T^a_{\alpha \beta }c_\beta ,\ \alpha , \beta =1,2,3
\end{align*}
with a constraint
\begin{align*}
 \sum_\alpha n_\alpha &= 1,\ n_\alpha =c_\alpha ^\dagger c_\alpha .
\end{align*}
Let us check here.

Using a useful relations
\footnote{ 
\begin{align*}
 T^a_{\alpha \beta } T^a_{\gamma \delta }
 &= -\epsilon_{a \alpha \beta } \epsilon_{a \gamma \delta }
 =-\delta_{\alpha \gamma }\delta _{\beta \delta }+\delta_{\alpha \delta }\delta _{\beta \gamma  }
 \\
 ( T^aT^b)_{\alpha \beta } ( T^aT^b)_{\gamma \delta }
 &=
 \epsilon_{a \alpha \kappa } \epsilon_{b \kappa \beta }
 \epsilon_{a \gamma \lambda } \epsilon_{\lambda \delta }
=
\epsilon_{a \alpha \kappa }
\epsilon_{a \gamma \lambda }
\epsilon_{b \kappa \beta }
\epsilon_{b\lambda \delta }
\\
&=
(\delta_ {\alpha \gamma }\delta_ {\kappa \lambda }-\delta _{\alpha \lambda }\delta _{\gamma \kappa })
(
\delta_{\kappa \lambda }\delta_{\beta \delta }
-\delta_ {\kappa \delta }\delta_ {\lambda \beta }
)
\\
&=
(\delta_ {\alpha \gamma }\delta_ {\kappa \lambda }-\delta _{\alpha \lambda }\delta _{\gamma \kappa })
\delta_{\kappa \lambda }\delta_{\beta \delta }
\\
& -
(\delta_ {\alpha \gamma }\delta_ {\kappa \lambda }-\delta _{\alpha \lambda }\delta _{\gamma \kappa })
\delta_ {\kappa \delta }\delta_ {\lambda \beta }
\\
&=
(\delta_ {\alpha \gamma }\delta_ {\lambda \lambda }-\delta _{\alpha \lambda }\delta _{\gamma \lambda })
\delta_{\beta \delta }
\\
& -
(\delta_ {\alpha \gamma }\delta_ {\delta \lambda }-\delta _{\alpha \lambda }\delta _{\gamma \delta })
\delta_ {\lambda \beta }
\\
&=
(3\delta_{\alpha \gamma }-\delta_{\alpha \gamma} )\delta_{\beta \delta }
\\
& -\delta_{\alpha \gamma}\delta_{\delta \beta}+\delta_{\alpha \beta}\delta_{\gamma \delta}
\\
&=
\delta_{\alpha \gamma }\delta_{\beta \delta }+\delta_{\alpha \beta}\delta_{\gamma \delta}
\end{align*}
}
\begin{align*}
 T^a_{\alpha \beta } T^a_{\gamma \delta }
&= \delta_{\alpha \delta }\delta _{\beta \gamma  }-\delta_{\alpha \gamma }\delta _{\beta \delta }
 \\
 ( T^aT^b)_{\alpha \beta } ( T^aT^b)_{\gamma \delta }
&= 
\delta_{\alpha \beta}\delta_{\gamma \delta}+\delta_{\alpha \gamma }\delta_{\beta \delta },
\end{align*}
and noting the constraint, we have
\begin{align*}
  \cmt{S^a}{S^b}
 &= 
  c_\alpha ^\dagger c_\beta c_\gamma ^\dagger c_\delta
  (
  {T^a}_{\alpha \beta }{T^b}_{\gamma \delta }
  -
  {T^b}_{\alpha \beta }{T^a}_{\gamma \delta }
  )
  \\
  &= 
  c_\alpha ^\dagger( \delta _{\beta \gamma }- c_\gamma ^\dagger c_\beta )
  c_\delta
  (
  {T^a}_{\alpha \beta }{T^b}_{\gamma \delta }
  -
  {T^b}_{\alpha \beta }{T^a}_{\gamma \delta })
  \\
  &= 
  c_\alpha ^\dagger
  c_\delta
  (
  {T^a}_{\alpha \beta }{T^b}_{\beta \delta }
  -
  {T^b}_{\alpha \beta }{T^a}_{\beta \delta })
  \\
  &= 
  c_\alpha ^\dagger
  c_\delta
(\cmt{T^a}{T^b})_{\alpha \delta }
=i \epsilon_{abc} S^c
\end{align*}
\begin{align*}
 \bm{S}^2 &= S^a S^a
=  c_\alpha ^\dagger c_\beta c_\gamma ^\dagger c_\delta
  {T^a}_{\alpha \beta}{T^a}_{\gamma \delta }
  \\
  &= 
  c_\alpha ^\dagger (\delta _{\beta \gamma }-c_\gamma ^\dagger c_\beta )
  c_\delta
  {T^a}_{\alpha \beta }{T^a}_{\gamma \delta  }
  \\
  &= 
  c_\alpha ^\dagger 
  c_\delta
  {T^a}_{\alpha \beta }{T^a}_{\beta \delta }=
  c_\alpha ^\dagger 
  c_\beta 
  ({T^a}^2)_{\alpha \beta }=2 \sum _\alpha n_\alpha =2
\end{align*}

As for the interaction between the sites $i,j$, it is written as
\begin{align*}
 \bm{S} _i\cdot \bm{S} _j
 &=
 c_{i, \alpha } ^\dagger
 T^a_{\alpha \beta }
 c_{i, \beta } 
 c_{j, \gamma } ^\dagger
 T^a_{\gamma \delta }
 c_{j, \delta  }
 \\
 &= 
 c_{i, \alpha } ^\dagger
 c_{i, \beta } 
 c_{j, \gamma } ^\dagger
 c_{j, \delta  }
 T^a_{\alpha \beta }
 T^a_{\gamma \delta }
 \\
 &= 
 c_{i, \alpha } ^\dagger
 c_{i, \beta } 
 c_{j, \gamma } ^\dagger
 c_{j, \delta  }
( \delta _{\alpha \delta }
 \delta _{\beta \gamma }
 -
\delta _{\alpha \gamma }
 \delta _{\beta \delta }
 )
 \\
&= 
 c_{i, \alpha } ^\dagger
 c_{i, \beta } 
 c_{j, \beta  } ^\dagger
 c_{j, \alpha  }
 -
 c_{i, \alpha } ^\dagger
 c_{i, \beta } 
 c_{j, \alpha  } ^\dagger
 c_{j, \beta  }
 \\
( \bm{S} _i\cdot \bm{S} _j)^2
 &=
( c_{i, \alpha } ^\dagger
 c_{i, \beta } 
 c_{j, \gamma } ^\dagger
 c_{j, \delta  }
 T^a_{\alpha \beta }
 T^a_{\gamma \delta }
 )
( c_{i,\alpha' } ^\dagger
 c_{i, \beta' } 
 c_{j, \gamma' } ^\dagger
 c_{j, \delta'  }
 T^b_{\alpha' \beta' }
 T^b_{\gamma' \delta' }
 )
 \\
 &= 
( c_{i, \alpha } ^\dagger
 c_{i, \beta } 
 c_{i,\alpha' } ^\dagger
 c_{i, \beta' } 
)(
 c_{j, \gamma } ^\dagger
 c_{j, \delta  }
 c_{j, \gamma' } ^\dagger
 c_{j, \delta'  }
 )
 T^a_{\alpha \beta }
 T^a_{\gamma \delta }
 T^b_{\alpha' \beta' }
 T^b_{\gamma' \delta' }
 \\
 &= 
 c_{i, \alpha } ^\dagger
(\delta _{\beta \alpha '}-
 c_{i,\alpha' } ^\dagger c_{i,\beta } )
 c_{i, \beta' } 
\cdot
 c_{j, \gamma } ^\dagger
(\delta_{\gamma ' \delta  }- c_{j \gamma' } ^\dagger
 c_{j, \delta  }
 )
 c_{j, \delta'  }
 \\
 &\qquad\qquad\qquad\qquad\qquad\qquad
 T^a_{\alpha \beta }
 T^a_{\gamma \delta }
 T^b_{\alpha' \beta' }
 T^b_{\gamma' \delta' }
 \\
 &= 
 c_{i, \alpha } ^\dagger
 c_{i, \beta' } 
\cdot
 c_{j, \gamma } ^\dagger
 c_{j, \delta'  }
\delta _{\beta \alpha '}
\delta_{\gamma ' \delta  }
 T^a_{\alpha \beta }
 T^a_{\gamma \delta }
 T^b_{\alpha' \beta' }
 T^b_{\gamma' \delta' }
 \\
 &= 
 c_{i, \alpha } ^\dagger
 c_{i, \beta' } 
 c_{j, \gamma } ^\dagger
 c_{j, \delta'  }
 T^a_{\alpha \beta }
 T^a_{\gamma \delta }
 T^b_{\beta \beta' }
 T^b_{\delta \delta' }
 \\
 &= 
 c_{i, \alpha } ^\dagger
 c_{i, \beta' } 
 c_{j, \gamma } ^\dagger
 c_{j, \delta'  }
 (T^a T^b)_{\alpha \beta '}
 ( T^a T^b)_{\gamma \delta' }
 \\
 &= 
 c_{i, \alpha } ^\dagger
 c_{i, \beta' } 
 c_{j, \gamma } ^\dagger
 c_{j, \delta'  }
 (\delta _{\alpha \beta '}\delta _{ \gamma \delta '}
 +
 \delta _{\alpha \gamma }\delta _{ \beta '\delta '})
 \\
 &= 
 c_{i, \alpha } ^\dagger
 c_{i, \alpha  } 
 c_{j, \gamma } ^\dagger
 c_{j, \gamma  }
 +
 c_{i, \alpha } ^\dagger
 c_{i, \beta } 
 c_{j, \alpha } ^\dagger
 c_{j, \beta  }
 \\
 &= 
 c_{i, \alpha } ^\dagger
 c_{i, \beta } 
 c_{j, \alpha } ^\dagger
 c_{j, \beta  } +1
\end{align*}

Note that the fermion number operators are written by the spin operators
as
\begin{align}
 n_{j, \alpha } &= 1-(S_j^\alpha )^2.
 \label{eq:fermion-spin}
\end{align}

Then omitting the constant, the Hamiltonian is given as
\begin{align*}
 H &= \sum_{i<j} 
 \cos \omega\,
 c_{i, \alpha } ^\dagger
 c_{i, \beta } 
 c_{j, \beta } ^\dagger
 c_{j, \alpha   }
 +
 (\sin \omega -\cos \omega )
 c_{i, \alpha } ^\dagger
 c_{i, \beta } 
 c_{j, \alpha } ^\dagger
 c_{j, \beta  }
 \\
 &= \cos \omega_S H^{(1)}(\{1\}) + (\sin \omega_S - \cos \omega_S ) H^{(2)}(\{1\})
\end{align*}
where (slightly extending the parameter space)
with $J^{1,2}_{\alpha \beta}=1  $ as
\begin{align*} 
 H^{(1)}(\{J^{(1)}_{i,\alpha ;j,\beta }\}) &=
 \sum_{i<j}
J^{(1)}_{i,\alpha j;j, \beta } c_{i \alpha } ^\dagger
 c_{i, \beta } 
 c_{j, \beta } ^\dagger
 c_{j, \alpha   }
 \\
 H^{(2)}
 ( \{J^{(2)}_{i,\alpha j;j, \beta } \})
 &=
 \sum_{i<j}
 J^{(2)}_{i,\alpha j;j, \beta }
 c_{i, \alpha } ^\dagger 
 c_{i, \beta }  
 c_{j, \alpha } ^\dagger
 c_{j, \beta  }
\end{align*}
Hermiticity implies
\begin{align*}
[H^{(1)}(J^{(1)}_{i,\alpha;j, \beta }) ] ^\dagger &= \sum_{i<j}
(J^{(1)}_{i,\alpha;j \beta })^*
 c_{j, \alpha   }^\dagger
 c_{j, \beta } 
 c_{i, \beta } ^\dagger
 c_{i, \alpha }
 \\
 &= \sum_{i<j}
(J^{(1)}_{i,\beta;j, \alpha })^*
 c_{i, \alpha   }^\dagger
 c_{i, \beta } 
 c_{j, \beta } ^\dagger
 c_{j, \alpha } \\
 J^{(1)}_{i,\alpha;j, \beta } &=
 J^{(1)}_{i,\beta j, \alpha }
\\
 [ H^{(2)}(\{
 J^{(2)}_{i,\alpha;j, \beta } 
  \}) ]^\dagger &=
 \sum_{i<j}
(J^{(2)}_{i,\alpha;j \beta })^*
 c_{j, \beta  }^\dagger
 c_{j, \alpha } 
 c_{i, \beta }^\dagger
 c_{i, \alpha }
 \\
 &= 
 \sum_{i<j}
 (J^{(2)}_{i,\beta;j, \alpha })^*
 c_{i, \alpha   }^\dagger
 c_{i, \beta } 
 c_{j, \alpha }^\dagger
 c_{j, \beta }
 \\
 J^{(2)}_{i,\alpha;j, \beta } &=
 J^{(2)}_{i,\beta j, \alpha }
\end{align*}
It is written as a matrix form
$({\bm{J}^{(1,2)}_{ij}}) ^\dagger = \bm{J}^{(1,2)}_{ij} $
where
$\big(\bm{J}^{(1,2)}_{ij} \big)_{\alpha, \beta }
=
J^{(1,2)}_{i,\alpha ;j,\beta }$.

\section{$SU(2)$ case 
\label{sec:su2}}

For the $N=2$ case,
let us perform a particle-hole transformation for the odd site $j$
as
\begin{align*}
 {\cal U}_\Theta &= \prod^\leftarrow_{j:\text{odd}} \xi_j=\cdots\xi_5\xi_3\xi_1
 , \ \   {\cal U}_\Theta ^{-1} = \prod^\rightarrow_{j:\text{odd}} \xi_j=\xi_1\xi_3\xi_5\cdots,
\end{align*}
\begin{align*}
 {\cal U}_\Theta  c_{j,\alpha } {\cal U}_ \Theta ^{-1} &= \mchss{c_{j,\alpha }}{j:\text{even}}{c_{j,\alpha } ^\dagger }{j:\text{odd}}.
\end{align*}
where $ \xi_j = c_j+c_j ^\dagger $ and 
$\xi_j^2=1$.
The Hamiltonian $H^{(2)}$ ($J^{2}_{j,\alpha \beta }=J_{j}^{(2)}$) is transformed as
\begin{align*}
 {\cal U}_\Theta  H^{(2)} (\{J_j^{(2)}\}) {\cal U} _ \Theta  
 &=
 \sum_{j:\text{odd},\alpha \beta }
 J_{j }^{(2)}
 c_{j, \alpha } 
 c_{j+1, \alpha } ^\dagger
 c_{j+1, \beta  }
 c_{j, \beta }^\dagger
 \\
 &+
 \sum_{j:\text{even},\alpha \beta }
 J_{j }^{(2)}
 c_{j, \alpha } ^\dagger
 c_{j+1, \alpha } 
 c_{j+1, \beta  }^\dagger
 c_{j, \beta }
 \\
 &=
 \sum_{j:\text{odd},\alpha \beta }
 J_{j }^{(2)}(
 \delta _{\alpha \beta }
 c_{j+1, \alpha } ^\dagger
 c_{j+1, \beta  }
\\&\ \ 
 - 
 c_{j, \beta }^\dagger
 c_{j+1, \alpha } ^\dagger
 c_{j+1, \beta  }
 c_{j, \alpha } )
 \\
 &
- \sum_{j:\text{even},\alpha \beta }
 J_{j }^{(2)}
 c_{j, \alpha } ^\dagger
 c_{j+1, \beta  }^\dagger
 c_{j+1, \alpha } 
 c_{j, \beta }
 \\
 &= -H^{(1)}(\{J_j^{(2)}\})+const.
\end{align*}

Note that $H^{(1)}$ is the $SU(2)$ Heisenberg model as confirmed 
by writing $c_{j, \alpha }=a_\alpha $ and $c_{j+1, \alpha }=b_\alpha $
\begin{align*}
 \sum_{\alpha \beta }& a_\alpha ^\dagger a_\beta b_\beta ^\dagger b_\alpha
 =
 a_{\uparrow} ^\dagger a_{\uparrow} b_{\uparrow} ^\dagger b_{\uparrow} 
 +
a_{\uparrow} ^\dagger a_{\downarrow} b_{\downarrow} ^\dagger b_{\uparrow} 
+
a_{\downarrow} ^\dagger a_{\uparrow} b_{\uparrow} ^\dagger b_{\downarrow} 
 +
 a_{\downarrow} ^\dagger a_{\downarrow} b_{\downarrow} ^\dagger b_{\downarrow}
 \\
&=  n^a_{\uparrow}  n^b_{\uparrow}  +
 n^a_{\downarrow}  n^b_{\downarrow}   +S^a_+S^b_-+S^a_-S^b_+
 = 2 \bm{S} ^a\cdot \bm{S} ^b+\frac 1 2 
\end{align*}

where $ 4S^a_zS^b_z
=(n^a_{\uparrow} -n^a_{\downarrow} )(n^b_{\uparrow} -n^b_{\downarrow} )
=
n^a_{\uparrow}n^b_{\uparrow}+n^a_{\downarrow}n^b_{\downarrow}
-n^a_{\uparrow}n^b_{\downarrow}-n^a_{\downarrow}n^b_{\uparrow}
=2(n^a_{\uparrow}n^b_{\uparrow}+n^a_{\downarrow}n^b_{\downarrow})-(n^a_{\uparrow} +n^a_{\downarrow} )(n^a_{\uparrow} +n^b_{\downarrow} )
=2(n^a_{\uparrow}n^b_{\uparrow}+n^a_{\downarrow}n^b_{\downarrow})-1
$ due to the constraint.

\section{Adiabatic approximation and the current}
\label{sec:adia}

{
Now let us here summarize a derivation of the current 
$ j = \langle G| \hat J | G \rangle $
in the adiabatic approximation \cite{Thouless83}.
\begin{eqnarray*}
 \hat J &=& 
 \hbar ^{-1} \partial_\theta H(\theta ).
\end{eqnarray*}
The many body state $|G \rangle $ is adiabatically
evolved from the snap shot ground state of a time dependent $H(t)$ as
\begin{eqnarray*}
 \mi \hbar | \dot{G}(t) \rangle &=& H(t)|G(t) \rangle,\quad 
 | G(0) \rangle = | g \rangle,
\end{eqnarray*}
where $|\alpha \rangle = |\alpha (t) \rangle $ is a orthonormalized eigen states of the
snap shot Hamiltonian and $|g \rangle$ 
is its ground state as
\begin{eqnarray*}
 H(t) | \alpha (t)\rangle &=& E_\alpha(t) |\alpha (t)\rangle,\quad
\langle \alpha | \beta \rangle = \delta _{\alpha \beta }.
\end{eqnarray*}
Writing as
$ |G \rangle = e^{-(\mi/ \hbar ) \int_0^t d t ^\prime \,  E_g (t ^\prime )} 
\sum_\alpha  |\alpha \rangle c_\alpha
$,
$c_g(0)=1$, $ c_\alpha (0) =0$, ($\alpha \ne g$),
the Schrodinger equation is written as
\begin{eqnarray*}
E_g \sum_\alpha | \alpha \rangle c_\alpha + \mi \hbar \sum_\alpha( | \alpha \rangle \dot c_\alpha +| \partial _t\alpha \rangle c_\alpha ) &=& \sum_\alpha
 E_\alpha | \alpha \rangle c_\alpha.
\end{eqnarray*}
Then multiplying $\langle g |$,
and noting that $|c_\alpha |\ll |c_g|$, $\alpha\ne g$, it reduces to
$\dot{c}_g  + c_g\langle g| \partial _t g \rangle \approx 0
$, that implies
\begin{eqnarray*}
 c_g &=& e^{\mi \gamma (t)},\quad
 \gamma (t) =
\mi \int_0^t d t ^\prime \, \langle g|\partial _t g \rangle .
\end{eqnarray*}
Also 
multiplying $\langle \alpha |$, ($\alpha \ne g$), one obtains
$
 E_g c_\alpha + \mi \hbar
\langle \alpha | \partial _t g \rangle c_g \approx 
E_\alpha c_\alpha
$, that
 implies
$
c_\alpha = \mi \hbar \frac {\langle \alpha | \partial _t g \rangle c_g }{E_\alpha -E_g}
$.
Now the time dependent ground state is given as
\begin{eqnarray*}
|G \rangle &=& 
e^{-(\mi /\hbar )\int_0^td t ^\prime E_g (t ^\prime ) }e^{\mi \gamma (t)}
\bigg[|g \rangle +
\mi \hbar
\sum_{\alpha\ne g} \frac {| \alpha \rangle \langle \alpha | \partial _t g \rangle }{E_\alpha -E_g}\bigg].
\end{eqnarray*}

Then the expectation value of $ \partial_\theta H $ is written as
\begin{eqnarray*}
 &&\langle G | \partial _\theta H |G \rangle
 =\langle g| \partial _\theta H |g \rangle
\\&& +\mi
 \hbar 
 \sum_{\alpha\ne g} \frac
   {\langle g|\partial _\theta H| \alpha \rangle \langle \alpha | \partial _t g \rangle
    -
 \langle \partial _t g | \alpha \rangle  \langle \alpha |
    \partial _\theta H
    | g \rangle 
   }
   {E_\alpha -E_g} .
\end{eqnarray*} 

Here let us remind a general relation
$
 \langle g | \partial _\theta H | \alpha \rangle =
 (E_\alpha -E_g )\langle g | \partial_\theta \alpha \rangle
$, that obeys from taking a derivative of the eigen equation
 $ H |\alpha \rangle = E_\alpha | \alpha \rangle $ as
 $
 \langle  \beta | \partial H | \alpha \rangle
 + E_\beta \langle \beta |\partial \alpha \rangle = 
 \partial E_\alpha \langle \beta | \alpha \rangle + E_\alpha \langle \beta | \partial \alpha \rangle
$.
\begin{eqnarray*}
 \delta j_x &=&\langle G | J |G \rangle- \langle g | J|g  \rangle,
 \\
&=& \mi
 \sum_{\alpha\ne g}
  \big( \langle g|\partial _\theta \alpha \rangle \langle \alpha | \partial _t g \rangle
    +
    \langle \partial_t g | \alpha \rangle  \langle \alpha |\partial _\theta g \rangle
    \big)
\\
&=& -
\mi
 \sum_{\alpha\ne g}
  \big( \langle \partial _\theta g| \alpha \rangle \langle \alpha | \partial _t g \rangle
    -
    \langle \partial _t g | \alpha \rangle  \langle \alpha |\partial _\theta g \rangle
    \big)
\\
&=& -\mi
 \sum_{\alpha}
  \big( \langle \partial _\theta g| \alpha \rangle \langle \alpha | \partial _t g \rangle
    -
    \langle \partial _t g | \alpha \rangle  \langle \alpha |\partial _\theta g \rangle
    \big)
\\
&=& -\mi
  \big( \langle \partial _\theta g| \partial _t g \rangle
    -
    \langle \partial _t g |\partial _\theta g \rangle
    \big)
    =-\mi B,
\end{eqnarray*}
where the field strength $B$ and the Berry connection $A_\mu $ is defined as
\begin{eqnarray*}
  B &=& \partial _\theta A_t   -\partial _t A_\theta ,\
    A_\mu= \langle g|\partial _\mu g \rangle ,\quad \mu =\theta,t.
\end{eqnarray*}
To summarize, in the adiabatic approximation,
we have
\begin{align*}
|G \rangle &= 
C \bigg[|g \rangle +
\mi \hbar
\sum_{\alpha\ne g} \frac {| \alpha \rangle \langle \alpha | \partial _t g \rangle }{E_\alpha -E_g}\bigg],
\\
 \langle G |\hat J| G \rangle &= 
 \langle g |\hat J| g \rangle -i B,
\end{align*}
where $C = e^{-(\mi /\hbar )\int_0^td t ^\prime E_g (t ^\prime ) }e^{\mi \gamma (t)}$.
 
\section{Berry phase and gauge fixing\cite {Hatsugai04gaugefix,Hatsugai07char,Hatsugai06qb,Hatsugai10}}
\label{sec:Berry-gauge}
Let us start
a $D$-dimensional Euclidean space
$x=(x_1,x_2,\cdots,x_D)\in \mathbb{R}^D$ as
a parameter space of 
the Hamiltonian $H(x)$.
As for the Berry phase associated with a loop $\ell$,
we further assume that
its ground state $|g(x) \rangle $,
\begin{align*}
 H|g \rangle &= |g \rangle E_g,\ \langle g|g \rangle =1,
\end{align*}
is gapped along the path $\ell$ 
\begin{align*}
  H(x) | n(x) \rangle &= | n(x) \rangle E_n(x),\ E_n(x)>E_g(x),\ n\ne g,\
 ^\forall \!x\in \ell.
\end{align*}
Note that the phase of the snapshot eigenstate is arbitrary
\begin{align*}
 | g \rangle &=  | g ^\prime  \rangle e^{i \Theta },\ \Theta \in\mathbb{R}
\end{align*} 
where $ H|g ^\prime  \rangle = |g ^\prime \rangle E_g$
and $\langle g ^\prime |g ^\prime  \rangle =1$.

As is well known the Berry connection
$ A_\mu = \langle g| \partial _\mu g \rangle $,
$ \mu=1,\cdots,D$
depends of the phase of the ground state as
\begin{align*}
 A_\mu &= \langle g| \partial _\mu g \rangle
 =A ^\prime  _\mu + i \partial _\mu \Theta,
\end{align*}
where
$A ^\prime _\mu = \langle g ^\prime | \partial _\mu g ^\prime  \rangle $ and
we assume
that the parameter dependence of $|g \rangle $ and $|g ^\prime  \rangle $
is smooth and differentiable.
The Berry phase is define as
\begin{align*}
 i \gamma _\ell &= \int_\ell dx_\mu A_\mu,
\end{align*}
which is gauge dependent (summation over $\mu $ is assumed).
To be specific, let us assume the closed path $\ell$ is parameterized by
$\theta \in[\theta _i,\theta _f]$
and take a different gauge when $\theta \in[\theta _1,\theta _2]$.
Then we have
\begin{align*}
  \gamma _\ell &=
 -i\int_{t_i}^{t_f} d \theta \, \dot x_\mu A_\mu,
 \quad \dot x_\mu =\frac {d x_\mu }{d \theta } 
 \\
 &= -i \big[ \int_{t_i}^{t_1} d \theta \, \dot x_\mu A_\mu
+ \int_{t_1}^{t_2} d \theta \, \dot x_\mu A_\mu
+ \int_{t_2}^{t_f} d \theta \, \dot x_\mu   A_\mu\big]
\\
&= \gamma _\ell ^\prime + \Delta \Theta,
\end{align*}
where
$\Delta \Theta = \Theta (x(\theta ))\big|^{\theta _2}_{\theta _1}$
and
\begin{align*} 
 \gamma ^\prime _\ell &= -i \big[\int_{t_i}^{t_1} d \theta \, \dot x_\mu A_\mu
+ \int_{t_1}^{t_2} d \theta \, \dot x_\mu A_\mu ^\prime 
+ \int_{t_2}^{t_f} d \theta \, \dot x_\mu   A_\mu\big].
\end{align*}
Since $\Delta \Theta $ is arbitrary, $\gamma_\ell$ does not have a definite
meaning unless one fixes the gauge globally.

The gauge is explicitly fixed by the scheme in Ref. \cite{Hatsugai04gaugefix}.
Let us start by taking an arbitrary state $|\phi \rangle $ as
a reference state.
Taking a constant $|\phi \rangle $ is simple but it may not
be necessarily constant but need to be single valued along the path $\ell$.
Then taking a gauge independent projection $P=| g \rangle \langle g|$,
the gauge fixing state by $\phi$ is given by
\begin{align*}
 | g_\phi \rangle &= P|\phi \rangle /\sqrt{N_\phi},
\end{align*}
where $N_\phi = \langle \phi|P|\phi \rangle =\eta_\phi^*\eta_\phi$ and 
$ \eta_\phi = \langle \phi| g \rangle$.
This gauge fixing is only allowed if $N_\phi=|\eta_\phi|^2\ne 0$.
Since $\eta_\phi\in\mathbb{C}$, this condition is always satisfied all over the 
(one-dimensional) loop $\ell$ by a suitable choice of $|\phi \rangle $
(if $\eta_\phi=0$ for $^\exists x\in \ell$, one may modify $\phi$ slightly).

Then by taking a different $|\phi_i \rangle $, $i=1,2$,
they are related with each other as
\begin{align*}
 |g_{\phi_1} \rangle &=  |g_{\phi_2} \rangle e^{i \Theta _{12}},
 \\
 e^{i \Theta _{12}}
 &=\sqrt{\frac {N_{\phi_2}}{N_{\phi_1}}}\frac {\langle g| \phi_1 \rangle }{\langle g| \phi_2 \rangle }
 = e^{i (\theta _1-\theta _2)},
\end{align*}
where ${\langle g| \phi_i \rangle }=|{\langle g| \phi_i \rangle }| e^{ i \theta _i}$, $i=1,2$.

Assuming that $| g _{\phi_1} \rangle $ and $| g _{\phi_2} \rangle $ are single valued on the loop $x\in\ell$,
Berry phases $\gamma _1$ and $\gamma _1$ are related as
\begin{align*}
 \gamma _1 &= \gamma _2 + \int_\ell d \Theta _{12} \equiv \gamma _2,\ \text{mod}\, 2\pi
\end{align*}
since $e^{i\Theta} $ is single valued over the loop
and $ \int_\ell d \Theta _{12}=
\int_{\theta _i}^{\theta _f} d \theta \, \dot x_\mu \partial _\mu \Theta (x)
=\int_{\theta _i}^{\theta _f} d \theta \, \frac {d }{d \theta } \Theta (x(\theta ))
=
\Theta (x(\theta ))\big|_{\theta _i}^{\theta _f} =2\pi n$, $n\in\mathbb{Z}$.

This ambiguity also clear from the discretized expression of the Berry phase
($\lim_{L\to\infty}\gamma _L=\gamma $)
\begin{align*}
  \gamma_L &\equiv
 \text{Arg}\,
 \langle g_{0}| g_{1} \rangle \cdots
 \cdots \langle g_n | g_{n+1} \rangle\cdots
 \langle g_{L-1}| g _{L} \rangle ,
\end{align*}
where
$\theta _n=\theta _i+\frac {n}{L} (\theta _f-\theta _i)$, $n=1,\cdots,L$ and
$| g_0 \rangle \equiv |g _L \rangle $.
 The expression is gauge invariant but the $\text{Arg}$ is well defined only in modulo $2\pi$.


\section{Dimer limit}
\label{sec:dimer}

Assuming $L$: even, let us consider a dimer limit
(1)$J_o<0,\ J_e=0$ or
(2)$J_o=0,\ J_e<0$ 
($o$: odd and $e$: even)
for the twisted Hamiltonian
Eq.(\ref{eq:dimerJ-twist}).
The Berry phase defined on the canonical path $\ell=\ell_{V_\alpha G V_{\alpha +1}}$, $\alpha =0,\cdots,Q$, 
$\gamma _Q=-i\int_\ell d \theta \, \langle g_{\tw}| \partial _ \theta g_{\tw} \rangle $
is quantized due to the $Z_\QN$ symmetry as discussed.

In the limit (1), it is trivially $\gamma _Q=0$
since the twist does not affect the
Hamiltonian.
Also inclusion of $J_e$ unless the finite gap closes, $\gamma _Q=0$ even for the
finite coupling case.
Only after the gap closing, the Berry phase may change.
This is a topological symmetry protection. 

As for the case (2), the Hamiltonian
 $H_\tw^{J_o=0}$ is decoupled for each dimers
and is written as {(See Eq.(\ref{eq:dimerJ-twist}))}
\begin{align*}
H_\tw^{J_o=0} &= J_e \sum_{\alpha, \beta }e^{{-}i(\varphi_\alpha -\varphi_\beta )}c_{L,\alpha } ^\dagger c_{1,\alpha } ^\dagger
 c_{1,\beta } c_{L,\beta } \\
 &\qquad\qquad
 + (\varphi_\alpha,\varphi_\beta \text{ independent terms}),
\end{align*}
where the last terms
 do not include $c_{L, \alpha }$. 

This is gauge out by the transformation
at the site $L$,  \cite{Hatsugai06qb,Hirano08Deg,Kariyado18}
\begin{align*}
 {\cal U}_L & =e^{{-}i \sum_\alpha \varphi_\alpha \hat n_{L,\alpha }},
\end{align*}
\begin{align*} 
{\cal U}_L c_{L,\alpha }{\cal U}_L ^\dagger &= e^{{+}i\varphi_\alpha }c_{L,\alpha },
\\
{\cal U}_L H^{J_o=0}_0 {\cal U}_L ^\dagger &= H_\tw^{J_o=0},
\end{align*}
where $H^{J_o=0}_0$ is without twist.

Then the ground state of the twisted Hamiltonian, $H_\tw$,
is given by $|g \rangle ={\cal U}_L | g _0 \rangle $ where $|g_0 \rangle $ is $\varphi_\alpha $ independent as
$H^0|g_0 \rangle = |g_0 \rangle E $.
The Berry phase is given by
\begin{align*}
 \gamma _Q &= -i\int_\ell d \theta \langle g| {\cal U}_L ^\dagger
 \partial _\theta {\cal U}_L |g \rangle 
 \\
 &={-} \int_\ell d \theta \partial _\theta \varphi_\alpha
 \langle g_0|\hat n_{L,\alpha} | g_0 \rangle
 = {-}\frac {1 }{\QN} \sum_\alpha\Delta \varphi_\alpha.
\end{align*}
since $ \langle g_0|\hat n_{L,\alpha} | g_0 \rangle = 1/\QN$ due to the $\SUQ$ invariance
of $H^0$.
Noting the discussion in Sec.\ref{sec:com},
 $\sum_\alpha \Delta \varphi_\alpha=
 {2\pi}(\QN-1)$ for a path
 $\ell_{V_0 G V_{1}}$ and
 $\sum_\alpha \Delta \varphi_\alpha=-2\pi $
 for a path, $\ell_{V_\beta  G V_{\beta +1}}$ ($\beta=2,\cdots,Q $),
\begin{align*}
 \gamma _Q &=  {+} \frac {2\pi } {\QN},\ \text{mod}\, 2\pi .
\end{align*}

\section{Symmetry of the paths}
\label{sec:s-ind}

Let us first discuss $N=3$ case to be simple.
The three paths are decomposed into $\ell_{V_jG}$ ($j=0,1,2$) and explicitly parameterized
by $\theta \in(0,\frac {2\pi} 3)$ as
\begin{align*}
 \ell_{V_0G} &= \{(\theta _1,\theta _2)=(\theta ,\theta )\},
 \\
 \ell_{V_1G} &= \{(\theta _1,\theta _2)=(2\pi-2 \theta ,\theta )\equiv (-2 \theta ,\theta ) \} ,
 \\
 \ell_{V_2G} &= \{(\theta _1,\theta _2)= (\theta ,2\pi-2 \theta )\equiv (\theta ,-2 \theta )\}.
\end{align*}
The modification associated with the twist is given by the
gauge transformation at the site $L$
\begin{align*}
 \bm{\theta } \in\ell_{V_0G} &\quad  (\varphi_1,\varphi_2,\varphi_3)=(\theta ,2\theta ,0 ),
 \\
\bm{\theta } \in\ell_{V_1G} &\quad  (\varphi_1,\varphi_2,\varphi_3)= (0 ,\theta ,2\theta ),
 \\
\bm{\theta } \in\ell_{V_2G} &\quad   (\varphi_1,\varphi_2,\varphi_3)=(2\theta, 0, \theta ),
\end{align*}
where ($\varphi_0=\varphi_3$)
\begin{align*}
 \theta_1 &= \varphi_1-\varphi_0
 \\
 \theta_2 &= \varphi_2-\varphi_1
 \\
 \theta_3 &= \varphi_0-\varphi_2.
\end{align*} 
Then $Z_3$ transformation of the Hamiltonian is as follows,
\begin{align*}
{\cal U}_{Z_3}  H(V_{0}G) {\cal U}_{Z_3} ^\dagger &= H(V_{1}G),\\
{\cal U}_{Z_3}  H(V_{1}G) {\cal U}_{Z_3} ^\dagger &= H(V_{2}G),\\ 
{\cal U}_{Z_3}  H(V_{2}G) {\cal U}_{Z_3} ^\dagger &= H(V_{0}G).
\end{align*}

In case of the generic $\SUQ$,
the $\QN$ paths are defined in a $d$-dimensional parameter space ($d=\QN-1$) and parameterized 
by $\theta \in(0,\frac {2\pi} \QN)$ as
\begin{align*}
 \ell_{V_0G} &= \{(\theta _1,\cdots,\theta _d)=(\theta ,\cdots,\theta )\},
 \\
 \ell_{V_jG} &= \{(\theta _1,\cdots,\theta _d)=(\theta ,\cdots,\overbrace{-(\QN-1) \theta}^{j},\cdots,\theta )\},\\ & \qquad\qquad j=1,\cdots,\QN-1=d
\end{align*}
\begin{align*}
 \bm{\theta } \in\ell_{V_0G} &\quad  (\varphi_1,\cdots,\varphi_N)=(\theta ,\cdots, \theta,-(\QN-1)\theta ),
 \\
 \bm{\theta } \in\ell_{V_jG} &\quad  (\varphi_1,\cdots,\varphi_\QN)=
(\theta ,\cdots,\overbrace{-(\QN-1) \theta}^{j},\cdots,\theta )\}.
\end{align*}
Again
$Z_\QN$ invariance, $c_{j,\alpha} \to c_{j,\alpha +1}$, of the Hamiltonian is written as
\begin{align}
 H(\theta ^\prime )
&=  {\cal U}^j_{Z_\QN}  H(\theta )({\cal U}_{Z_\QN}^j) ^\dagger,\ j=1,\cdots,\QN=d+1,
 \label{eq:transH}
 \end{align} 
 where $\theta \in\ell_{V_{0}G}$ and $\theta ^\prime \in\ell_{V_{j}G}$.
It implies a relation between the ground states
\begin{align*}
 |g (\theta') \rangle &= {\cal U}^j _{Z_\QN}| g (\theta ) \rangle.
\end{align*}

Note that at the vertices $V_j$'s and $G=(\frac {2\pi}{N},\cdots) $, the Hamiltonian is
invariant as
\begin{align}
 \cmt{H(V_j)}{{\cal U} _{Z_\QN}} &= 0,
\label{eq:invV} \\
 \cmt{H(G)}{{\cal U} _{Z_\QN}} &= 0.
\label{eq:invG} 
\end{align}

\section{Discontinuity of the center of mass: $\Delta P=\pm \frac 1 2 $.}
\label{sec:disc}
Noting that $N=\sum_j\hat \rho_j$,
$\hat\rho_j=(-1)^j\hat n_j$ commutes with the Hamiltonian as
\begin{align*}
 \cmt{H}{N_\rho} &= 0, \ \ N_\rho = \sum_j \hat \rho_j,
\end{align*}
let us assume that
a (generic) level crossing of the ground state $|g(t) \rangle $ at $t=t_i$
between $|g_ - \rangle $ and $|g_+ \rangle $ as
\begin{align*}
 \langle g_+| N |g_+ \rangle
 -
 \langle g_-| N |g_- \rangle
 &= 
 \sum_j(\rho_j^+-\rho_j^-)=\pm 1,
 \\
 \rho_j^\pm &=  \langle g_\pm| \hat \rho_j |g_\pm \rangle . 
\end{align*}
We may further
assume that this is due to
the edge state localized near $j\sim L$ with a localization length $\xi$
This can be justified by the low energy spectrum as discussed
in Sec.\ref{sec:spec-finite}.
Then it implies
\begin{align*}
\Delta \rho_j &=  \rho_j^+- \rho_j^-\to C e^{j/\xi },\ (L\to\infty).
\end{align*}
The normalization constant is evaluated as $\pm C ^{-1} =\sum_je^{j\xi ^{-1} }=e^{1/\xi }\frac {e^{L/\xi}-1}{e^{1/\xi}-1}= e^{L/\xi} {+{\cal O}(1) } $ and
\begin{align*}
 \Delta P &=
 \langle g_+| P |g_+ \rangle - \langle g_-| P |g_- \rangle
 \\
 &=\sum_j \frac {1}{L} (j-j_0)\Delta \rho_j
 =C L ^{-1} \sum_j (j-j_0)
 e^{j\xi ^{-1} }
 \\
 &\to C L ^{-1} (\frac {d }{d \xi ^{-1} } e^{L \xi ^{-1} }-j_0e^{L\xi ^{-1} })
 = \pm \frac 1 2 {+ {\cal O}(L ^{-1} ) },\ L\to\infty.
\end{align*}

\bibliographystyle{apsrev4-2}
%

\vfill
.\eject\vfill
\end{document}